\documentclass[twocolumn]{aastex631}

\newcommand{\revise}[1]{{\color{black}{#1}}}

\usepackage{graphicx}	% Including figure files
\usepackage{amsmath}	% Advanced maths commands

\def\arraystretch{0.9}

\begin{document}

\title[Neutron star LMXB outbursts]{Catalog of outbursts of neutron star LMXBs}
%% you desire. In v6.31 you can include a footnote in the title.

%% The \author command is the same as before except it now takes an optional
%% argument which is the 16 digit ORCID. The syntax is:
%% \author[xxxx-xxxx-xxxx-xxxx]{Author Name}
%%
%% Use \affiliation for affiliation information. The old \affil is now aliased
%% to \affiliation. AASTeX v6.31 will automatically index these in the header.
%% When a duplicate is found its index will be the same as its previous entry.
%%
%% Use \email to set provide email addresses. Each \email will appear on its
%% own line so you can put multiple email address in one \email call. A new
%% \correspondingauthor command is available in V6.31 to identify the
%% corresponding author of the manuscript. It is the author's responsibility
%% to make sure this name is also in the author list.
%%
%% While authors can be grouped inside the same \author and \affiliation
%% commands it is better to have a single author for each. This allows for
%% one to exploit all the new benefits and should make book-keeping easier.

%\correspondingauthor{August Muench}
%\email{greg.schwarz@aas.org, gus.muench@aas.org}

\author[0000-0003-3944-6109]{Craig O. Heinke}
\email{heinke@ualberta.ca}
\affiliation{Physics Dept., CCIS 4-183, University of Alberta, Edmonton, AB, T6G 2E1, Canada}

\author[0009-0001-2157-3866]{Junwen Zheng}
\affiliation{Dept. of Physics, 360 Huntington Avenue, Northeastern University, Boston, 02115-5000, USA}
\affiliation{Physics Dept., CCIS 4-183, University of Alberta, Edmonton, AB, T6G 2E1, Canada}

\author[0000-0003-0976-4755]{Thomas J. Maccarone}
\affiliation{Department of Physics and Astronomy, Texas Tech University, Lubbock TX 79409, USA}

\author[0000-0002-0092-3548]{Nathalie Degenaar}
\affiliation{Anton Pannekoek Institute for Astronomy, University of Amsterdam, Science Park 904, 1098 XH, Amsterdam, the Netherlands}

\author[0000-0003-2506-6041]{Arash Bahramian}
\affiliation{International Centre for Radio Astronomy Research, Curtin University, Bentley, WA 6102, Australia}

\author[0000-0001-6682-916X]{Gregory R. Sivakoff}
\affiliation{Physics Dept., CCIS 4-183, University of Alberta, Edmonton, AB, T6G 2E1, Canada}

\author{Simrat Toor}
\affiliation{Physics Dept., CCIS 4-183, University of Alberta, Edmonton, AB, T6G 2E1, Canada}

\begin{abstract}
Many X-ray binaries are transiently accreting. Having statistics on their recurrence times is helpful to address questions related to binary evolution and populations, as well as the physics of binary systems.
We compile a catalog of known outbursts of %86 
\revise{87}
transient neutron star (identified through bursts or pulsations) low-mass X-ray binaries, until %late 2023. 
\revise{mid-2025.}
Most outbursts are taken from the literature, but we also identify some outbursts from public X-ray monitoring lightcurves. 
We find 109 outbursts not previously identified in the literature; most are from the frequent transients GRS 1747-312, and the Rapid Burster MXB 1730-335, though we suspect that two outbursts from Liller 1 may be from another transient, besides the Rapid Burster.  
We also find new outbursts for 10 other systems, and verify substantial quiescent intervals for XMM J174457-2850.3, XMMU J174716.1-281048, and AX J1754.2-2754. 
Outburst detection has been relatively efficient since 1996 for outbursts above $F_X$(2-10 keV)$=3\times10^{-10}$ ergs \revise{cm$^{-2}$ s$^{-1}$}. 
While several systems have many known outbursts, 40 of the %86 
\revise{87} 
systems we track have zero or one recorded outburst between 1996 and 2023.
This suggests that many faint Galactic Center X-ray binaries may be neutron star X-ray binaries, though we cannot completely rule out the proposition that most neutron star X-ray binaries undergo frequent outbursts below all-sky monitor detection limits.
\end{abstract}


\section{Introduction} \label{sec:intro}

The total number of mass-transferring neutron star low-mass X-ray binaries in the Galaxy is thought to be of the order of $10^4$ \citep{PortegiesZwart97,Kiel06,Jonker11,vanHaaften15}. However, only a small fraction of these systems have been detected, typically during outbursts in which their X-ray luminosity increases by a factor of $10^3$--$10^7$ compared to quiescence. Understanding the full population of these systems, and their binary evolution, is helpful for us to understand the physics of binary systems \citep[e.g.][]{Podsiadlowski02,Postnov14,Van19}, and also to understand the products of their evolution, such as millisecond pulsars \citep{Alpar85,Verbunt93,Tauris99,Tauris12}, which may produce the excess gamma-rays seen from the Galactic bulge \citep{Wang05,Gonthier18}. 
One question of current interest is the nature of X-ray sources of intermediate luminosity ($L_X$(2-10 keV)$\sim10^{31}-10^{33}$ erg/s) detected near the Galactic Centre \citep{Muno05,Hailey18,Mori21,Maccarone22}.  The objects in this region that show no outbursts  have been argued to be mostly black hole LMXBs, based on an argument that neutron star LMXBs do not have long enough recurrence times.   Having detailed statistics on X-ray outbursts from neutron stars can be helpful in beginning to answer questions such as these.

Large numbers of transient X-ray binaries have been detected by a wide range of X-ray surveys. For many of these X-ray binaries, the nature of the accreting object is not confidently known. However, many transient X-ray binaries have secure identifications of the accretor as neutron stars, typically through the identification of thermonuclear X-ray bursts from the neutron star surface \citep{Lewin93}, or through detection of pulsation from the neutron star spin \citep{Wijnands98}. Verification of the opposite case, that the accretor is a black hole, is harder, requiring optical or infrared spectroscopy to measure, through radial velocities, the mass of the unseen companion, and show it is above the generally accepted maximum neutron star mass of 3 M$_{\odot}$ \citep{Casares14}. 

A number of catalogs of X-ray binaries have been produced over the years. The catalogs of \citet{Liu_vanParadijs2001,Liu_vanParadijs2007} have been particularly well-cited, but are now 17 years old.
\citet{RitterKolb03} have an extensive and updated catalog of X-ray binaries and cataclysmic variables with known periods, but do not include objects without known (or suggested) orbital periods. 
More recent catalogs have been published of black hole X-ray binaries, one focused on compiling detailed lightcurves \citep{Tetarenko16}, and one focusing on multiwavelength studies \citep{CorralSantana16}. We are interested in the particular problem of understanding the frequency of X-ray outbursts from neutron star LMXBs, but found that we lacked a catalog to answer this question.  The catalogs of \citealt{Avakyan23} and \citet{Fortin24} were released when this work was largely complete, and while they are very useful, they also do not list outbursts. \citet{YanYu15} and \citet{Lin19} construct tables of outburst frequency for a number of X-ray binaries with known orbital periods, but these works only include 14 and 19 NS LMXBs respectively. \citet{YanYu15} do include a detailed listing of the outbursts they include, which is helpful. We identify several more outbursts from 4U 1608-52, XTE J1709-267, MXB 1730-33, XTE J1739-285, 1A 1744–361, SAX J1748.9-2021, and Aquila X-1 in the time range these authors cover, typically fainter outbursts. 

\S~2 gives details of the construction of our catalog. In \S~3, we make a first use of this catalog to address the question of the nature of intermediate luminosity X-ray sources near the Galactic Centre. 

\section{Catalog}

To construct such a catalog, we used two clear methods to verify the presence of a NS in an LMXB; presence of thermonuclear bursts, and/or presence of coherent millisecond pulsations.  
We took lists of confirmed neutron star LMXBs from the MINBAR X-ray burst catalog \citep{Galloway2020}, which identified 115 known X-ray bursters using data from BeppoSAX, RXTE, and INTEGRAL (1996-2012), plus the literature. 

We omit %42 
\revise{41}
NS LMXBs in the MINBAR catalog that appear to be persistent.
\revise{Identifying whether a NS LMXB is persistent can be tricky. Seven known NS transients have had outbursts lasting over a decade (EXO 0748-676, KS 1731-260, XB 1733-30, IGR J17062-6143, HETE J1900.1-2455, 1H 1905+000, and 4U 2129+47), and it is possible that systems we list as persistent will turn off in the future.  Other systems are persistently active at a low level, then brighten substantially for weeks or months (e.g. 4U 1705-32, \citealt{Negoro21}; \citealt{intZand07}). 
UW CrB (1E 1603.6+2600) at a distance of $4.6^{+1.1}_{-2.5}$ kpc \citep[Gaia parallax,][]{BailerJones21} appears unusually faint for an actively accreting system, at $L_X\sim3\times10^{33}$ erg s$^{-1}$, but this is thought to be due to extreme obscuration of its intrinsic X-ray luminosity \citep[see e.g.][]{Hakala05}.  }
We list the \revise{systems we consider } persistent, verified bursting \revise{or pulsing} NS XRBs in 
%Table~\ref{tab:persistent}
Table 10, in Appendix D.

We omit a few transient bursters; the position of 
XB 1940-04 \citep{Murakami83} is very poorly known. 
The eccentric X-ray binary Cir X-1 (or 3A 1516-569) has outbursts that typically reflect its orbital period, not disk outbursts \citep{Murdin80}. During the period from MJD=55000--58500, it was unusually under-luminous, and showed occasional outbursts that were not at exact multiples of its orbital period, suggesting outbursts from a disk. We catalog those outbursts in Appendix B, but due to Cir X-1's differences in accretion mode compared to other LMXBs, we omit it from most of our compiled analyses. 
%AX J1824.5-2451 was a burst from the globular cluster M28 in quiescence, which we address along with IGR J18245-2452. 
We exclude 1A 1742-289 (or A1742-289), as it is not clear that the burst activity seen from MXB 1743-29 in the Galactic Center in 1976 \citep{Lewin76_GalCenter} can be attributed to this radio-bright \citep{Davies76} source (we note that this is definitely not the same object as AX J1745.6-2901; \citealt{Kennea96}).
This leaves 70 sources from the MINBAR catalog (71 when including Cir X-1).

We add to the MINBAR list the known recent bursters\footnote{as updated by Jean in 't Zand; %\url{https://personal.sron.nl/~jeanz/bursterlist.html}}
\url{https://burst.sci.monash.edu/wiki/index.php?n=MINBAR.SourceTable}}
4U 1730-22, 
MAXI J1816-195, 
MAXI J0556-332, 
 MAXI J0911-655, 
Terzan 6 X-2 (see Appendix \S~\ref{frequent}), 
and SRGA J144459.2-604207
\footnote{We include SRGA J144459.2-604207 in our list for completeness, but %otherwise our summary of outbursts stops in
\revise{our statistical calculations include data only through }
 2023.}.

We also add known transient accreting millisecond pulsars  that are not known bursters, from the \citet{PatrunoWatts} catalog and more recent literature, 
XTE J0929-314,
NGC 6440 X-2,
XTE J1751-305,
Swift J1756.9-2508,
XTE J1807-294,
IGR J17494-3030,
IGR J16597-3704,
and MAXI J1957+032. 
%We add, to our list of transients,  nine transient accreting millisecond pulsars without known bursts (XTE J1751-305, XTE J0929-314, XTE J1807-294, Swift J1756.9-2508, NGC 6440 X-2, MAXI J1816-195--now burster, MAXI J1957+032, IGR J17494-3030, and IGR J16597-3704 in NGC 6256).
Finally, we include the transient slow (2.14 Hz) accreting pulsar GRO J1744-28. (We also list the two persistent LMXBs that are pulsars in %Table~\ref{tab:persistent}.)
Table~10.)
This gives us a total of %85 
\revise{86}
transient neutron star low-mass X-ray binaries, %86
\revise{87} 
if including Cir X-1. 

We performed literature searches for this list of transients, and identify the known X-ray outbursts from each system. For each outburst, we attempt to determine the peak observed 2-10 keV X-ray luminosity. This will have some significant uncertainty, since not every outburst is caught at its peak, some outbursts are  observed only in other energy ranges, and many distances are not well-known. We attempt to use the best-fit spectrum to convert measurements to 2-10 keV fluxes, but we only attempt this with simple power-law spectra (we do not try to convert complex Comptonized models). Where information is lacking, we use the best literature $N_H$ estimate and an assumed photon index of 2.1 to make conversions, or use literature estimates of the 2-10 keV flux in mCrab (assuming 1 Crab = $2\times10^{-8}$ ergs \revise{cm$^{-2}$ s$^{-1}$} ). 
A few bursters have recorded bursts without any recorded outbursts \citep{Gotthelf97,Cornelisse02}; we list these in Table~\ref{tab:no_outbursts} in Appendix C.
%, as bursts were recorded without an accompanying outburst; these systems are AX J1824.5-2451 \citep[][observed with ASCA from M28]{Gotthelf97}, and three "burst-only"  systems discovered by BeppoSAX \citep{Cornelisse02} (see Table~\ref{tab:no_outbursts}). 

%In AASTeX v6.31 all deluxetables are float tables and thus if they are longer than a page will spill off the bottom. Long deluxetables should begin with the {\tt\string\startlongtable} command. This initiates a longtable environment.  Authors might have to use {\tt\string\clearpage} to isolate a long table or optimally place it within the surrounding text.

%Tables longer than 250 data lines and complex tables should only have a short example table with the full data set available in the machine readable format.  The machine readable table will be available in the HTML version of the article with just a short example in the PDF. Authors are required to indicate in the table comments that the data in machine  readable format in the full article. Authors are encouraged to create their own machine readable tables using the online tool at \url{http://authortools.aas.org/MRT/upload.html} but the data editors will review and edit all submissions prior to publication.

%Full details on how to create the different types of tables are given in the AASTeX guidelines at \url{http://journals.aas.org/authors/aastex.html}

We present the catalog of known outbursts of neutron star LMXBs in Table \ref{tab:outbursts}, for which we provide a sample here.   
The full table is in %the electronic version of the journal 
\revise{(Table~\ref{tab:outbursts_app}) in Appendix A)}. \footnote{We also maintain an updated version at \url{https://tinyurl.com/5euncwmh}.} 
We include a line for each recorded outburst, with a literature reference for each outburst if available. Several objects with repeated bright outbursts can be easily spotted in the public lightcurves of the RXTE PCA Bulge Scan  \citep{Swank01}\footnote{\url{https://asd.gsfc.nasa.gov/Craig.Markwardt//galscan/main.html}}, the RXTE All-Sky Monitor \citep{Levine96}\footnote{\url{https://heasarc.gsfc.nasa.gov/docs/xte/asm_products.html}} %comments at \url{http://xte.mit.edu/XTE/xte\_anno.html}} 
and/or in MAXI GSC observations \citep{Matsuoka09,Suguzaki11}\footnote{\url{http://maxi.riken.jp/top/index.html}}. We have added some obvious outbursts visible in these surveys to the list (e.g. outbursts from XTE J1709-267 in 2005, 2009, 2018/19, and 2022). 
The line for defining an `outburst' can be unclear; for simplicity we require an increase of a factor of 10 from the quiescent flux (if known), and $L_X>10^{34}$ erg/s. 

We separate the objects with many outbursts into separate tables, placed in Appendix B. For the most frequently repeating objects with bright outbursts (of order once per year or more), 4U 1608-52, MXB 1730-33 (the Rapid Burster in Liller 1), Aquila X-1, GRS 1747-312 (the eclipser in Terzan 6), and Cir X-1, we use the three public lightcurves above, as well as Swift/XRT public lightcurves  \citep{Evans07}\footnote{\url{https://www.swift.ac.uk/user\_objects/}}, and the INTEGRAL JEM-X Galactic Bulge Monitoring lightcurves \citep{Kuulkers07}\footnote{\url{http://integral.esac.esa.int/BULGE/}} to develop the best possible list of outbursts from 1996 to 2023, along with literature identification of outbursts before 1996. For the objects with the most frequent outbursts, we also record the MJD date of the outburst peak (as best we can tell).  We place the outbursts of NGC 6440 X-2 in a separate table in Appendix B as well, as their high frequency requires MJD dates to differentiate them.

 %We also add the two known transitional millisecond pulsars without major outbursts (PSR J1023+0038 and XSS J12270-4859); although they have experienced no outbursts, they are low-mass X-ray binaries that show X-ray (and radio) pulsations proving they contain neutron stars \citep{Archibald15,Papitto15}. --skip
%These frequent outbursters are 4U 1608-52, 
%XTE J1709-267, the Rapid Burster (MXB 1730-335, in Liller 1), 
%GRS 1741.9-2853, GRS 1747-312 (in Terzan 6), and Aquila X-1.
%NGC 6440 X-2, 
%and SAX J1808.4-3658.  

When identifying additional outbursts, we generally search for multiple points that are above 3 sigma. The MAXI baseline may often vary substantially, so we additionally impose that the potential outburst must rise significantly above the quiescent baseline. 

We find additional outbursts, not mentioned in the literature, for 12 systems; XTE J1709-267 (4 outbursts), XTE J1701-407 (1), XTE J1739-285 (2), 1A 1744-361 (2), XMMU J174716.1-281048 (1), AX J1754.2-2754 (3), SAX J1810.8-2609 (1), 4U 1608-52 (3), MXB 1730-335 (aka the Rapid Burster, 52), GRS 1747-312 (37), NGC 6440 X-2 (2), and Aquila X-1 (1), for a total of 109 newly identified outbursts. We also identify 42 additional outbursts from Cir X-1 not mentioned in the literature between 2006 and 2018, though we list these separately as their nature is somewhat ambiguous.

We give comments on each X-ray binary's outburst history in Appendices A and B, including the reference we use for the distance, and any discrepancies we identify in the literature regarding their outburst histories.  Here we highlight a few key points;

$\bullet$ XMM J174457-2850.3, which showed frequent low-level outbursting behavior between 2001 and 2012, has seen its activity subside, with only one short outburst since 2012.

$\bullet$ We use {\it Chandra}, {\it XMM-Newton}, and Swift/XRT observations to verify that XMMU J174716.1-281048 declined from a $\sim$10-year-long outburst in 2012, reached full quiescence by 2018, showed another (short, $<$3 months) outburst in 2021, and then returned again to quiescence.

$\bullet$ We use Swift/XRT and {\it Chandra} observations to show that AX J1754.2-2754 has had $L_X$(2-10 keV)$<2\times10^{33}$ erg/s for much of the past eight years, in contrast with its quasi-persistent behavior in 1999-2017 \citep{Bassa08_AXJ1754}.
%earlier. 

$\bullet$ SAX J1806.5-2215's long outburst, which started in Feb. 2011, ended sometime between July 2017 and January 2018.

$\bullet$ We catalog 52 outbursts from MXB 1730-335 (aka the Rapid Burster) that have not previously been identified in the literature. Some of these may not be produced by the Rapid Burster, but by other LMXBs in the host cluster Liller 1; another such outbursting transient has been verified \citep{Homan18}, and we do identify two unusually short intervals (28 and 37 days) between outbursts that are good candidates for other transients. However, the relatively regular outbursts, and detection of Type II bursts in most cases where instruments are sensitive to them, indicate most outbursts are due to the Rapid Burster. \citet{Masetti02} noted that the average outburst interval decreased from 200 days to 100 days around 2000, with the peak flux also dropping by a factor of 2. Since 2000, the average outburst interval has varied between 100 and 170 days, and the peak $L_X$ has remained around $2\times10^{37}$ erg/s.

$\bullet$ We catalog 37 previously unidentified outbursts from GRS 1747-312 (in the globular cluster Terzan 6). The outburst interval averages 138 days, with significant scatter, but regular enough to indicate that most outbursts are indeed due to a single source. 
A second transient bursting source, Terzan 6 X-2, has recently been identified in Terzan 6 \citep{Saji16,vandenBerg24}, but appears not to reach such high fluxes as GRS 1747-312.
%However, one outburst in particular (in Sept. 2009) was unusual in appearing to lie in the middle of an interval; in its high peak flux; in its rapid flux decline; and in lacking an eclipse at the predicted orbital ephemeris \citep{Saji16}, all of which indicate that this outburst was likely produced by a second transient LMXB in this cluster. --indeed, produced by another LMXB outside cluster.

In \S 3 we summarize some useful information on monitoring instruments, use the information in these tables to draw some general conclusions about our sensitivity to X-ray outbursts, and apply this information to consideration of interesting X-ray sources near the Galactic Center.

%\onecolumn

% 
\begin{deluxetable*}{lcccccccr}
%	\centering
%	\begin{tabular} %
\tablehead{		\colhead{NS XRB} & \colhead{Begin} & \colhead{End} & \colhead{Peak $F_X$}  & \colhead{$N_H$} & \colhead{Index} & \colhead{$L_X$} & \colhead{d}  &  \colhead{References}\\ 
% \tablehead{ 
\colhead{}   & \colhead{} & \colhead{} & \colhead{2-10 keV} & \colhead{cm$^{-2}$} & \colhead{} & \colhead{2-10 keV} & \colhead{kpc}  & \colhead{} } 
\startdata
		IGR J00291+5934 & 1998/11 & 1998/11 & 4.5e-10 &  - & -& 9.6e35 & 4.2  &  \citet{Remillard04} \\
		IGR J00291+5934 & 2001/9 & 2001/9 & 3.9e-10 &  - & - & 8.2e35 & 4.2  & \citet{Remillard04} \\
		IGR J00291+5934 & 2004/12 & 2004/12 & 6.1e-10  & 7e21 & 1.7 & 1.3e36 & 4.2  & \citet{Eckert04,Galloway2005}\\
		IGR J00291+5934 & 2008/8 & 2008/10 & 6.1e-10  & - & 1.7 & 1.3e36 & 4.2  &  \citet{Chakrabarty08,Lewis10} \\
		IGR J00291+5934 & 2015/8 & 2015/8 & 5.6e-10  & 1e22 & 1.7 & 1.2e36 & 4.2  & \citet{Cummings15,deFalco17} \\
  \hline 
        MAXI J0556-332 & 2011/1 & 2012/5 & 1.9e-09 & - & - & 4.4e38 & 43.6 & \citet{Matsumura11,Sugizaki13} \\
        MAXI J0556-332 & 2012/10  & 2012/12 & 4e-10 & 3.2e20 &  - &  9e37 & 43.6 &  \citet{Sugizaki12,Parikh17} \\
        MAXI J0556-332 & 2016/1 & 2016/3 & 7.7e-10 & - & - & 1.8e38 & 43.6 &  \citet{Negoro16,Russell16} \\
        MAXI J0556-332 & 2020/4 & 2020/11 & 8.6e-10 & - & - & 2.0e38 & 43.6 & \citet{Negoro20_J0556,Page22}\\
\hline
        EXO 0748-676    & 1984/7 & 2008/8 & 1.1e-9  & 1.2e21 & 1.9 & 7e36 & 7.4  & \citet{Parmar86,Degenaar09b} \\ 
%\hline 
%        GS 0836-429 & 1970/12 & 1972/1 & 1.1e-9 & $<$3.5e22 & & 1.1e37 & $<$9.2 &  \citet{Markert77} \\
%        GS 0836-429 & 1990/10 & 1991/5 & 3.8e-10 & 2.2e22 & 1.5 & 3.9e36 & $<$9.2 &  \citet{Aoki92,Belloni93} \\
%        GS 0836-429 & 2003/1 & 2003/5 & 1.0e-9 & 2.7e22 & 1.5 & 1.0e37 & $<$9.2  &   \citet{Aranzana16} \\
%        GS 0836-429 & 2003/9 & 2004/6 & 1.3e-9 & 2.7e22 & 1.5 & 1.3e37 & $<$9.2  &   \citet{Aranzana16} \\
%\hline
%        MAXI J0911-655 & 2016/2 & 2023/10 & 1.7e-10 & 2.7e21  & 1.27 & 1.9e36 & 9.5 &  \citet{Serino16_0911,Sanna17} \\
%\hline
%        XTE J0929-314 & 2002/4 & 2002/4 & 6.7e-10 & 7.6e20 & 1.7-2  & $>$2.9e36 & $>$6 &  \citet{Galloway02,Juett03}\\
%\hline
%        SAX J1324.5-6313 & 2015/8 & 2015/8 & 1.3e-10 & 1.5e22 & - & 1.0e36 & $<$6.2 &   \citet{Negoro15,Cornelisse02_chandra} \\
%\hline
%        MAXI J1421-613 & 2014/01 & 2014/1 & 1.9e-9 & 4.8e22 & 2.1 & 2e36  & 3 &   \citet{Serino15,Nobukawa20} \\
%\hline
%        SRGA J144459.2-604207 & 2022/1 & 2022/1 & 4e-10 & - & - & 3.4e36 & 8.5 & \citet{Negoro24} \\
%        SRGA J144459.2-604207 & 2023/12 & 2023/12 & 6e-10 & - & - & 5.2e36 & 8.5 & \citet{Negoro24,Sguera24} \\
%        SRGA J144459.2-604207 & 2024/2 & 2024/3 & 2.7e-9 & 2.9e22 & 1.8 & 2.3e37 & 8.5 & \citet{Molkov24,Ng2024}\\
%\hline
%        Cen X-4 & 1969/07 & 1969/7 & 6e-7 & - & - &   1.2e38 & 1.3 &  \citet{Evans70,Kuulkers09} \\
%        Cen X-4 & 1979/05 & 1979/5 & 9.6e-8 & 9e20 & - &  1.9e37 & 1.3 &  \citet{Kaluzienski80,Kuulkers09}\\
\enddata
%\end{tabular}
\caption{Published outbursts of verified neutron star LMXBs. Max $F_X$ is highest reported flux in each outburst, transformed to 2-10 keV, generally using the power-law spectral index and $N_H$ noted. We assume 1 Crab is $F_X$(2-10 keV)=$2\times10^{-8}$ erg cm$^{-2}$ s$^{-1}$. Peak observed intrinsic $L_X$ in 2-10 keV, for the assumed distance (listed). Reference to an instrument (e.g. MAXI) indicates we have made use of that instrument's archive to generate some information for that line. The full Table is \revise{Table~\ref{tab:outbursts_app}} in Appendix A; %(for arXiv), or online (for journal); 
a portion is reproduced here.}
\label{tab:outbursts}
\end{deluxetable*}

\section{Discussion}

\subsection{Monitoring instruments}

 We first review key information about some relevant monitoring instruments.

The first effective all-sky X-ray monitor was a scintillation 3-12 keV X-ray detector on Vela 5B, which flew from 1969-1979. Its limited collimation and thus high background limited its sensitivity to of order 0.3 Crabs \citep{PriedhorskyHolt87}. Ariel V (1974-1980) had a slightly more sensitive (perhaps 0.2 Crabs) 3-6 keV all-sky monitor, operated as a pinhole camera \citep{Holt76}.  In 1987-1991, Ginga's ASM operated from 1-20 keV with sensitivity down to 50 mCrab  \citep{Tsunemi89}. 

RXTE/ASM monitoring produced a dramatic increase in detected outbursts, from January 1996 until 2009-2011 \citep{Levine96}\footnote{The RXTE/ASM sensitivity decayed from 2009 until RXTE was turned off in 2011.}. RXTE/ASM achieved roughly 15 mCrab ($F_X=3\times10^{-10}$ ergs \revise{cm$^{-2}$ s$^{-1}$}, 3-sigma) sensitivity per day, in 2-10 keV, for known sources. 
The RXTE/ASM team could identify new sources as faint as 50 mCrab with one day's data, while integrating for a week could delve down to 7 mCrab, in uncrowded regions \citep{Bradt01}. 
Indeed, detections from 1996 on did reach $3\times10^{-10}$, though 
%may have been a little higher, perhaps 
 the completeness limit was higher, perhaps
$10^{-9}$ ergs \revise{cm$^{-2}$ s$^{-1}$}, or 50 mCrab. 
The RXTE/ASM covered most sources daily, except for about 30 days each year when the Sun was nearest for sources in the ecliptic. 

From 1996 to 2002, in each spring and fall, BeppoSAX conducted regular monitoring of the Galactic Bulge with its Wide-Field Camera (WFC), with a 40$^o$ by 40$^o$ field of view, 5' angular resolution, and sensitivity of about 15 mCrab in 2-28 keV \citep{Heise99,intZand04_review}. Monitoring of the Galactic Center was performed weekly for two months in the spring and fall \citep{Heise99}.  
\citet{Jager97} estimated the typical observation sensitivity as 10 mCrab at high galactic latitudes, and 30 mCrab at the Galactic Center. 
BeppoSAX's WFC discovered a large number of faint ($L_X<10^{37}$ erg/s) X-ray transients in the Galactic Bulge, and also several X-ray bursts from transients that were too faint for the WFC to detect individually \citep{Cornelisse02}.

From 1999 to 2011, RXTE also conducted monitoring of the Galactic Bulge using slews of the PCA instrument \citep{Swank01}.  These Bulge scans could detect new uncrowded sources down to 0.5 mCrab \citep{Swank01}, but in practice, in this crowded region, they reached a sensitivity of, at best, 5 mCrab, or $10^{-10}$ ergs \revise{cm$^{-2}$ s$^{-1}$} in 2.5-10 keV. Notably, the RXTE/PCA was able to detect several faint ($\sim$5 mCrab) outbursts of NGC 6440 X-2 in 2009-2011  \citep{Heinke10_6440paper}, but these were only barely visible in the Bulge Scan lightcurves.  The PCA Bulge scans initially covered about 250 square degrees twice weekly \citep{Markwardt02}, extended to 500 square degrees in 2003 \citep{Markwardt06b}, and for 2009-2011 added the Aquila region, to cover over 700 square degrees \citep{Markwardt10}. These scans were paused for the 2 months when the Sun is closest to the Galactic Center \citep{Swank01}. 

The INTEGRAL satellite (launched 2002, \revise{decommissioned in Feb.\ 2025}) has two relevant coded-aperture instruments: IBIS \citep{Ubertini03} has peak sensitivity in the 20-60 keV range with a 20 degree field of view (at half sensitivity) and 12' angular resolution, while JEM-X \citep{Lund03} observes in 3-35 keV, with a 7.5 degree field of view and 3.4' angular resolution. 
The INTEGRAL Galactic Bulge hard X-ray monitoring program has been  particularly productive, with 5-15 mCrab %daily 
sensitivity
per 1800 s observation  
 (every three days), and summed detections reaching down to 2.4 mCrab \citep{Kuulkers07}. INTEGRAL could observe the Bulge only %40\% 
1/3 
of the year, for 2 months in each spring and fall. %Twelve 
\revise{Thirteen }
of our transients have principal names starting with IGR (only XTE %has more, with 13), 
\revise{ matches this number),} 
suggesting INTEGRAL's impact. There appears (Fig.~\ref{fig:per_year}) to be a small decline in detections in the past few years, which may be due to declining sensitivity of INTEGRAL's IBIS/ISGRI instrument.\footnote{The Integral Galactic Bulge Monitoring Program's website, \url{http://integral.esac.esa.int/BULGE/}, presents 18-40 keV IBIS/ISGRI lightcurves for sources, which show increasing noise from 2016 and stop in 2018.}

Swift/BAT's hard X-ray (15-50 keV) monitoring program  has operated since 2006, with a daily 16 mCrab detection limit \citep{Krimm13}.
Swift's XRT instrument (narrow-field imaging in 0.3-10 keV) has also been vital in tracking transients, due to its flexible scheduling, ease of ToO requests, good sensitivity, and simplicity of data analysis \citep[e.g.][]{Evans07}.  
% Seven Swift-named sources found by BAT, one by Swift/XRT monitoring.

The MAXI \revise{instrument (on the International Space Station)} has two camera groups, of which the Gas Slit Cameras, or MAXI/GSC \citep{Mihara11,Sugizaki11} has been significantly more productive. MAXI/GSC has a field of view of 1.5$^o$ by 160$^o$, and monitors $\sim$85\% of the sky every 92 minutes, reaching 95\% of the sky each day \citep{Negoro16_survey}.  
MAXI/GSC  
has monitored the sky since 2009 daily in 2-20 keV, with the best sensitivity in \revise{3-10 keV. 
Typical sensitivities for 4-day integrations reach 7 mCrab \citep{Negoro16_survey}.}  Positions of isolated sources are typically determined to 0.1$^o$,  
%However, as individual photons have an angular precision of 1.5 degrees (in the scan direction), 
\revise{but} MAXI/GSC's power to resolve the Galactic Center is limited.  MAXI's efficient and automated nova search algorithms \citep{Negoro16_survey} have been highly productive, identifying transients which can then be followed up by other missions to get precise positions. 
%As MAXI is most sensitive in 2-10 keV, while INTEGRAL/ISGRI has peak sensitivity at 20-40 keV, it may not be a coincidence that the peak number of detected outbursts occurred in 2011, when both were active.

 The SRG spacecraft scans the sky twice per year, and contains two instruments: ART-XC in 4-30 keV, and eROSITA in 0.2-8 keV \citep{Sunyaev21}. 
eROSITA completed 4 of 8 planned all-sky surveys (2019-2022), with sensitivity per survey (200 s exposure) reaching $F_X=10^{-13}$ ergs \revise{cm$^{-2}$ s$^{-1}$} in 2-4 keV \citep{Sunyaev21}, before the Ukraine war terminated Russian/German cooperation. 
ART-XC continues operating, and reaches 1 mCrab sensitivity in 4-12 keV each survey sweep \citep{Sunyaev21}. ART-XC detected a new X-ray binary, SRGA J144459.2-604207 \citep{Molkov24}, during the preparation of this catalog.

Two new instruments capable of transient LMXB identifications began surveying the sky within the last year.
The Einstein Probe is a Chinese soft X-ray mission launched on January 9, 2024. Its Wide-Field X-ray Telescope (WXT) monitors \revise{the entire night sky} on a daily basis, using lobster-eye optics to achieve a 3600 square degree instantaneous field of view, a spatial resolution of 5', and a nominal 0.5-4 keV sensitivity of 0.8 mCrab in 1000 s \citep{Yuan22}. 
The Follow-up X-ray Telescope (FXT) has 30" angular resolution, and sensitivity similar to Swift/XRT, reaching $F_X$(0.5-2 keV)$=10^{-14}$ erg/s in 1500 s \citep{Yuan22}.

Finally, the Space Variable Objects Monitor (SVOM) is a joint Chinese/French mission launched on 22 June 2024, with a science focus on gamma-ray bursts. Its wide-field hard X-ray (4-150 keV) detector, ECLAIRS, has a sensitivity of 50 mCrab per orbit, with a PSF of 52' \citep{Godet14}. It also hosts a narrow-field soft X-ray (0.2-10 keV) detector, MXT, with angular resolution of 4.5' and a sensitivity reaching to $F_X$(0.2-10 keV)$10^{-12}$ with 10 ks of integration. SVOM's mission plan of generally pointing away from the Galactic Plane will tilt its science toward GRBs, but it is likely to also detect LMXB outbursts \citep{Wei16}.

Both INTEGRAL/IBIS and Swift/BAT are sensitive to harder X-rays than the 2-10 keV range we focus on here. Thus, they will be more likely to detect LMXBs in hard states, vs. soft states at the same 2-10 keV flux.  The ratio of Crab-normalized Swift/BAT 15-50 keV vs.\ MAXI/GSC 2-10 (or 4-10) keV transient NS LMXB fluxes is about 1 in the hard state, and often 0.1 in the soft state (e.g. \citealt{Asai15}, \citealt{Bahramian14}; cf. \citealt{Tetarenko16} for BH LMXBs, which can reach lower ratios in the soft state). 
Most transient NS LMXBs observed near the limit of all-sky monitor detection ($\sim$5 mCrab) at distances of 10 kpc (thus, $L_X$(2-10)$\sim10^{36}$ erg/s) are found to be in the hard spectral state \citep{Maccarone03,Asai15}. %with spectra representable by a power-law with photon index typically 1.7-2 \citep{Wijnands15}, with a cutoff around 20 keV \citep{Church14}. 
However, INTEGRAL/IBIS and Swift/BAT would have a 2-10 keV $L_X$ detection threshold roughly 10 times higher, for soft-state transients.

%Fig.~\ref{fig:all_outbursts} clearly shows that since 1996, detection of outbursts above $F_X$(2-10)=$3\times10^{-10}$ erg cm$^{-2}$ s$^{-1}$ appear to be quite regular, perhaps reaching down to $1\times10^{-10}$ erg cm$^{-2}$ s$^{-1}$ since 2008 or so. For a typical distance of 8 kpc, this gives a completeness limit of $L_X$(2-10)=$1.5\times10^{36}$ erg/s. 
%Thus, we can see that our outburst history is far more thorough since 1996, suggesting selecting this time range as key for major analyses.

\subsection{General trends}
\begin{figure}
    \centering
    \includegraphics[width=\columnwidth]{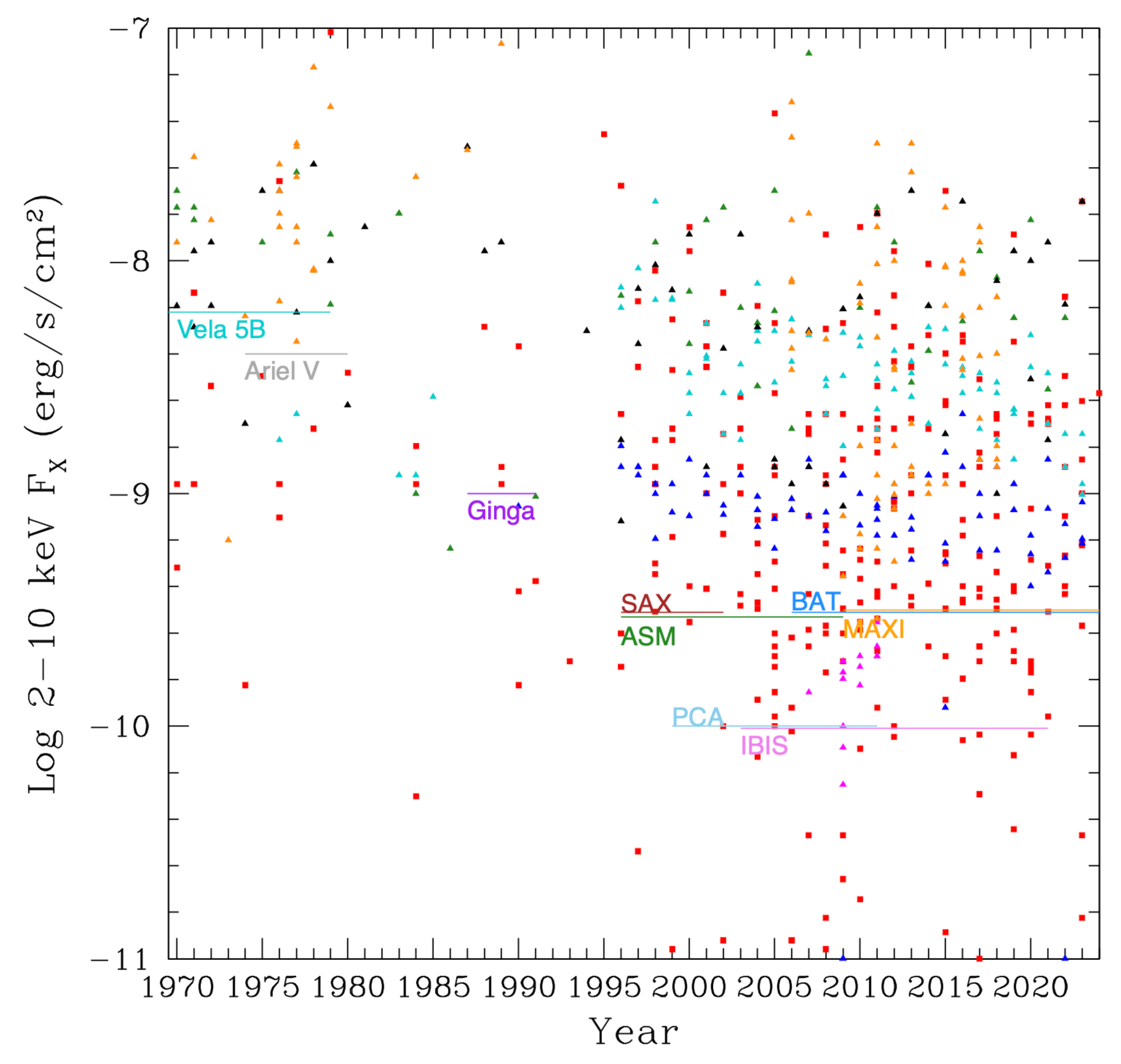}
 \caption{Peak fluxes by year from transient neutron star LMXBs. 
Green, cyan, blue, magenta, orange, and black triangles indicate outbursts by 4U 1608-52, MXB 1730-335, GRS 1747-312, NGC 6440 X-2, Cir X-1, and Aquila X-1. 
 Red squares indicate all other transient LMXB outbursts. Horizontal colored lines indicate rough sensitivity limits for various X-ray monitoring instruments. 
 }
    \label{fig:all_outbursts}
\end{figure}

\begin{figure}
    \centering
    \includegraphics[width=\columnwidth]{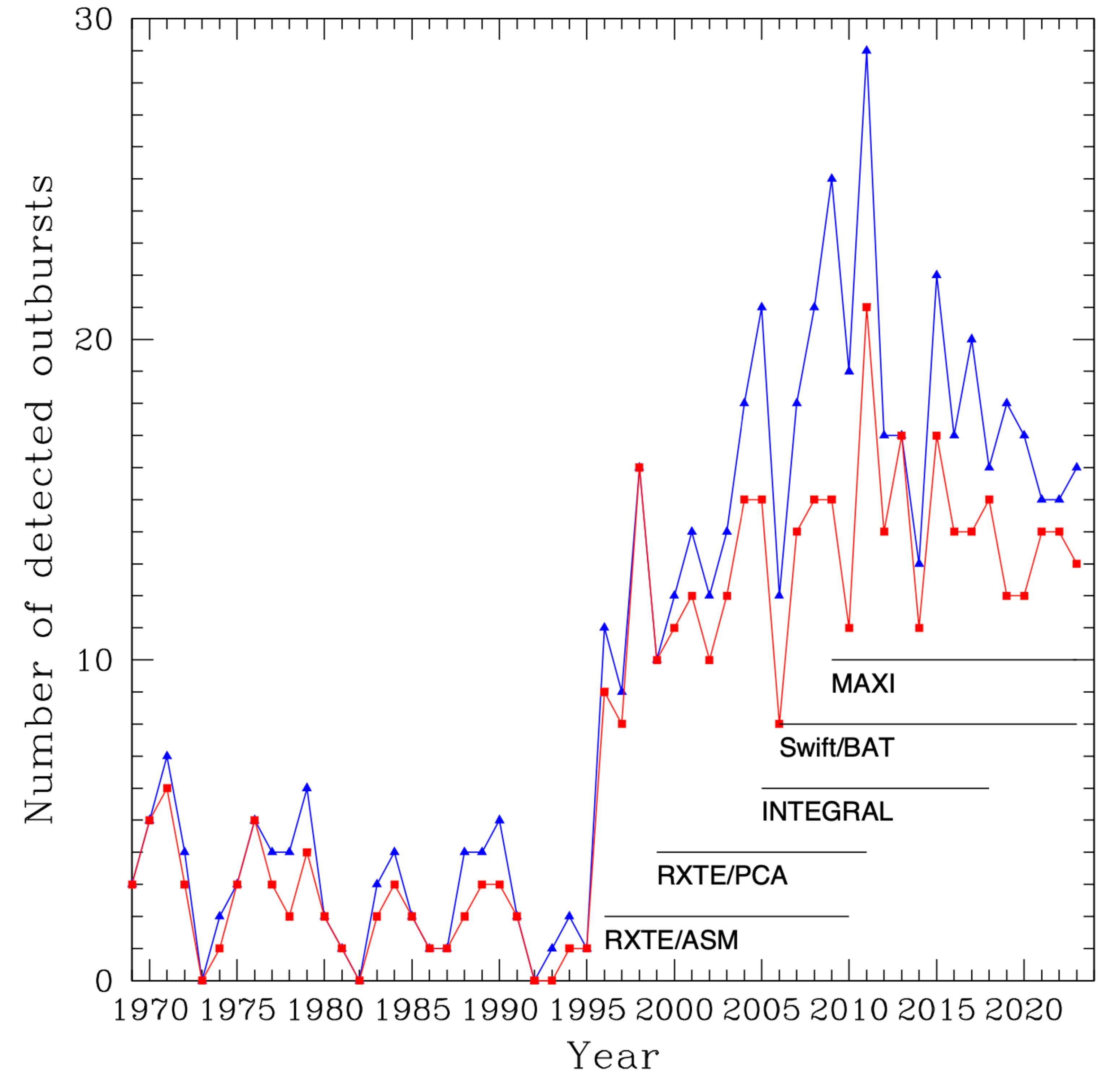}
 \caption{Number of total outbursts per year (blue triangles) and outbursts with $F_X>3\times10^{-10}$ ergs \revise{cm$^{-2}$ s$^{-1}$} per year (red dots), from transient neutron star LMXBs. 
The effective periods in which the RXTE/ASM, RXTE/PCA bulge scans, INTEGRAL galactic bulge monitoring, Swift/BAT monitor, and MAXI/GSC monitoring were effective are indicated. Since 1996, detections of bright outbursts have been roughly constant in time.
 }
    \label{fig:per_year}
\end{figure}

In Fig.~\ref{fig:all_outbursts} we show the peak fluxes from all transient neutron star LMXBs (a few lie outside our bounds), with the most frequent outbursters indicated with different colors.
In Fig.~\ref{fig:per_year}, we show the number of outbursts per year (omitting Cir X-1's many outbursts in some years, which can be rather distorting).

From Fig.~\ref{fig:per_year}, we see that 1996 marks a dramatic change in detected outburst frequency.  A significant peak appears around 2009-2011 in all outbursts.  Fig. 2 of \citet{Tetarenko16}, transient outbursts of BH LMXBs, reveals a similar peak in 2007-2012. However, this peak is not significant in brighter NS LMXB outbursts with $F_X>3\times10^{-10}$ ergs \revise{cm$^{-2}$ s$^{-1}$}, suggesting that fainter outbursts were more detectable in that time frame (e.g. by the RXTE PCA bulge monitoring program; which detected, e.g., the NGC 6440 X-2 outbursts at $1-2\times10^{-10}$ ergs \revise{cm$^{-2}$ s$^{-1}$}; \citealt{Heinke10_6440paper}).  

\begin{figure}
    \centering
    \includegraphics[width=\columnwidth]{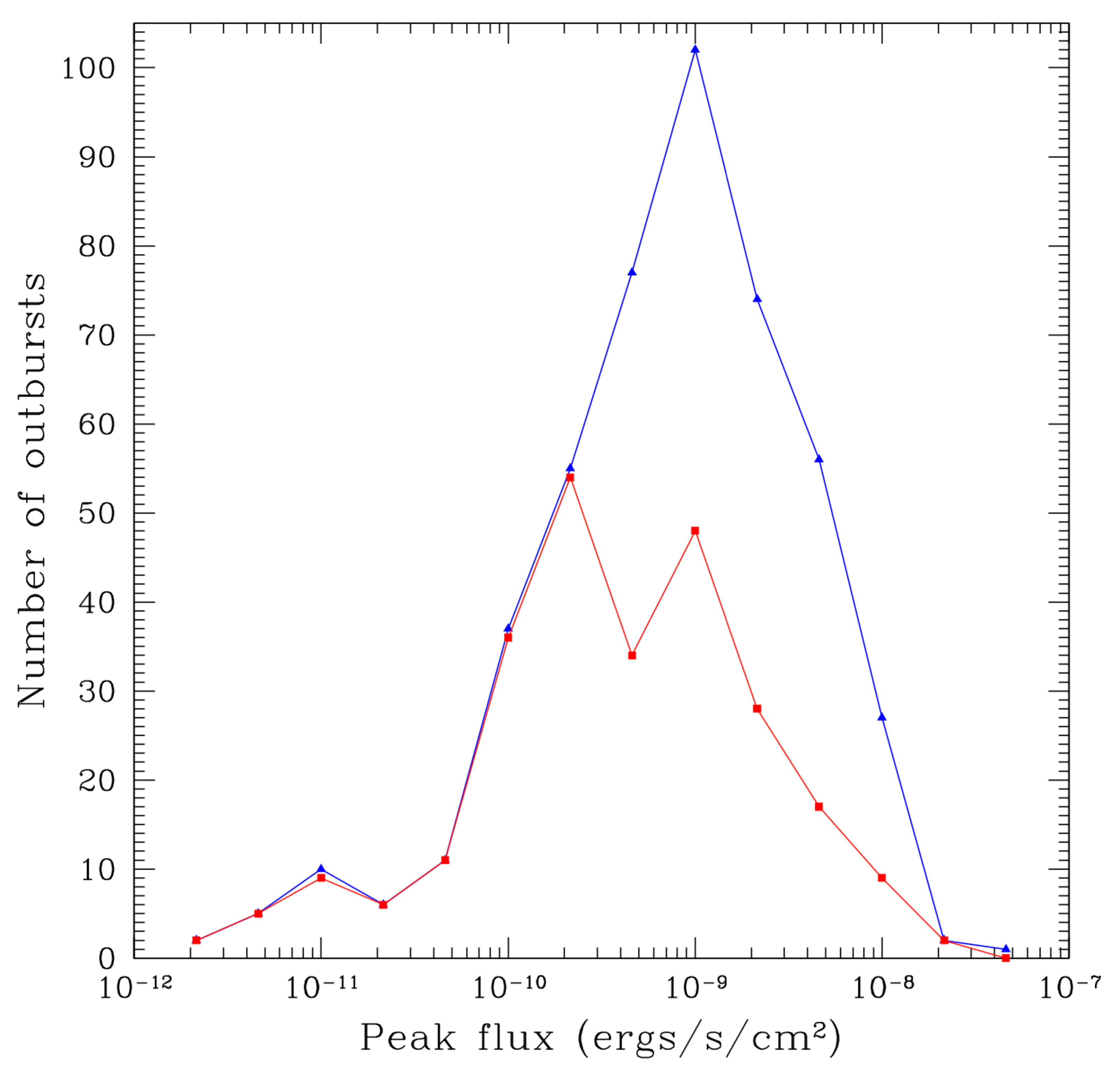}
 \caption{Number of outbursts by peak flux since 1996, for all transient neutron star LMXBs (blue triangles), and for all except the four bright regular outbursting systems: Aquila X-1, 4U 1608-52, MXB 1730-335, and GRS 1747-312 (red squares). 
 }
    \label{fig:fluxes}
\end{figure}

We then construct a histogram of fluxes from all outbursts since 1996 (Fig.~\ref{fig:fluxes}, blue triangles). This is sharply peaked at $F_X=10^{-9}$ ergs \revise{cm$^{-2}$ s$^{-1}$}, which might suggest that our data becomes highly incomplete below  $F_X=10^{-9}$ ergs \revise{cm$^{-2}$ s$^{-1}$}. However, if we exclude the four most frequent outbursters (Aquila X-1, 4U 1608-52, MXB 1730-335, and GRS 1747-312), all of which have relatively long orbital periods, high mass transfer rates, and high outburst fluxes, we see a rather different histogram (Fig.~\ref{fig:fluxes}, red squares), which does not peak until the bin $F_X=2-4\times10^{-10}$ ergs \revise{cm$^{-2}$ s$^{-1}$}. This suggests that significant incompleteness is happening in the next lower bin, covering $F_X=1-2\times10^{-10}$ ergs \revise{cm$^{-2}$ s$^{-1}$}. Referring to Fig.~\ref{fig:all_outbursts}, it seems that the majority of monitors have a limit around $F_X=2\times10^{-10}$ ergs \revise{cm$^{-2}$ s$^{-1}$}, so the peak here is likely due to sensitivity limits. Below $F_X=1\times10^{-10}$ ergs \revise{cm$^{-2}$ s$^{-1}$}, most detections use pointed instruments (e.g. Swift/XRT, or RXTE in pointing mode). 
 
\begin{figure}
    \centering
    \includegraphics[width=\columnwidth]{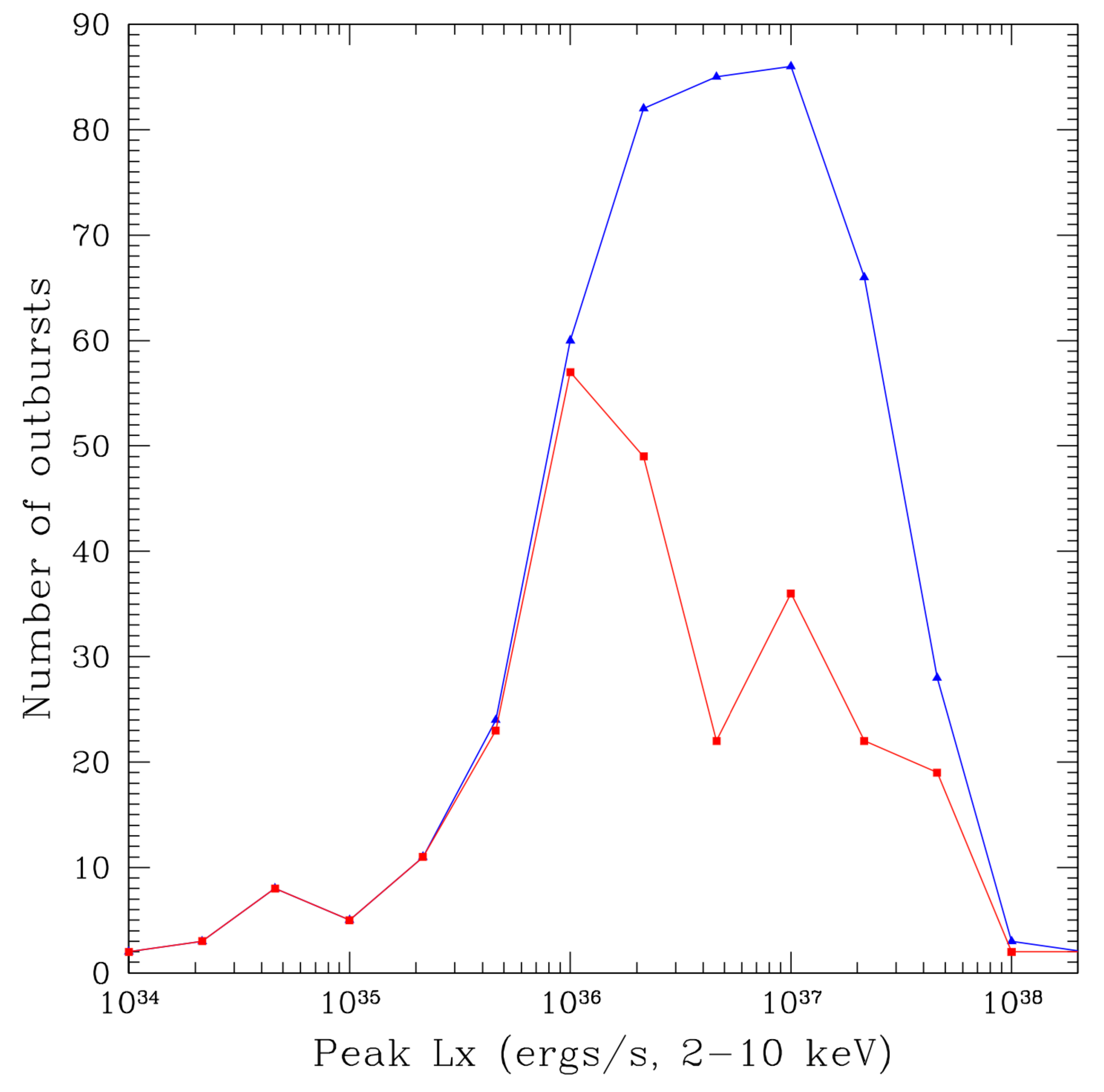}
 \caption{Number of outbursts by peak $L_X$ since 1996, for all transient neutron star LMXBs (blue triangles), and for all except the four bright regular outbursting systems: Aquila X-1, 4U 1608-52, MXB 1730-335, and GRS 1747-312 (red squares). 
 }
    \label{fig:Lx}
\end{figure}

We can also look at the distribution of peak $L_X$ in detected outbursts (Fig.~\ref{fig:Lx}). (We must keep in mind that the majority of these $L_X$ values are likely overestimates, due to the assumed distances being often upper limits, if the observed bursts were not actually Eddington-limited as assumed.) The $L_X$ distribution looks generally similar to the distribution of fluxes, due to the concentration of transient LMXBs in the Galactic Bulge, but due to the variation in distance, the peak is less sharp. 
An equivalent plot for BH LMXB systems (Fig.~15 in \citealt{Tetarenko16}\footnote{Including a shift down by roughly a factor 3 to move from bolometric to 2-10 keV luminosity.}) peaks close to $L_X\sim 10^{39}$ erg/s, with a tail extending down to $10^{35}$ erg/s. The %difference between BHs and NSs 
\revise{tendency for BHs to have higher $L_X$ }
may be attributed to \revise{BHs'} bias towards longer orbital periods, which are correlated with higher peak $L_X$ \citep{Wu10}.
%Black hole X-ray binaries may have longer orbital periods with main sequence donor stars, because unstable mass transfer sets in for mass ratios greater than about 0.8-1.0 for typical stellar structures on the lower main sequence. 
At very short periods, even the expected peak luminosities leave black holes in radiatively inefficient states, leading to a strong selection effect against %these objects
\revise{low-luminosity black holes }
(e.g. \citealt{Arur18}). Another selection effect is that verifying a black hole in an LMXB typically requires phase-resolved spectroscopy, which is more difficult in a short-period, thus fainter, system; while verifying a neutron star uses X-ray bursts or pulsations, which aren't tightly correlated with peak $L_X$.  

Intriguingly, there is a sharper peak in the $1-2\times10^{36}$ erg/s bin than appears in the flux histogram. If this peak were due entirely to limited sensitivity below it, we would expect this peak to be less sharp in the $L_X$ histogram than the $F_X$ histogram. It is possible (but not certain) that the number of outbursts falls off below $10^{36}$ erg/s ("very faint" X-ray binary outbursts). A gap in the (persistent) X-ray luminosity function between $10^{34}$ and $10^{36}$ erg/s has previously been identified in globular clusters \citep{Hertz83,Verbunt06,Bahramian2020}, where the sources above, and just below, this gap consist predominantly of neutron star LMXBs.  A dropoff below $L_X\sim10^{36}$ erg/s has also been suggested in the bulge of M31 \citep{Voss07}, in globular clusters in nearby galaxies \citep{Zhang11},
and possibly (but with inadequate data) for the Milky Way \citep{Sazonov06,Bahramian21}. On the other hand, extensive Swift, XMM-Newton and Chandra monitoring of the Galactic Center identified 28 outbursts from 13 LMXBs, between 2000 and 2014, including 11 between $L_X=10^{34}$ erg/s and $10^{35}$ erg/s, which argues strongly against any cutoff \citep{Wijnands06,Degenaar15}.  This apparent discrepancy may be reconciled if most persistently-bright systems have $L_X>10^{36}$ erg/s, but transient outbursts are often fainter, so that a given snapshot would find a gap, but a list of peak outburst luminosities would not.  Following the standard disk-instability model, the only LMXBs that could remain persistently below $10^{36}$ erg/s will be a small range of periods for ultracompact systems \citep[e.g.][]{intZand07,vanHaaften13}.  
If this is indeed the case, then more sensitive X-ray monitoring programs (e.g. proposed for STROBE-X, \citealt{Ray19}) would find substantially more X-ray outbursts. 

 %Smaller increases in sensitivity, down to a combined completeness limit of perhaps $2\times10^{-10}$ ergs/cm$^2$/s and detection limit of $10^{-10}$ ergs/cm$^2$/s, are visible in 2005 and 2006 as INTEGRAL and Swift's BAT monitor (respectively) began operating \citep{Kuulkers07, Krimm13}. \footnote{We do not attempt to find additional outbursts using only Swift/BAT or INTEGRAL/IBIS data, as these instruments only explore higher energy ranges.} 

\begin{figure}
    \centering
    \includegraphics[width=\columnwidth]{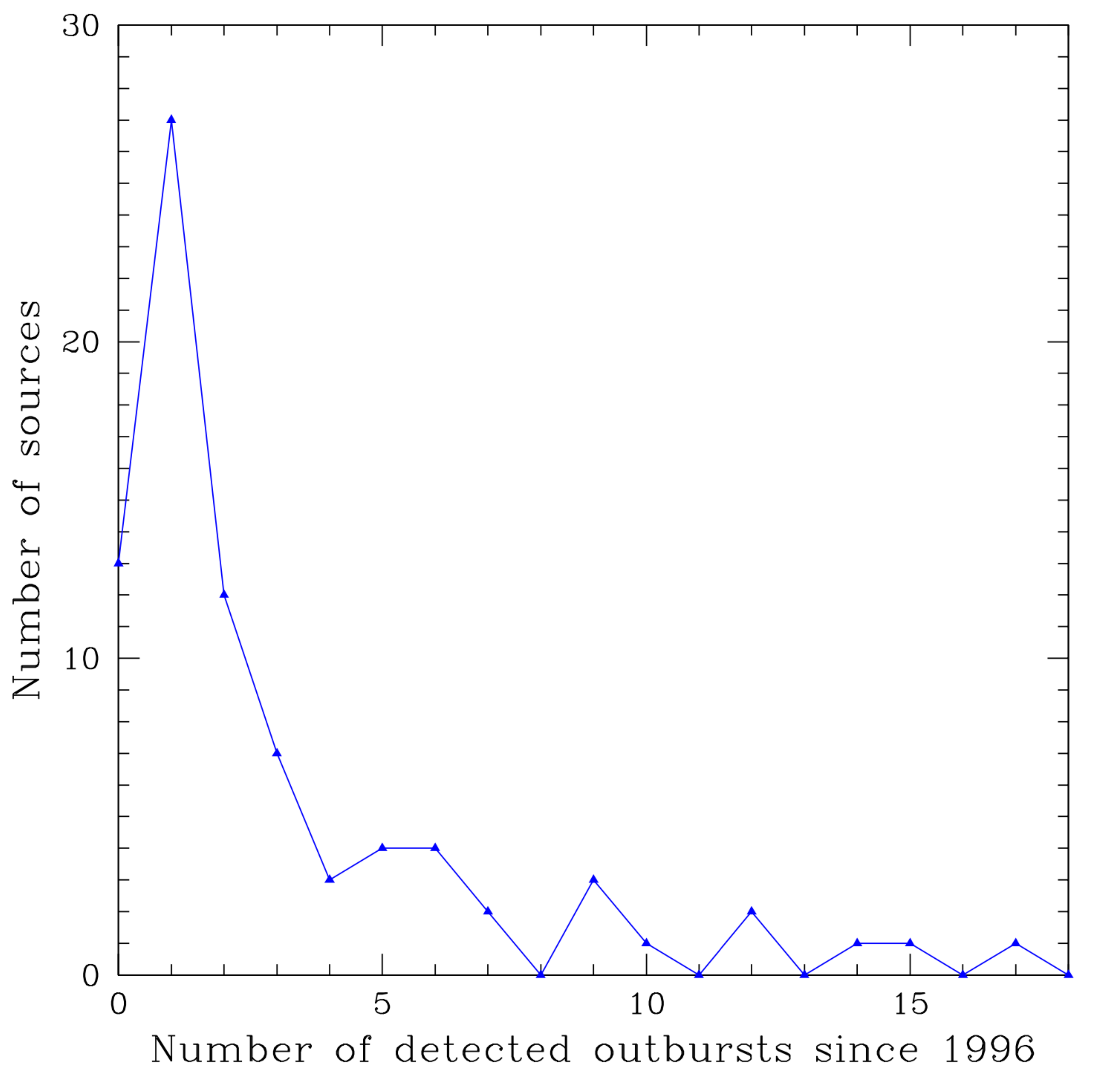}
 \caption{Number of neutron star LMXBs with a given number of total outbursts since 1996. Off the scale to the right are 4U 1608-52 (26), Aquila X-1 (35), GRS 1747-312 (69), and MXB 1730-335 (74).
 }
    \label{fig:per_source}
\end{figure}

\subsection{Application to the Galactic Center}

\citet{Hailey18} identified a dozen {\it Chandra} X-ray sources within a few parsecs of the Galactic Center that have non-thermal X-ray spectra (best fit with a power law with photon index between 1.5 and 2.9), which along with their $L_X$ values indicates these are quiescent LMXBs. \citet{Hailey18} and \citet{Mori21} suggest that these objects must be black hole LMXBs, arguing that all neutron star LMXBs have short recurrence times, generally less than 5 years, while black hole LMXBs can have longer recurrence times. \citet{Maccarone22} argued that a number of known neutron star LMXB systems do have longer recurrence times, and that the non-thermal quiescent Galactic Center X-ray sources may be dominated by neutron star LMXBs. \citet{Mori21} and \citet{Mori22} argued that most neutron star LMXBs undergo frequent outbursts, based on the history of the known NS transients studied in the Galactic Center. This argument assumes that the confirmed neutron star transients are representative of the underlying population, and not subject to biases; for instance, that the more frequent outbursters are easier to confirm as neutron stars  (e.g. they produce more X-ray bursts).

Clearer conclusions on the recurrence times of neutron star LMXBs require a thorough compilation of known outbursts, which we performed in this paper, allowing us to comment on the fractions of (observed) neutron star LMXBs with different recurrence times. These conclusions are, however, subject to various systematic uncertainties, due principally to the limited sensitivity and angular resolution of X-ray all-sky monitors.

A key question we can ask is what fraction of transient neutron star LMXBs have a recurrence time greater than 10 years, from the data available. 
We can answer this by checking for gaps of at least 10 years between 1996 and 2022 (the time frame with the best sensitivity). 
We identify, for each object, the probability (based on the recorded 1996-2022 outbursts) of not seeing outbursts during a continuous 10-year period.  (We allow a continuous time interval to wrap from 2022 to 1996, for this purpose.) 
\revise{For instance, Cen X-4 has a 100\% chance of not seeing any outburst in a random 10-year period of this range, while MXB 1659-29 has only a 22\% chance of having no outbursts in a 10-year block, and SAX J1808.4-3658 has a 0\% chance of seeing no outbursts in a 10-year block.}
Among 84 systems we study (omitting Terzan 6 X-2, \revise{ Terzan 5 X-2 and X-3,} due to confusion), only 24 do not show any 10-year-long gap, suggesting that \revise{the majority}  %could hide
\revise{have a chance of hiding }
for 10 years or longer. %Naive direct application of this to the Galactic Centre suggests that for every NS LMXB with a shorter recurrence time (e.g. the 6 known in the Galactic Centre), there are {\it at least} three times more systems with longer recurrence periods. 
Randomly selecting a contiguous ten-year period from a random neutron star in our sample \revise{gives an average} 37\% probability of finding no outbursts. 
%Thus, a significant  reservoir of so-far undetected neutron star systems with longer recurrence periods are likely to exist in the Galactic Center region. 

This calculation does not include systems that have been quiescent over the past 30+ years, such as the quiescent LMXBs observed in many globular clusters \citep[e.g.][]{Heinke03_qLMXBs,Guillot11,Bahramian15}. As only 15 neutron star LMXBs have been observed to outburst from globular clusters \citep{Bahramian23}, but of order 200 quiescent neutron star LMXBs are estimated in Galactic globular clusters \citep{Heinke05_new_qLMXBs}, this would suggest a neutron star LMXB population over ten times larger than observed. 
\revise{Some evolutionary calculations support the likely existence of such binaries \citep{MeyerHofmeister24}.}
%(discussed further below).  

%This is a very rough estimate, subject to numerous caveats. 
The principal caveat is that  
%Most importantly, 
all-sky monitors have limited sensitivity. We infer that we are missing most outbursts below $L_X\sim2.3\times10^{36}*(d/(8 {\rm kpc}))^2$ erg/s;  even brighter outbursts for $\sim$10\% of the year when the Sun is nearby; and some (not easily quantifiable) due to crowding with bright nearby sources. The Swift campaigns on the Galactic Center show that  the three closest verified neutron star LMXBs  show at least one fainter outburst as well as bright outbursts; \citet{Degenaar15} list 4 outbursts with peak $L_X>2.3\times10^{36}$ erg/s, and 11 outbursts with peaks below this. Since 1996, our larger databases give 8 and 22, respectively.  \citet{Mori22} argue that the  much larger survey frequency of the Galactic Center gives it the only well-constrained outburst histories, particularly regarding very faint X-ray transient behavior. 

However, there are also nine other transients, showing very-faint X-ray outbursts (peak $L_X<10^{36}$ erg/s), that have been detected in the Galactic Center region since 2000, which have not been confirmed as neutron stars through X-ray bursts, and have also not shown previous evidence for being black holes \citep{Degenaar15}. 
These 9 very-faint transients have each had only one or two outbursts within the past 23 years, suggesting relatively long recurrence times ($>$20 years), and that a larger population of such transients may exist there. 
It is generally difficult to attribute X-ray bursts observed near the Galactic Center with RXTE \citep{Galloway08}, and very faint X-ray binaries may be expected to burst only rarely \citep{Peng07,Lin19}.  %It  seems likely that some (or most) of these very faint X-ray binaries are neutron stars, given that 
The fact that known black hole outbursts tend to be brighter than known neutron star outbursts \citep{YanYu15} suggests that some, or most, of these very faint X-ray binaries are neutron stars.  
However, we have observed few short-period black hole systems, and it is possible that short-period black hole LMXBs are relatively faint; cf. \citealt{Knevitt14}. 

\begin{figure}
    \centering
    \includegraphics[width=3.4in]{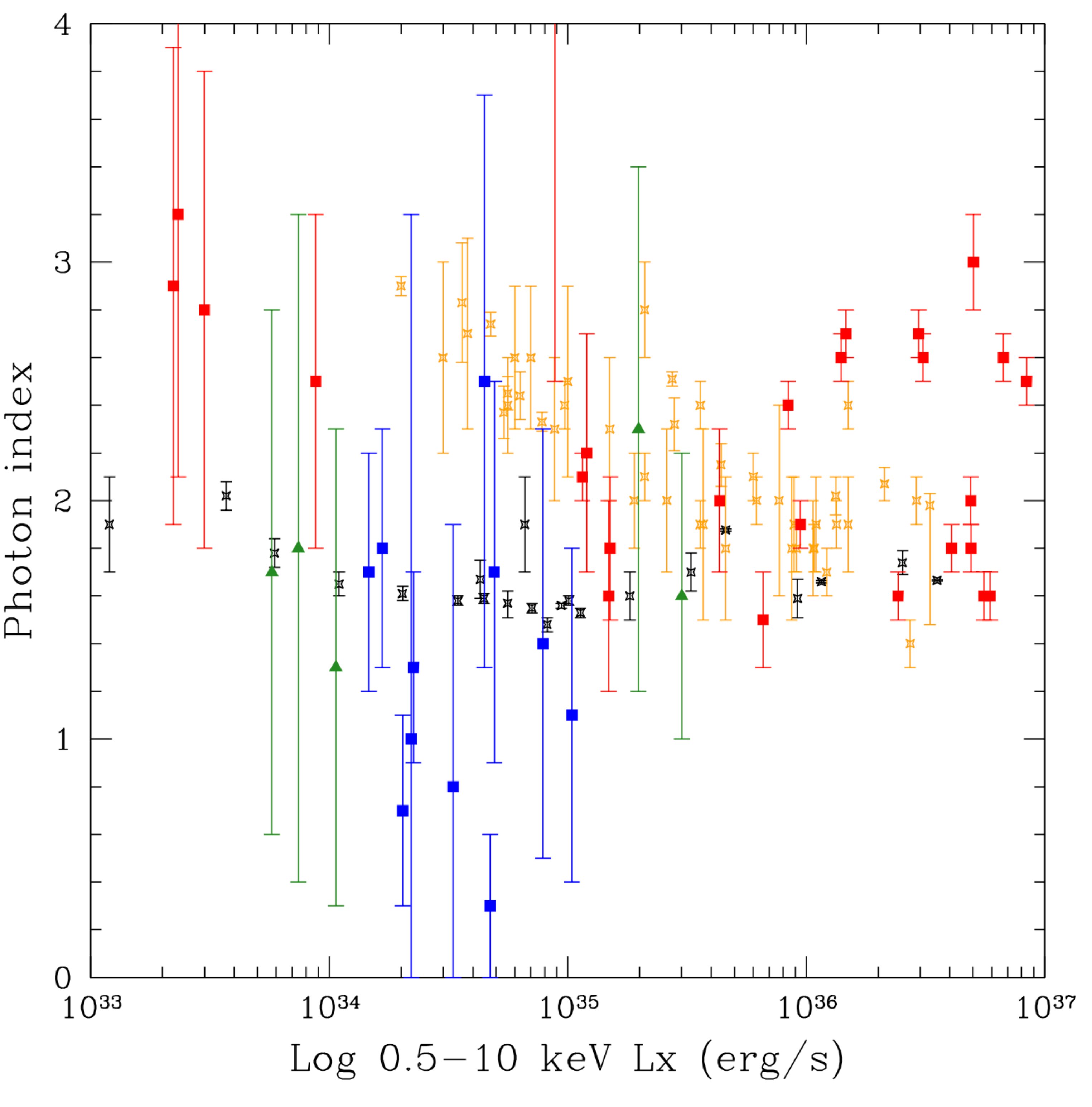}
 \caption{Fitted power-law photon indices vs. 0.5-10 keV $L_X$, for very faint X-ray transients in the Galactic Center region (blue squares; \citealt{Degenaar10,Degenaar12,Degenaar15}), vs. known neutron star LMXBs at similarly high $N_H$ (red squares; \citealt{Degenaar10,Degenaar12,Wijnands15}). Low-$N_H$ neutron star LMXBs are plotted as orange crosses, and low-$N_H$ black hole LMXBs are plotted as black crosses, both from \citet{Wijnands15} which introduced the dichotomy. The neutron star LMXB XMM J174457-2850.3 (green triangles), \revise{ and the}  other 
 very faint X-ray transients \revise{(blue),} have unusually hard X-ray spectra, similar to those of black holes.
 }
    \label{fig:photonindices}
\end{figure}

One empirical test between neutron star and black hole LMXBs, introduced by \citet{Wijnands15}, compares the photon index of a power-law spectrum fitted to a transient LMXB in the very-faint regime ($L_X=10^{34-36}$ erg/s), where black hole LMXBs have significantly harder spectra than most neutron star LMXBs.  High extinction weakens the power of this test, but it may still be useful. We have plotted the low-extinction neutron star and black hole systems used by \citet{Wijnands15} to introduce this test in Fig.~\ref{fig:photonindices}, along with the measurements published for several high-extinction known neutron star systems (AX J1745.6-2901, GRS 1741.9-2853, SAX J1747.0-2853, KS 1741-293, GRO 1744-28, XMMU J1747161-281048, and XMM J174457-2850.3). All but the last of these systems conform to the trend of high photon indices for neutron star systems with $L_X$ between $10^{34-35}$ erg/s. Plotting the very-faint X-ray transients, however, we see that their photon indices align with those of black hole LMXBs. Although this is extremely interesting, we follow \citet{Wijnands15} in pointing out that a number of neutron star LMXBs also show equally hard spectra in this $L_X$ range (e.g. IGR J18245-2452 in M28, \citealt{Linares14}; EXO 1745-248, \citealt{RiveraSandoval18}; and XMM J174457-2850.3,  \citealt{Degenaar14_XMMJ1744}). Although some of these are, or may be, transitional millisecond pulsars \citep{Degenaar14_XMMJ1744}, not all such hard-spectrum neutron star LMXBs are, and it seems unlikely that there would be so many transitional millisecond pulsars in active states in such a small region (given that only $\sim$2\% of quiescent neutron star LMXBs appear as active-state transitional millisecond pulsars, \citealt{Bahramian2020}).

For most neutron star transients elsewhere, we cannot constrain their history of fainter outbursts peaking at $L_X<2\times10^{36}$ erg/s.  Only for three nearby neutron stars accreting from low-mass companions are we able to rule out outbursts since 1996 above $L_X>7\times10^{34}$ ergs s$^{-1}$; the LMXB Cen X-4 (at 1.3 kpc), and the two transitional millisecond pulsars PSR J1023+0038 \citep{Archibald09} and PSR J12270-4859 \citep{Roy15}, both at 1.4 kpc. It is unclear if these two transitional millisecond pulsars undergo regular, or even any, major accretion outbursts, as IGR 18245-2452 did.  
These three objects suggest that most neutron stars accreting from low-mass companions remain in a  quiescent state for long periods.

Cen X-4's X-ray properties are fully consistent with the nonthermal Galactic Center sources described by \citet{Mori21}; Cen X-4's thermal emission would be completely absorbed by the $N_H>10^{23}$ at the Galactic Center, while its harder, variable power-law tail with photon index 1-1.7 and 2-10 keV $L_X=5-50\times10^{31}$ \citep{Cackett10,Chakrabarty14} resembles the properties of the Galactic Center sources. 
Incidentally, the same is true for many other qLMXBs, including some detected in globular clusters (e.g. EXO 1745-248 and Swift J174805.3-244637 in Terzan 5; \citealt{Wijnands05_Ter5,Bahramian14}). While it is true that most identified globular cluster qLMXBs that have not experienced outbursts lack significant hard spectral tails \citep{Mori22}, this is likely a selection effect, as globular cluster qLMXBs without detected outbursts are generally selected by their extremely soft spectra \citep[e.g.][]{Heinke03_qLMXBs,Guillot11}. The deepest cluster X-ray observations, of 47 Tuc, indeed identified two strong candidate qLMXBs with significant power-law tails (index $\sim$2, $L_X=2-3\times10^{31}$ erg/s), and no recorded outbursts \citep{Heinke05_new_qLMXBs}. 47 Tuc is far enough from the ecliptic that it is never blocked from view for all-sky X-ray monitors, and at 4.5 kpc, 20 mCrab comes to $10^{36}$ erg/s, such that even relatively faint outbursts would have been observed. 

Among the other transient neutron star LMXBs with long ($>$10-year)  periods between known outbursts, we can identify several more with substantial power-law components in their quiescent spectra;
XTE J0929-314 \citep{Wijnands05}, SAX J1324.5-6313 \citep{Cornelisse02_chandra,Bahramian15atel}, XB 1733-30 (CX2 in Terzan 1; \citealt{Cackett06,Bozzo20_Integral}, GRO J1744-28 \citep{Doroshenko20}, and 1RXS J180408.9-342058 \citep{Parikh17_1804}. XTE J0929-314 and probably XB 1733-30 are ultracompact systems, which would not be dynamically disrupted even in the Galactic Center region.

The XB-NEWS group has conducted frequent optical monitoring of Cen X-4 since 2008 without a proper outburst, although they did observe a slow optical brightening culminating with a failed outburst in late 2020, peaking at $L_X\sim3\times10^{33}$ erg/s \citep{Baglio22}; this serves to illustrate that the XB-NEWS optical monitoring \citep{Lewis08,Russell19} is effective at identifying outbursts. 
 XB-NEWS monitors 50 systems weekly, including a number of other NS LMXB transients detectable in the optical (such as IGR J00291+59, XTE J2123-058, XTE J0929-314, XTE J1814-338, and HETE J1900.1-2455; \citealt{Lewis08}). Careful analysis of the XB-NEWS archive may enable  determination of how many fainter outbursts we are missing.

\section{Conclusions}

We have compiled a list of known neutron star transient LMXBs, from the lists of known bursters and accreting pulsars. We have searched the literature to produce a list of known outbursts from these systems, for each of which we record the date, approximate 2-10 keV peak flux and (inferred, given the best-estimate distance) luminosity, and one (or a few) relevant references. For a few objects, we have identified outbursts not identified in the literature, using public (generally all-sky) lightcurves. The purpose of this list of outbursts is to facilitate more in-depth analyses of the outburst records. Such studies could be done using only the information from these tables, in which case it is important that such investigators bear in mind the non-uniformity and limited sensitivity of these lists, or could use the information and references here as a jumping-off point for more detailed study of one or more objects.

We use the information assembled to make a few new conclusions. For instance, we consider arguments about the distance to 4U 1608-52, and identify periods of prolonged quiescence in AX J1754.2-2754, XMM J174457-2850.3, and XMMU J174716.1-281048 (aka IGR J17464-2811). We then look at the distribution of outbursts with time, and flux, to identify a rough empirical flux completeness limit for X-ray binary outbursts since 1996 of $F_X=3\times10^{-10}$ ergs \revise{cm$^{-2}$ s$^{-1}$} (though a number of outbursts have been identified below this limit, and some above have been missed). 

Finally, we look at the distribution of the frequency of recorded outbursts from these transients. While four transients average roughly one to three outbursts per year, nearly half (40 of %86 
\revise{87} 
systems) have zero or one recorded outburst between 1996 and 2023. It is certainly possible that many of these have shown fainter outbursts below our detection level during this period. However, relatively stringent limits on outbursts from nearby transient LMXBs, such as Cen X-4, PSR J1023+0038, and PSR J12270-4859, for several decades suggest that there are transient neutron star LMXBs with properties consistent with the intriguing Galactic Center X-ray binaries identified by \citet{Hailey18}.

{\it Note added in proof:} \citet{Ni25} report a new X-ray transient detected by the Einstein Probe. \citet{Wu25b} report a refined position, an eclipse, and identify two flares in its X-ray lightcurve as X-ray bursts, indicating that EP J171159.4-333253 is almost certainly a new neutron star X-ray transient.

%\begin{acknowledgments}
\section*{Acknowledgements}

We thank the referee for an exceptionally thoughtful and useful report. 
Swift/XRT lightcurves supplied by the UK Swift Science Centre at the University of Leicester \citep{Evans07}. 
RXTE/ASM results provided by the ASM/RXTE teams at MIT and at the RXTE SOF and GOF at NASA's GSFC. 
RXTE/PCA results from NASA Goddard Space Flight Center, \url{https://asd.gsfc.nasa.gov/Craig.Markwardt//galscan/main.html}.
INTEGRAL/JEM-X lightcurves from \citet{Kuulkers07} using the OSA10.1 software, distributed by the ISDC \citep{Courvoisier03}.
 MAXI results provided by RIKEN, JAXA and the MAXI team \citep{Matsuoka09}.
 
 This paper made very extensive use of the Astrophysical Data Service (\url{https://ui.adsabs.harvard.edu}), ArXiv (\url{http://arxiv.org}), and the Astronomer's Telegram (\url{http://www.astronomerstelegram.org}).
CH is supported by NSERC Discovery Grant RGPIN-2023-04264. JZ has been supported by a Mitacs Globalink Research Internship. 

%\end{acknowledgments}

%%%%%%%%%%%%%%%%%%%%%%%%%%%%%%%%%%%%%%%%%%%%%%%%%%
%\section*{Data Availability (not necessary?????)}

%The data shown here is principally taken from literature reports, along with publicly available lightcurves (see Acknowledgements for details). Any data products are available from the primary author on reasonable request.

\vspace{5mm}
\facilities{RXTE(ASM and PCA), Swift(XRT), MAXI(GSC), Integral(JEM-X), CXO, XMM-Newton}

This paper employs a list of Chandra datasets, obtained by the Chandra X-ray Observatory, contained in the Chandra Data Collection (CDC) 346:~\dataset[doi:10.25574/cdc.346]{https://doi.org/10.25574/cdc.346}. 



\appendix

In Appendix A we give the full table (Table~\ref{tab:outbursts_app}) of data on neutron star LMXB outbursts, except for the frequent outbursters, which are given in Appendix B. Appendix C tabulates information on bursters without recorded outbursts, and Appendix D tabulates the neutron star X-ray binaries that we consider to be persistent systems (not further discussed in this paper). 

\section{Outburst details}

\startlongtable
\begin{longrotatetable}
    %\centerwidetable
	\begin{deluxetable}{lcccccccl} %
 %\tabletypesize{8pt}
\tablehead{		\colhead{NS XRB} & \colhead{Begin} & \colhead{End} & \colhead{Peak $F_X$}  & \colhead{$N_H$} & \colhead{Index} & \colhead{$L_X$} & \colhead{d}  &  \colhead{References}\\ 
% \tablehead{ 
\colhead{}   & \colhead{} & \colhead{} & \colhead{2-10 keV} & \colhead{cm$^{-2}$} & \colhead{} & \colhead{2-10 keV} & \colhead{kpc}  & \colhead{} } 
 \startdata
		IGR J00291+5934 & 1998/11 & 1998/11 & 4.5e-10 &  - & -& 9.6e35 & 4.2  &  \citet{Remillard04} \\
		IGR J00291+5934 & 2001/9 & 2001/9 & 3.9e-10 &  - & - & 8.2e35 & 4.2  & \citet{Remillard04} \\
		IGR J00291+5934 & 2004/12 & 2004/12 & 6.1e-10  & 7e21 & 1.7 & 1.3e36 & 4.2  & \citet{Eckert04,Galloway2005}\\
		IGR J00291+5934 & 2008/8 & 2008/10 & 6.1e-10  & - & 1.7 & 1.3e36 & 4.2  &  \citet{Chakrabarty08,Lewis10} \\
		IGR J00291+5934 & 2015/8 & 2015/8 & 5.6e-10  & 1e22 & 1.7 & 1.2e36 & 4.2  & \citet{Cummings15,deFalco17} \\
  \hline 
        MAXI J0556-332 & 2011/1 & 2012/5 & 1.9e-09 & - & - & 4.4e38 & 43.6 & \citet{Matsumura11,Sugizaki13} \\
        MAXI J0556-332 & 2012/10  & 2012/12 & 4e-10 & 3.2e20 &  - &  9e37 & 43.6 &  \citet{Sugizaki12,Parikh17_MAXIJ0556} \\
        MAXI J0556-332 & 2016/1 & 2016/3 & 7.7e-10 & - & - & 1.8e38 & 43.6 &  \citet{Negoro16,Russell16} \\
        MAXI J0556-332 & 2020/4 & 2020/11 & 8.6e-10 & - & - & 2.0e38 & 43.6 & \citet{Negoro20_J0556,Page22}\\
\hline
        EXO 0748-676    & 1984/7 & 2008/8  & 1.1e-9  & 1.2e21 & 1.9 & 7e36 & 7.4  & \citet{Parmar86,Degenaar09b} \\
        EXO 0748-676    &  2024/5 & 2025/3 & 3.0e-10 & 1.5e21 & 2 & 2.0e36 & 7.4 & \citet{D'Elia24,Mihara24,Degenaar25}\\
\hline 
        GS 0836-429 & 1970/12 & 1972/1 & 1.1e-9 & $<$3.5e22 & & 1.1e37 & $<$9.2 &  \citet{Markert77} \\
        GS 0836-429 & 1990/10 & 1991/5 & 3.8e-10 & 2.2e22 & 1.5 & 3.9e36 & $<$9.2 &  \citet{Aoki92,Belloni93} \\
        GS 0836-429 & 2003/1 & 2003/5 & 1.0e-9 & 2.7e22 & 1.5 & 1.0e37 & $<$9.2  &   \citet{Rodriguez03,Aranzana16} \\
        GS 0836-429 & 2003/9 & 2004/6 & 1.3e-9 & 2.7e22 & 1.5 & 1.3e37 & $<$9.2  &   \citet{Chelovekov05,Aranzana16} \\
\hline
        MAXI J0911-655 & 2016/2 & 2023/10 & 1.7e-10 & 2.7e21  & 1.27 & 1.9e36 & 9.5 &  \citet{Serino16_0911,Sanna17,Ng21} \\
\hline
        XTE J0929-314 & 2002/4 & 2002/4 & 6.7e-10 & 7.6e20 & 1.7-2  & $>$2.9e36 & $>$6 &  \citet{Remillard02,Galloway02,Juett03}\\
\hline
        SAX J1324.5-6313 & 2015/8 & 2015/8 & 1.3e-10 & 1.5e22 & - & 1.0e36 & $<$6.2 &   \citet{Negoro15,Cornelisse02_chandra} \\
\hline
        MAXI J1421-613 & 2014/1 & 2014/1 & 1.9e-9 & 4.8e22 & 2.1 & 2e36  & 3 &   \citet{Morooka14,Serino15,Nobukawa20} \\
\hline
        SRGA J144459.2-604207 & 2022/1 & 2022/1 & 4e-10 & - & - & 3.4e36 & 8.5 & \citet{Negoro24} \\
        SRGA J144459.2-604207 & 2023/12 & 2023/12 & 6e-10 & - & - & 5.2e36 & 8.5 & \citet{Negoro24,Sguera24} \\
        SRGA J144459.2-604207 & 2024/2 & 2024/3 & 2.7e-9 & 2.9e22 & 1.8 & 2.3e37 & 8.5 & \citet{Mereminskiy24,Molkov24,Ng2024}\\
\hline
        Cen X-4 & 1969/07 & 1969/7 & 6e-7 & - & - &   1.2e38 & 1.3 &  \citet{Evans70,Kuulkers09} \\
        Cen X-4 & 1979/5 & 1979/5 & 9.6e-8 & 9e20 & - &  1.9e37 & 1.3 &  \citet{Kaluzienski80,Kuulkers09}\\
\hline
        MAXI J1621-501 & 2017/9 & 2019/1 & 3.1e-9 & 6.1e22 & 2.7 & 2.6e37 & 8.4 & \citet{Hashimoto17,Gorgone19,Chenevez18} \\
\hline
        MAXI J1647-227 & 2012/6 & 2012/6 & 9.2e-10 & 4.3e21 & 1.5 & 1.1e37 & 10 &   \citet{Negoro12,Onodera14,Kennea12j1647} \\
\hline
        MXB 1659-298 & 1976/10 & 1979/7 & 1.9e-9 & - & - & 2.3e37 & 10 &  \citet{Lewin76_MXB1659,Cominsky83} \\
        MXB 1659-298 & 1999/9 & 1999/9 & 6.5e-10 & - & - & 7.8e36 & 10 &  \citet{intZand99_MXB1659,Heise99,Oosterbroek01} \\
        MXB 1659-298 & 2015/9 & 2015/9 & 5.5e-10 & 2.7e21 & - & 6.6e36 & 10 &  \citet{Negoro15,Sharma18} \\ 
\hline
        IGR J16597-3704 & 2017/10 & 2017/10 & 2.2e-10 & 1.5e22 & 1.46 & 2.2e36 & 9.1 &  \citet{Bozzo17_IGRJ16597,Sanna18} \\
\hline 
        XTE J1701-407 & 2008/6 & 2011/7 & 7.3e-10 & 3.4e22 & 2.0 & 3.3e36 & 6.2 &  \citet{Markwardt08_xteJ1701,Falanga09,Linares09} \\
        XTE J1701-407 & 2011/9 &  2011/9 & 3.8e-10 & 3e22 & 1.4 & 1.8e36 & 6.2  & \citet{Degenaar11atel} \\
        XTE J1701-407 & 2016/8 & 2017/4 & 3.5e-10 & - & - & 1.6e36 & 6.2  & Swift/XRT \\
        XTE J1701-407 & 2018/3 & 2018/3 & 3.5e-10 & - & - & 1.6e36 & 6.2  & \citet{Barthelmy18gcn},Swift/XRT \\        
        XTE J1701-407 & 2019/4 & 2019/4 & 4e-10 & 3.2e22 & 1.9 & 1.8e36 & 6.2  & \citet{Iwakiri19} \\
        XTE J1701-407 & 2020/1 & 2020/3 & 1.8e-10 & - & - & 8.2e35 & 6.2  & \citet{Onori20,Palmer20}\\ %(2016-2020 same?)\\
  %   {\bf Write Swift prop}, swift analysis requested? \\
\hline
        XTE J1701-462 & 2005/12 & 2007/8 & 4.3e-8 & 2.7e22 & 1.9 & 4e38 & 8.8  & \citet{Remillard06_xteJ1701,Lin09a,Kennea06} \\
        XTE J1701-462 & 2022/9 & 2023/3 & 1.4e-8 & 3e22 & - & 1.3e38 & 8.8 & \citet{Iwakiri22_1701,Cocchi23},Swift/XRT \\ 
\hline
        IGR J17062-6143 & 2005/12 & 2022/9* & 1.4e-10 & 2.3e21 & - & 8.7e35 & 7.3  & \citet{Churazov07,Remillard08,Bult21} \\
\hline
        XTE J1709-267 & 1990/3 & 1990/8 & 4.3e-9 & - & - & 3.7e37 & 8.5 &  \citet{Yamauchi90,Yamauchi10} \\
        XTE J1709-267 & 1997/1 & 1997/4 & 6.7e-9 & -& -& 5.8e37 & 8.5 &  \citet{Marshall97,Cocchi98} \\
        XTE J1709-267 & 2002/3 & 2002/4 & 7.3e-9 & 4.4e21 & -& 6.3e37 & 8.5 &  \citet{Jonker03} \\
        XTE J1709-267 & 2004/5 & 2004/5 & 7.7e-10 & -& -& 6.6e36 & 8.5  & \citet{Markwardt04,Jonker04_xteJ1709} \\
        XTE J1709-267 & 2005/5 & 2005/5 & 2e-10 &- & -& 1.7e36 & 8.5  & RXTE PCA Scan \\
        XTE J1709-267 & 2007/11 & 2008/9 & 1.9e-9 & -& -& 1.6e37 & 8.5  & \citet{Remillard07brief}, RXTE PCA Scan \\
        XTE J1709-267 & 2009/3 & 2009/5 & 2.5e-10 & -& -& 2.2e36 & 8.5  & RXTE PCA Scan \\
        XTE J1709-267 & 2010/7 & 2010/7 & 1.9e-9 & -&- & 1.6e37 & 8.5  & \citet{Negoro10}  \\
        XTE J1709-267 & 2012/8 & 2012/8 & 3.7e-9 & -&- & 3.2e37  & 8.5  & \citet{SanchezFernandez12,Degenaar13_xteJ1709} \\
        XTE J1709-267 & 2013/8 & 2013/8 & 1e-9 & -& -& 8.9e36 & 8.5  & \citet{Negoro13} \\
        XTE J1709-267 & 2016/6 & 2016/6 & 4.5e-9 & -& -& 3.9e37 & 8.5  & \citet{Nakahira16,Ludlam17} \\
        XTE J1709-267 & 2017/8 & 2017/8 & 3.9e-10 & -& -& 3.4e36 & 8.5  & \citet{Negoro17} \\
        XTE J1709-267 & 2018/12 & 2019/1 & 2.2e-9 & -& -& 1.9e37 & 8.5  & MAXI \\
        XTE J1709-267 & 2020/9 & 2020/12 & 2.2e-9 & -& -& 1.9e37 & 8.5  & \citet{Adachi20} \\
        XTE J1709-267 & 2022/6 & 2022/6 & 1.3e-9 & -& -& 1.1e37 & 8.5  & MAXI \\
        XTE J1709-267 & 2023/10 & 2023/11 & 1.4e-9 & -& -& 1.2e37 & 8.5 &  \citet{Negoro23},MAXI \\
        XTE J1709-267 & 2025/5 & 2025/6* & 3.5e-10 & 3.9e21 & 2.0 & 3.0e36 & 8.5 & \citet{Wu25} \\
\hline
        2S 1711-339 & 1975/2 & 1976/9  & 3.2e-9 & 1.5e22 & 2.2 & 2.2e37 & $<$7.5 &  \citet{Carpenter77,Cornelisse02} \\
        2S 1711-339 & 1998/7 & 1999/5 & 1.1e-9 & -& -& 7.2e36 & $<$7.5 &  \citet{Remillard98} \\
        2S 1711-339 & 2004/3  &  2004/3  & 4.5e-10 & -& -& 3.0e36 & $<$7.5 &  \citet{Markwardt04b} \\
        2S 1711-339 & 2010/2  & 2010/2  & 4.3e-10 & -& -& 2.9e36 & $<$7.5 &  \citet{Negoro10_1711} \\
        2S 1711-339 & 2016/11  & 2016/11  & 1.1e-9 & -& -& 7.7e36 & $<$7.5 &   \citet{Negoro16_1711} \\
\hline	
        1M 1716-315 & 1971/9 & 1984/6 & 1.1e-9 & 2.1e21 & 2 & 6.2e36 & 6.9 &  \citet{Markert76,Makishima81_1715,Jonker07} \\
\hline
        IGR J17191-2821 & 2007/3 & 2007/5 & 1.8e-9 & 3.4e21 & 1.75 & 2.7e37 & $<$11 &  \citet{Turler07,Altamirano10,Markwardt07} \\
        IGR J17191-2821 & 2012/6 & 2012/6 & 9.3e-10 & -&- & 1.3e37 & $<$11 &  \citet{Sugimoto12b} \\
\hline
        XTE J1723-376 & 1999/2 & 1999/3 & 1.9e-9 & 4e22 & -& 1.4e37 & 8 & \citet{Marshall99a,Marshall99b} \\ 
\hline
        4U 1730-22 & 1972/8 & 1972/8 & 2.9e-9 & -& -& 1.7e37 & 6.9 &  \citet{Cominsky78,Chen97} \\
        4U 1730-22 & 2021/6 & 2024/3 & 2.4e-9 & 3.9e21 & -& 1.4e37 & 6.9  & \citet{Kobayashi21,Chen23,Bult22}, MAXI, Swift/XRT \\
        % check Swift new observation
\hline
        KS 1731-260 & 1988/10 & 2001/2 & 5.2e-9 & 1.1e22 & -& 3.0e37 & 7 & \citet{Sunyaev90_KS1731,Wijnands01} \\
\hline
        XB 1733-30 & 1980/7 & 1999/4 & 3.3e-9 & 1.8e22 & 2.1 & 1.3e37 & 5.7 &   \citet{Makishima81_Terzan,Guainazzi99} \\
        XB 1733-30 & 2020/4 & 2025/5* & 2e-9 & 2.2e22 & 2 & 7.7e36 & 5.7  & \citet{Mereminskiy20_Ter1,Negoro23_Ter1,Homan23}  \\
\hline 
        Swift J1734.5-3027 & 2013/5 & 2013/6 & 3e-10 & - & - & 2e36 & 7.2 &  \citet{LaParola13,Bozzo15} \\
        Swift J1734.5-3027 & 2013/9 & 2013/9 & 3e-10 & 1e22 & 1.7 & 1.9e36 & 7.2  & \citet{Kennea13_1734,Bozzo15} \\
        Swift J1734.5-3027 & 2013/10 & 2014/4 & 3.6e-10 & 1e22 & 2.36 & 2.2e36 & 7.2 & \citet{Bozzo15} \\ 
\hline
        IGR J17379-3747 & 2004/2 & 2004/2 & 1.3e-10 & -& -& 1.1e36 & 8.5 &  \citet{Markwardt08_1737,Chelovekov10} \\
        IGR J17379-3747 & 2005/3 & 2005/3 & 2.2e-10 & -& -& 1.9e36 & 8.5 &  \citet{Bult19} \\        
        IGR J17379-3747 & 2008/9 & 2008/9 & 2.7e-10 & -& -& 2.3e36 & 8.5 &  \citet{Markwardt08_1737,Krimm08} \\
        IGR J17379-3747 & 2014/2 & 2014/2 & 2.2e-10 & -& -& 1.9e36 & 8.5 &  \citet{Negoro18_17379,Bult19} \\
        IGR J17379-3747 & 2018/3 & 2018/3 & 8.0e-10 & 9e21 & 1.88 & 6.9e36 & 8.5 &  \citet{Negoro18_17379,Bult19} \\
        IGR J17379-3747 & 2020/9 & 2020/9 & 9.2e-11 & -& 2.1 & 8e35 & 8.5 &  \citet{Mereminskiy2020} \\
\hline
        XTE J1739-285 & 1999/9 & 1999/9 & 5.6e-9 &  1.7e22 & -& 3.6e37 & $<$7.3 &  \citet{Markwardt99,Bult21b} \\
        XTE J1739-285 & 2001/5 & 2001/5 & 1.0e-9 &  1.7e22 & -& 6.4e36 & $<$7.3 &  \citet{Kaaret07},RXTE/PCA \\
        XTE J1739-285 & 2003/10 & 2003/10 & 1.0e-9 & 1.7e22 & -& 6.4e36 & $<$7.3 &  \citet{Kaaret07},RXTE/PCA \\
        XTE J1739-285 & 2005/8 & 2006/3 & 2.7e-9  & 1.7e22 & -& 1.7e37 & $<$7.3 &  \citet{Bodaghee05} \\
        XTE J1739-285 & 2012/8 & 2013/3 & 1.9e-9  & 1.7e22 & -& 1.2e37 & $<$7.3 &  \citet{SanchezFernandez12,Kuulkers13} \\
        XTE J1739-285 & 2014/1 & 2014/12 & 4.8e-9  & 1.7e22 & -& 3.0e37 & $<$7.3 &  MAXI \\
        XTE J1739-285 & 2019/9 & 2020/9 & 4.5e-9 &  1.7e22 &- & 2.9e37 & $<$7.3 &  \citet{Mereminskiy19_xteJ1739,Negoro20_1739b,Beri23}, MAXI \\
        XTE J1739-285 & 2021/5 & 2021/8 & 2e-9  & 1.7e22 & 2 & 1.3e37 & $<$7.3 &  Swift/XRT, MAXI \\
        XTE J1739-285 & 2022/12 & 2023/2 & 5.4e-10 &  1.7e22 &- & 3.4e36 & $<$7.3 &  \citet{Kashyap23},MAXI \\
\hline
        GRS 1741.9-2853 & 1990/3 & 1990/4 & 1.5e-10 & & & 1.5e36 & 9.0 & \citet{Sunyaev90,Pavlinsky94}  \\
        GRS 1741.9-2853 & 1994/9 & 1994/9 & 3.2e-12 & 1.1e23 & 2 & 3.1e34 & 9.0 &  \citet{Sakano02} \\
        GRS 1741.9-2853 & 1996/9 & 1996/9 & 1.8e-10 & 1.1e23 & 2.36 & 1.7e36 & 9.0 &  \citet{Sakano02}  \\
        GRS 1741.9-2853 & 2000/10 & 2000/10 & 4.0e-10 & 9.7e22 & 1.8 & 3.9e36 & 9.0 &  \citet{Muno03}  \\
        GRS 1741.9-2853 & 2002/10 & 2002/10 & 1.2e-11 & 1.1e23 & & 1.2e36 & 9.0 &  \citet{Trap09} \\ 
        GRS 1741.9-2853 & 2005/4 & 2005/6 & 1.1e-10 & -& -& 1.1e36 & 9.0 &  \citet{Wijnands05_GalCen,Wijnands06,Trap09}  \\
        GRS 1741.9-2853 & 2006/9 & 2006/9 & 1.2e-11 & -& -& 1.2e35 & 9.0 &  \citet{Degenaar09,Degenaar10}  \\
        GRS 1741.9-2853 & 2007/3 & 2007/4 & 2.6e-10 & -& -& 2.5e36 & 9.0 &  \citet{Kuulkers07_GRS1741,Degenaar09,Trap09} \\
        GRS 1741.9-2853 & 2009/10 & 2009/10 & 2.2e-9 & -& -& 2.1e37 & 9.0 &  \citet{Chenevez09,Degenaar10} \\
        GRS 1741.9-2853 & 2010/7 & 2010/7 & 2.6e-10 & -& -& 2.5e36 & 9.0 &  \citet{Degenaar14}  \\
        GRS 1741.9-2853 & 2013/8 & 2013/8 & 4.3e-9 & -& -& 4.2e37 & 9.0 &  \citet{Degenaar14}  \\
        GRS 1741.9-2853 & 2016/3 & 2016/3 & 8.7e-11 & -& -& 8.4e35 & 9.0 &  \citet{Degenaar16_GRS_1741}  \\
        GRS 1741.9-2853 & 2017/10 & 2017/10 & 1.5e-9 & -& -& 1.5e37 & 9.0 &  \citet{Degenaar17}  \\
        GRS 1741.9-2853 & 2020/4 & 2020/10 & 1.9e-10 & -& -& 1.8e36 & 9.0 &  \citet{Degenaar20_GRS1741,Pike21}  \\
        % add any more?
        \hline
        KS 1741-293 &  1989/8 & 1989/8 & 1.1e-9 & -& -& 5.0e36 & $<$6.2 &  \citet{intZand91} \\
        KS 1741-293 & 1998/3 & 1998/9 & 1.1e-9 & -& -& 5.0e36 & $<$6.2 &  \citet{intZand98,Sakano02} \\
        KS 1741-293 & 2003/3 & 2003/3 & 3.3e-10 & -& -& 1.5e36 & $<$6.2 &  \citet{deCesare07} \\
        KS 1741-293 & 2004/3 & 2004/3 & 3.2e-10 & 3.5e23 & 2 & 1.5e36 & $<$6.2 &  \citet{deCesare07} \\
        KS 1741-293 & 2005/7 & 2005/9 & 2.5e-10 & 1.7e23 & 1.8 & 1.2e36 & $<$6.2 & \citet{Kuulkers07} \\
        KS 1741-293 & 2007/9 & 2007/9 & 3.4e-11 & 1.7e23 & 1.8 & 1.6e35 & $<$6.2 &  \citet{Degenaar12} \\
        KS 1741-293 & 2008/5 & 2008/8 & 4.9e-10 & 1.7e23 & 1.8 & 2.3e36 & $<$6.2 &  \citet{Degenaar08_GalCen,Degenaar12} \\
        KS 1741-293 & 2010/2 & 2010/3 & 3.4e-10 &  -&  -& 1.6e36 & $<$6.2 &  \citet{Chenevez10_atel} \\
        KS 1741-293 & 2011/9 & 2012/2 & 1.6e-8 &  -&  -& 7.3e37 & $<$6.2 &  \citet{Linares11_KS1741,Chenevez11,Chenevez12}\\
        KS 1741-293 & 2013/2 & 2013/4 & 5.7e-10 & 1.7e23 & 1.8 & 2.7e36 & $<$6.2 &  \citet{Kuulkers13_KS1741}, Swift/XRT  \\
        KS 1741-293 & 2016/8 & 2016/8 & 3.4e-10 & -& -& 1.6e36 & $<$6.2 &  \citet{Chenevez16}  \\
        KS 1741-293 & 2017/5 & 2017/5 & 1.3e-09 & 1.7e23 & 1.8 & 6.0e36 & $<$6.2 & \citet{Bahramian21} \\
        KS 1741-293 & 2019/5 & 2019/5 & 7.5e-11 & -& -& 3.4e35 & $<$6.2 &  \citet{Heinke19} \\
        \hline
            1A 1744-361 & 1976/2 & 1976/4 & 1.1e-09 & -& -& 8.3e36 & 8 &  \citet{Carpenter77} \\
    1A 1744-361 & 1989/8 & 1989/8 & 1.3e-09 & -& -& 9.9e36 & 8 &  \citet{intZand04_A1744} \\
    1A 1744-361 & 2003/11 & 2003/11 & 2.6e-09 & -& -& 2.0e37 & 8 &  \citet{Remillard03_A1744} \\
    1A 1744-361 & 2004/4 & 2004/4 & 3.4e-10 & -& -& 2.6e36 & 8 &    \citet{Markwardt04_A1744} \\
    1A 1744-361 & 2005/7 & 2005/7 & 5.1e-10 & -& -& 3.9e36 & 8 &  \citet{Swank05} \\
    1A 1744-361 &  2008/6 & 2008/6 & 2.2e-09 & -& -& 1.7e37 & 8 &  \citet{Remillard08_A1744} \\
    1A 1744-361 & 2009/11 & 2009/11 & 1.4e-09 & -& -& 1.1e37 & 8 &  \citet{Remillard09_A1744} \\
    1A 1744-361 & 2010/3  & 2010/3  & 5.8e-10 & -&- & 4.4e36 & 8 &  RXTE/PCA \\
    1A 1744-361 & 2011/9  & 2011/9  & 1.7e-09 & -& -& 1.3e37 & 8 &  RXTE/PCA \\
    1A 1744-361 & 2013/8 & 2013/8 & 2.1e-09 & 5.1e21 & 1.5 & 1.6e37 & 8 &  \citet{Bahramian13_A1744}    \\
    1A 1744-361 & 2022/5 & 2022/8 & 2.4e-09 & -&- & 1.9e37 & 8 &   \citet{Kobayashi22_A1744,Hughes22,Ng23} \\
    1A 1744-361 & 2024/7 & 2025/2* & 5.9e-9 & 5.7e21 & 1.4 & 4.5e37 & 8 & \citet{Xu24,Cheng24} \\
    \hline
    IGR J17445-2747 & 2004/2 & 2004/3 & 7.4e-11 & -& 2.2 & 5.7e35 & 8 &  \citet{Bird06,Mereminskiy17} \\ 
    IGR J17445-2747 & 2017/4 & 2017/4 & 9.2e-11 & 4.9e22 & 2.2 & 7.1e35 & 8 &    \citet{Heinke17,Mereminskiy17} \\
    IGR J17445-2747 & 2019/5 & 2019/5 & 4.6e-12 & -& 2.2 & 3.6e34 & 8 &  \citet{Heinke19} \\
    IGR J17445-2747 & 2023/2  & 2023/3  & 1.5e-11 & 5.6e22 & 2 & 1.1e35 & 8 &  \citet{Kashyap23} \\
    \hline   
    % done checking for first-reports to here
    XMM J174457-2850.3 & 2001/9 & 2001/9 & 6.2e-12 & 5.9e22 & 1.9 & 3.1e34 & 6.5  & \citet{Sakano02} \\ 
    XMM J174457-2850.3 & 2005/6 & 2005/6 & 1.0e-10 & -&- & 5.2e35 & 6.5  & \citet{Wijnands05_GalCen,Wijnands06} \\
    XMM J174457-2850.3 & 2008/6 & 2008/7 & 2.5e-10 & 1.1e23 & 1.43 & 1.3e36 & 6.5  & \citet{Degenaar10,Degenaar14_XMMJ1744} \\ %confirmed
    XMM J174457-2850.3 & 2009/9 & 2009/10 & 2.2e-11 & 7.5e22 & 2.3 & 1.1e35 & 6.5  & \citet{Degenaar10,Degenaar14_XMMJ1744}  \\
    XMM J174457-2850.3 & 2010/7 & 2010/8 & 3.8e-11 & 7.5e22 & 2.0 & 1.9e35 & 6.5  & \citet{Degenaar14_XMMJ1744,Degenaar15} \\
    %XMM J174457-2850.3 & 2011/05 & 2011/11 & 4e-12 & & & 2e34 & 6.5 & \citet{Degenaar16_174457}  \\
    XMM J174457-2850.3 & 2012/8 & 2012/8 & 1.9e-10 & -&- & 9.8e35 & 6.5 & \citet{Degenaar14_XMMJ1744,Degenaar15} \\
    XMM J174457-2850.3 & 2016/9 & 2016/10 & 3.3e-11 & 1.1e23 & 1.6 & 1.7e35 & 6.5 &  \citet{Degenaar16_174457} \\ %confirmed
    \hline
    AX J1745.6-2901 & 1993/10 & 1994/10 & 1.9e-10 & 2.8e23 & 2.4 & 1.5e36 & 8 &  \citet{Maeda96}  \\
    AX J1745.6-2901 & 1997/3 & 1997/3 & 2.9e-11 & 1.3e23 & 1.44 & 2.2e35 & 8 &  \citet{Sakano02}  \\
    AX J1745.6-2901 & 2006/2 & 2006/6 & 1.2e-10 & 2.4e23 & 2.4 & 9.2e35 & 8 & \citet{Degenaar10}   \\
    AX J1745.6-2901 & 2007/2 & 2008/9 & 8e-10 & 2.4e23 & 2.7 & 6.1e36 & 8 &  \citet{Degenaar10}  \\
    AX J1745.6-2901 &  2010/6 & 2010/11 & 8e-11 & 2.3e23 & 3.4 & 6.1e35 & 8 &   \citet{Degenaar10_AXJ1745}  \\
    AX J1745.6-2901 & 2013/7 & 2016/6 & 3.3e-10 & 2e23 & 2.6 & 6.7e36 & 8 &  \citet{Degenaar14}  \\
    AX J1745.6-2901 & 2017/4 & 2017/4 & 1e-11 & 2e23 & 2 & 8e34 & 8 &  \citet{Degenaar17_AXJ1745a}   \\
    AX J1745.6-2901 &  2017/10 & 2017/10 & 5.1e-11 & 3.4e23 & 1.9 & 4e35 & 8 &  \citet{Degenaar17_AXJ1745b}  \\
    AX J1745.6-2901 & 2019/9 & 2020/2 & 3.6e-11 & 2e23 & 3.3 & 3e35 & 8 &  \citet{Degenaar19_AXJ1745,Degenaar20_AXJ1745}  \\
    AX J1745.6-2901 & 2021/7 & 2021/7 & 1.1e-10 & 2.2e23 & 3.5 & 8.4e35 & 8 &  \citet{Degenaar21_AXJ1745}   \\
    AX J1745.6-2901 & 2023/6 & 2025/5* & 3.4e-11 & 2.2e23 & 1.0 & 2.6e35 & 8 &  \citet{Reynolds23_AXJ1745,Degenaar25} \\
    \hline   
    GRO J1744-28 & 1995/12 & 1996/5 & 3.5e-8 & 3.5e22 & 1.2 & 6.7e37 & 4 &  \citet{Fishman95,Giles96} \\
    GRO J1744-28 & 1996/12 & 1997/4 & 2.1e-8 & -&- & 4.0e37 & 4 & \citet{Kouveliotou97,Woods99} \\
    GRO J1744-28 & 2014/1 & 2014/1 & 9.7e-9 & 8.6e22 & 0.99 & 1.9e37 & 4 & \citet{Negoro14_GROJ1744,Degenaar14_GRO_J1744} \\
    GRO J1744-28 & 2017/2 & 2017/2 & 4.4e-10 & 7e22 & 1.2 & 8.4e35 & 4 & \citet{Mereminskiy17_GROJ1744,Konig20,Sanna17_atel} \\
    GRO J1744-28 & 2021/6 & 2021/6 & 1.2e-9 & 9e22 & 1.1 & 2.3e36 & 4 & \citet{Mereminskiy21}  \\
    \hline 
    EXO 1745-248 & 2000/7 & 2000/8 & 1.4e-8 & -&- & 6e37 & 5.9 &   \citet{Markwardt00_T5,Heinke03_ter5} \\
    EXO 1745-248 & 2011/10 & 2011/11 & 2.1e-9 & 1.2e22 & 2.25 & 8.8e36 & 5.9 &     \citet{Altamirano11_T5,Serino12}\\
    EXO 1745-248 & 2015/3 & 2015/6 & 2.0e-8 & 3.0e22 & 0.93 & 8.4e37 & 5.9 &  \citet{Altamirano15,Tetarenko16b} \\
    \hline
    IGR J17480-2446 & 2010/10 & 2010/11 & 1.4e-8 & 1.2e22 & 2.1 & 5.1e37 & 5.9 &  \citet{Bordas10,Patruno12,Miller11} \\
    \hline
    Swift J174805.3-244637 & 2012/7 & 2012/8 & 1.1e-8 & -& 1.6 & 4.6e37 & 5.9 &  
    \citet{Wijnands12,Bahramian14} \\
    Swift J174805.3-244637 & 2023/2 & 2023/3 & 2.5e-9 & 2.2e22 &- & 1.0e37 & 5.9 &   \citet{Negoro23_T5X3,Heinke23_Ter5,Sanna23_ter5} \\
    \hline 
    IGR J17473-2721 & 2005/5 & 2005/5 & 1.2e-9 & -&- & 4.3e36 & 5.5 &  \citet{Grebenev05,Markwardt05_XTE_J1747} \\
    IGR J17473-2721 & 2008/3 & 2008/8 & 1.3e-8 & 3.8e22 & -& 4.7e37 & 5.5 &  \citet{DelMonte08,Altamirano08_XTE_J1747,Chenevez11_IGR_J17473} \\
    \hline 
    XMMU J174716.1-281048 & 2003/3 & 2012/10 & 7.4e-12 & 9e22 & 2.25 & 6.3e34 & $<$8.4 &  \citet{Sidoli03,Degenaar07_174716,Kaur17} \\
    XMMU J174716.1-281048 & 2021/4 & 2021/4 & 5.9e-12 & -& -& 5e34 & $<$8.4 & Swift/XRT \\
    \hline 
    EXO 1747-214 & 1984/4 & 1985/4 & 1.6e-9 & 1.1e22 & 2.4 & 2.4e37 & $<$11 &  \citet{Parmar85,Tomsick05} \\
    \hline 
    SAX J1747.0-2853 & 1976/6  & 1976/6  & 7.9e-10 &  -& -& 5.3e36 & 7.5 &   \citet{Proctor78,Grebenev02}    \\
    SAX J1747.0-2853 & 1991/10 & 1991/10 & 4.2e-10 & 7e22 & 2.4 & 2.8e36 & 7.5 &  \citet{Grebenev02}    \\
    SAX J1747.0-2853 & 1998/3 & 1998/3 & 5e-10 & 7e22 & 2.4 & 3.4e36 &  7.5 &  \citet{intZand98_SAX_J1747,Sidoli98}    \\
    SAX J1747.0-2853 & 1999/3 & 1999/3 & 3.4e-9 & 8.8e22 & 2.4 & 2.3e37 &  7.5  & \citet{Werner04}    \\
    SAX J1747.0-2853 & 2000/2 & 2000/9 & 1.1e-8 & 7e22 & 2 & 7.4e37 &  7.5 &  \citet{Markwardt00_SAX_J1747,Wijnands02_SAX_J1747,Werner04} \\
    SAX J1747.0-2853 & 2001/9 & 2001/9 & 5.4e-9 & 7e22 & 2 & 3.6e37 &   7.5 &  \citet{Wijnands02_SAX_J1747,Werner04} \\
    SAX J1747.0-2853 & 2004/3 & 2004/3 & 6.4e-9 & & -& 4.3e37 & 7.5 &   \citet{Markwardt04b} \\
    SAX J1747.0-2853 & 2005/10 & 2006/9 &  3.9e-10 & 9.5e22 & 2 & 2.6e36 & 7.5 &  \citet{Wijnands05_SAX_J1747,Degenaar12} \\
    SAX J1747.0-2853 & 2007/10 & 2007/10 & 2.2e-9 & -& -& 1.5e37 & 7.5 &  \citet{Brandt07} \\
    SAX J1747.0-2853 & 2009/3 & 2009/3 & 3.4e-11 & -& -& 2.3e35 & 7.5 &  \citet{Chenevez09_SAXJ1747,Campana09_SAX_J1747} \\
    SAX J1747.0-2853 & 2011/2 & 2011/2 & 6e-9 & -& -& 4e37 & 7.5 &  \citet{Kennea11_SAX_J1747},Swift/XRT \\
    SAX J1747.0-2853 & 2012/2 & 2012/2 & 7.1e-9 & -& -& 4.8e37 & 7.5 &  \citet{Chenevez12} \\
    SAX J1747.0-2853 & 2013/4 & 2013/4 & 3.5e-9 & -& -& 2.4e37 & 7.5 &  \citet{Negoro13_SAX_J1747} \\
     % Early 2014 MAXI flare from diff source!!!
    SAX J1747.0-2853 & 2014/10 & 2014/10 & 4.1e-10 & -& -& 2.7e36 & 7.5 &  \citet{Chenevez14} \\
    SAX J1747.0-2853 & 2016/6 & 2016/6 & 4.8e-9 & -& -& 3.2e37 & 7.5 &   \citet{Clavel16} \\
    SAX J1747.0-2853 & 2017/4 & 2017/10 & 5.4e-10 & -& -& 4e36 & 7.5 &  \citet{Heinke17},Swift/XRT \\
    SAX J1747.0-2853 & 2019/2 & 2019/9 & 1.3e-8 & - & -& 9e37 & 7.5 &  \citet{Negoro19_SAX_J1747,Mereminskiy19},Swift/XRT \\
    SAX J1747.0-2853 & 2022/4 & 2022/8 & 7e-9 & -&- & 5e37 & 7.5 &  \citet{Iwakiri22,Kennea22_SAX_J1747},Swift/XRT \\
    SAX J1747.0-2853 & 2023/5 & 2023/6 & 1.4e-9 & -&- & 9.4e36 & 7.5 &  \citet{Negoro23_1747,Li23_SAX_J1747},MAXI,Swift/XRT \\
    \hline
    SAX J1748.9-2021? & 1971/12 & 1972/1 & 7.3e-9 & 5.3e22 & 2.1 & 6.3e37 & 8.5 &   \citet{Markert75,Forman76} \\
    SAX J1748.9-2021 & 1998/8 & 1998/8 & 1.3e-9 & 8.2e21 & 1.44 & 1.1e37 & 8.5 &  \citet{intZand98_6440,intZand99} \\
    SAX J1748.9-2021 & 2001/8 & 2001/11 & 4.3e-9 & -& -& 3.7e37 & 8.5 &  \citet{intZand01} \\
    SAX J1748.9-2021 & 2005/5 & 2005/7 & 5.4e-9 & -& -& 4.7e37 & 8.5 &  \citet{Markwardt05_6440,Altamirano08_6440} \\
    SAX J1748.9-2021 & 2009/12 & 2010/1 & 5.4e-9 & -& -& 4.7e37 & 8.5 &  \citet{Suzuki09,Patruno10_6440} \\
    SAX J1748.9-2021 & 2015/2 & 2015/4 & 4e-9 & 5.7e21 & 2.3 & 3.4e37 & 8.5 &  \citet{Kuulkers15,Pintore16} \\
    SAX J1748.9-2021 & 2017/10 & 2017/10 &  1.9e-10 & 5.1e21 & 1.6 & 1.6e36 & 8.5 &  \citet{Negoro17_6440,Pintore18} \\
    SAX J1748.9-2021 & 2021/11 & 2021/12 & 4.9e-10 & 6e22 & 1.8 & 4.2e36 & 8.5 &  \citet{Negoro21_6440,Pike21_6440a,Pike21_6440b} \\
    \hline 
    IGR J17494-3030 & 2012/3 & 2012/3 & 1.0e-10 & 1.8e22 & 1.9 & 7.7e35 & 8 &  \citet{Boissay12,ArmasPadilla13}  \\
    IGR J17494-3030 & 2020/10 & 2020/10 & 1.4e-10 & -& -& 1.0e36 & 8 &  \citet{Ducci20,Ng21_IGR_J17494}\\
    \hline
    Swift J1749.4-2807 & 2006/6 & 2006/6 & 9.5e-11 & 3e22 &  & 5e35 & 6.7  & \citet{Schady2006,Wijnands09} \\
    Swift J1749.4-2807 & 2010/4 & 2010/4 & 5.2e-10 & -& -& 2.8e36 & 6.7  &   \citet{Pavan10,Ferrigno11,Altamirano11} \\
    Swift J1749.4-2807 & 2021/3 & 2021/3 & 3.1e-10 & 4.5e22 &- & 1.7e36 & 6.7  & \citet{Mereminskiy21_J1749.4,Marino22} \\
    \hline 
    IGR J17498-2921 & 2011/8 & 2011/9 & 5.1e-10 & 1.2e22 & 1.8 & 2.0e36 & 5.7 &  \citet{Gibaud11,Falanga12,Linares11}   \\
    IGR J17498-2921 & 2023/4 & 2023/5 & 6.1e-10 & 1.2e22 & 1.8 & 2.4e36 & 5.7 &  \citet{Grebenev23,Sanna23_IGR_J17498}, NICER  \\
    \hline 
    SAX J1750.8-2900 & 1997/3 & 1997/3 & 3.5e-9 & 6e22 & 2.4 & 1.9e37 & 6.79 &   \citet{Bazzano97,Natalucci99} \\
    SAX J1750.8-2900 & 2001/3 & 2001/3 & 3.5e-9 & -& -& 4.2e36 & 6.79 &  \citet{Kaaret02}  \\
    SAX J1750.8-2900 & 2008/3 & 2008/10 & 5.1e-9 & 2.3e22 & 1.2 & 2.8e37 & 6.79 &  \citet{Markwardt08_J1750,Linares08_J1750.8}   \\
    SAX J1750.8-2900 & 2011/2 & 2011/2 & 3.6e-10 & 3.2e22 & 2 & 2.0e36 & 6.79 &  \citet{Fiocchi11,Natalucci11} \\
    SAX J1750.8-2900 & 2015/9 & 2016/3 & 2.4e-9 & 2.3e22 & 2 & 1.3e37 & 6.79 &   \citet{Cummings15_SAX_J1750.8,SanchezFernandez15}, IGR/JEM-X \\
    SAX J1750.8-2900 & 2018/9 & 2018/9 & 4.6e-10 & -&- & 2.5e36 & 6.79 &  \citet{Bozzo18_SAX_J1750.8} \\
    SAX J1750.8-2900 & 2019/4 & 2019/6 & 3.8e-10 & 3e22 & 2 & 2.1e36 &  6.79 &  \citet{Heinke19may,Heinke19} \\
    \hline 
    XTE J1751-305 & 1998/6 & 1998/6 & 1.7e-9 & -& 1.8 & 1.5e37 & 8.5 &  \citet{Markwardt02}  \\
    XTE J1751-305 & 2002/4 & 2002/4 & 1.1e-9 & -& 1.8 & 9.9e36 & 8.5 &  \citet{Markwardt02_IAUC,Markwardt02} \\
    XTE J1751-305 & 2005/3 & 2005/3 & 1.8e-10 & -& -& 1.6e36 & 8.5 &  \citet{Grebenev05_XTEJ1751,Swank05_XTE_J1751} \\
    XTE J1751-305 & 2007/4 & 2007/4 & 2.2e-10 & -&- & 1.9e36 & 8.5 &  \citet{Markwardt07_XTE_J1751} \\
    XTE J1751-305 & 2009/10 & 2009/10 & 3.2e-10 & 9.8e21 & -& 2.8e36 & 8.5 &  \citet{Chenevez09,Miller03} \\
    \hline  
    IGR J17511-3057 & 2009/9 & 2009/9 & 4.5e-10 & 5e21 & 1.2 & 2.6e36 & 6.9 &   \citet{Baldovin09,Bozzo10} \\
    IGR J17511-3057 & 2015/3 & 2015/4 & 5e-10 & 1.2e22 & 1.5 & 2.8e36 & 6.9 &  \citet{Bozzo15_J17511,Papitto16}   \\
    IGR J17511-3057 & 2025/2 & 2025/2* & 3e-10 & - & - & 2e36 & 6.9 & \citet{Sguera25,Sguera25b} \\
    \hline 
    SAX J1753.5-2349? & 2006/4 & 2006/4 & 2.4e-10 & -& -& 1.8e36 & 8 &  \citet{Turler06} \\
    SAX J1753.5-2349 & 2008/10 & 2008/10 & 1.7e-10 & 1.9e22 & 2.1 & 1.3e36 & 8 &  \citet{Markwardt08_SAXJ1753.5,DelSanto10}  \\
    SAX J1753.5-2349 & 2010/3 & 2010/3 & 2.8e-10 & -& 2.1 & 2.2e36 & 8 &  \citet{Chenevez10_J1753} \\
    \hline 
    AX J1754.2-2754 & 1999/10 & 2008/5 & 1.1e-11 & 2.1e22 & 2.54 & 5.7e34 & 6.6 &  \citet{Sakano02,Bassa08_AXJ1754}  \\
    AX J1754.2-2754 & 2008/10 & 2012/5 & 1.1e-11 & 2.3e22 & 2.9 & 5.7e34 & 6.6 &  \citet{Jonker08} \\
    AX J1754.2-2754 & 2014/2 & 2015/3 & 3e-12 & 2.1e22 & 2.5 & 1.6e34 & 6.6 &  Swift/XRT \\
    AX J1754.2-2754 & 2015/8 & 2015/8 & 4.6e-12 & 2.1e22 & 2.5 & 2.4e34 & 6.6 &  Swift/XRT \\
    AX J1754.2-2754 & 2015/10 & 2017/3 & 1.3e-11 & 2.1e22 & 2.5 & 7e34 & 6.6 &  \citet{Chenevez17}, Swift/XRT \\
    \hline
    Swift J1756.9-2508 & 2007/6 & 2007/6 & 3.7e-10 & 6.2e22 & 2.2 & 2.8e36 & 8 &  \citet{Krimm07,Krimm07_atel} \\
    Swift J1756.9-2508 & 2009/8 & 2009/8 & 5.7e-10 & -&- & 4.4e36 & 8 &   \citet{Markwardt09,Patruno09_SwiftJ1756.9} \\
    Swift J1756.9-2508 & 2018/4 & 2018/4 & 2.5e-10 & 4.3e22 &- & 1.9e36 & 8 &  \citet{Mereminskiy18,Rai19} \\
    Swift J1756.9-2508 & 2019/6 & 2019/6 & 2.6e-10 & 4.2e22 & -& 2.0e36 & 8 &   \citet{Sanna19,Li21_Swift_J1756.9} \\
    \hline
   IGR J17591-2342 & 2018/8 & 2018/10 & 3.2e-10 & 4.4e22 & 2.4 & 2.5e36 & 8 &   \citet{Ducci18,Gusinskaia20} \\
   \hline 
    \revise{IGR J17597-2201} & 2001/2 & 2008/3 & 4.6e-10 & 4.5e22 & 1.7 & 3.5e36 & 8 & \citet{Lutovinov03,Markwardt03_17597,Bodaghee12} \\
   \hline 
    2S 1803-245 & 1976/5 & 1976/5 & 2.2e-8 & 1.47e22 & -& 1.4e38 & $<$7.3 &  \citet{Jernigan76,Cornelisse07} \\
    2S 1803-245 & 1998/4 & 1998/4 & 9.1e-9 & -& -& 5.8e37 & $<$7.3 &  \citet{Muller98,Marshall98_2S_1803} \\
    \hline 
    1RXS J180408.9-342058 & 1990 & 1990 & 2.0e-12 & -& 2.3 & 2.4e34 & 10.0 &  \citet{Voges99} \\
%    1RXS J180408.9-342058 & 2012/4 & 2012/4 & 4.1e-13 & 4.6e21 & 3.6 & 4.8e33 & 10.0 &  \citet{Chenevez12_1RXS_J1804} \\
    1RXS J180408.9-342058 & 2015/1 & 2015/6 & 2.5e-9 & 3.3e21 & -& 3.0e37 & 10.0 &  \citet{Krimm15,Degenaar16_RXS_J1804,Parikh17_1804}  \\
    \hline 
    SAX J1806.5-2215 & 1996/3 & 1997/10 & 2.5e-10 & -&- & 1.9e36 & $<$8 & \citet{Cornelisse02_chandra} \\
    SAX J1806.5-2215 & 2011/2 & 2017/7 & 1.5e-9 & 4.2e22 & 1.7 & 1.1e37 & $<$8 &  \citet{Altamirano11_SAX_J1806,DelSanto12,Shaw17}, Swift/XRT \\
    \hline 
    MAXI J1807+132 & 2017/3 & 2017/3 & 2.2e-10 & -& -& 4e36 & $<$12.4 &   \citet{Negoro17_1807,Shidatsu17} \\
    MAXI J1807+132 & 2019/9 & 2019/9 & 1.9e-10 & 1.3e21 & 2 & 3.5e36 & $<$12.4 &  \citet{Shidatsu19,Albayati21} \\
    MAXI J1807+132 & 2023/7 & 2023/10 & 2.7e-10 & 1.3e21 & 2 & 4.9e36 & $<$12.4 &    \citet{Saikia23,Illiano23_MAXI_1807,Rout25}, Swift/XRT \\
    \hline 
    SAX J1808.4-3658 & 1996/9 & 1996/10 & 2.2e-9 & -&- & 1.9e36 & 2.7  &   \citet{intZand98_SAX_J1808,Revnivtsev03}   \\
    SAX J1808.4-3658 & 1998/4 & 1998/4 & 1.3e-9 & -& -& 1.1e36 & 2.7 &  \citet{Marshall98_SAXJ1808,Chakrabarty98,Bult15} \\
    SAX J1808.4-3658 & 2000/1 & 2000/4 & 2.8e-10 & 1.3e21 & 2.16 & 2.4e35 & 2.7 &   \citet{vanderKlis00,Wijnands01_SAX_J1808} \\
    SAX J1808.4-3658 & 2002/10 & 2002/11 & 1.8e-9 & -&- & 1.6e36 & 2.7 &  \citet{Markwardt02_1808,Bult15} \\
    SAX J1808.4-3658 & 2005/5 & 2005/7 & 1.3e-9 & -& -& 1.1e36 & 2.7 &  \citet{Markwardt05_1808,Bult15} \\
    SAX J1808.4-3658 & 2008/9 & 2008/11 & 1.2e-9 & -&- & 1.0e36 & 2.7 &  \citet{Markwardt08_1808,Bult15}   \\
    SAX J1808.4-3658 & 2011/11 & 2011/11 & 1.7e-9 & -&- & 1.5e36 & 2.7 &  \citet{Markwardt11,Bult15}  \\
    SAX J1808.4-3658 & 2015/4 & 2015/4 & 1.2e-9 & 1e21 & 1.6 & 1.0e36 & 2.7 &  \citet{Sanna15_1808_initial,Sanna15_1808}  \\
    SAX J1808.4-3658 & 2019/8 & 2019/8 & 2.1e-10 & 2.1e21 & 2 & 1.8e35 & 2.7 &  \citet{Goodwin19_1808,Russell19_1808,Bult19_1808} \\
    SAX J1808.4-3658 & 2022/8 & 2022/10 & 3.7e-10 & -& -& 3.2e35 & 2.7 &   \citet{Sanna22_1808,Imai22,Illiano23} \\
    \hline 
    XTE J1807-294 & 2003/2 & 2003/6 & 1.3e-9 & 5.6e21  & -& 9.6e36 & 8 &   \citet{Markwardt03,Falanga05}  \\
    \hline 
    XTE J1810-189 & 2008/3 & 2008/6 & 1.1e-9 & 3.8e22 & 1.7 & 4.0e36 & 5.5 &  \citet{Markwardt08_XTEJ1810,Chelovekov09,Weng15}    \\
    XTE J1810-189 & 2009/1 & 2009/3 &  1.9e-10 & 3.8e22 & 1.7 & 6.9e35 & 5.5 &  \citet{Chelovekov09},RXTE/PCA    \\
    XTE J1810-189 & 2011/6 & 2011/6 & 2.9e-9 & 4.9e22 & 2.6 & 1.0e37 & 5.5 &  \citet{Rangelov14}  \\
    XTE J1810-189 & 2012/12 & 2013/1 & 8.6e-10 & -& -& 3.1e36 & 5.5 &   \citet{Negoro13_XTE_J1810} \\
    XTE J1810-189 & 2020/8 & 2020/11 & 5.2e-10 & 6e22 & 1.8 & 1.9e36 & 5.5 &  \citet{Bozzo20_Swift,Bozzo20_Integral,Manca23}  \\
    XTE J1810-189 & 2023/2 & 2023/10 & 1e-9 & -&- & 3.6e36 & 5.5 &   \citet{Kobayashi23}, MAXI \\
    \hline
    SAX J1810.8-2609 & 1998/3 & 1998/3 & 3.1e-10 & -& -& 1.2e36 & $<$5.7 &  \citet{Ubertini98,Natalucci00} \\
    SAX J1810.8-2609 & 2002/4 & 2002/4 & 1.0e-10 & 5e21 & 2 & 3.9e35 & $<$5.7 &  RXTE/PCA,RXTE/ASM \\
    SAX J1810.8-2609 & 2007/8 & 2007/10 & 8e-10 & 5e21 & 2 & 3e36 & $<$5.7 &  \citet{Parsons07,Degenaar07_1810,Degenaar13_1810}   \\
    SAX J1810.8-2609 & 2012/5 & 2012/5 & 5.2e-9 & -&- & 2.0e37 & $<$5.7 &  \citet{Degenaar13_1810} \\
    SAX J1810.8-2609 & 2018/4 & 2018/4 & 1.8e-9 & -&- & 7e36 & $<$5.7 &  \citet{Negoro18_1810} \\ 
    SAX J1810.8-2609 & 2021/5 & 2021/10 & 2.1e-9 & -&- & 8.2e36 & $<$5.7 &  \citet{Iwakiri21,Hughes23} \\
    \hline 
    XTE J1812-182 & 2003/1 & 2003/3 & 1.9e-9 & 1.3e23 & 2.47 & 4.5e37 & 14 &  \citet{Cackett06} \\
    XTE J1812-182 & 2008/8 & 2008/10 & 2.2e-9 & -&- & 5.1e37 & 14 &  \citet{Markwardt08_XTEJ1812,Goodwin19} \\ 
    XTE J1812-182 & 2020/3 & 2020/3 & 2.0e-9 & -&- & 4.7e37 & 14 &  \citet{Chenevez20_1812}  \\
    \hline 
    XTE J1814-338 & 1984/9 & 1984/9 & 5e-11 & -&- & 3.8e35 & 8 &     \citet{Wijnands03_1814} \\
    XTE J1814-338 & 2003/6 & 2003/7 & 3.7e-10 & 1.6e21 & 2.1 & 2.8e36 & 8 &    \citet{Markwardt03_XTEJ1814,Papitto07,Krauss05} \\
    \hline 
    MAXI J1816-195 & 2022/6 & 2022/7 & 3.2e-9 & 2.4e22 & 2.6 & 1.5e37 & $<$6.3 &   \citet{Negoro22_1816,Kennea22_MAXI_J1816,Bult22_MAXI_J1816}\\
    \hline 
    Swift J181723.1-164300 & 2017/8 & 2017/8 & 3.6e-10 & 8e22 & 1.5 & 2.1e36 & 7 & \citet{Barthelmy17,Parikh17}  \\
    \hline 
    IGR J18245-2452 & 2013/3 & 2013/4 & 1.2e-9 & 4e21 & 1.26 & 4.3e36 & 5.5 &  \citet{Eckert13,Papitto13,Linares14}  \\
    \hline 
    SAX J1828.5-1037 & 1993/4 & 1993/4 & 7.2e-12 & 4.1e22 & 1.7 & 3.3e34 & $<$6.2 &   \citet{Cornelisse02} \\
    SAX J1828.5-1037 & 2001/3 & 2001/3 & 7.5e-12 & 5e22 &- & 6e34 & $<$6.2 &  \citet{Hands04}  \\
    SAX J1828.5-1037 & 2008/10 & 2008/11 & 1.5e-11 & 4.1e22 & 1.6 & 6.9e34 & $<$6.2 &   \citet{Campana09,Degenaar08_SAXJ1828}, Swift/XRT \\
    SAX J1828.5-1037 & 2011/11 & 2011/11 & 2.9e-10 & 4.1e22 & 1.6 & 1.3e36 & $<$6.2 &  \citet{Asada11,Serino16}, Swift/XRT \\
    \hline 
    Swift J185003.2-005627 & 2011/5 & 2011/7 & 1.2e-10 & 1.1e22 & 1.6 & 2.0e35 & $<$3.7 &  \citet{Beardmore11,Degenaar12_two_new} \\
    \hline 
    Swift J1858.6-0814 & 2018/10 & 2020/5 & 2.1e-9 & 1.8e23 & -& 4.1e37 & 12.8 & \citet{Krimm18,Negoro20_SwiftJ1858,Hare20}    \\
    \hline 
        HETE J1900.1-2455 & 2005/6 & 2015/10 & 8e-10 & 2e21 & -& 2.1e36 & 4.7 & \citet{vanderspek05,Degenaar17_HETE_J1900,Kaaret06}  \\
    \hline 
    1H 1905+000 & 1974/11 & 1985/9 & 1.5e-10 & 1.9e21 & 2.1 & 1.0e36 & 7.5 & \citet{Seward76,Jonker06} \\
    \hline
    Swift J1922.7-1716 & 2004/11 & 2006/7 & 1.3e-10 & 2.9e21 & 2.1 & 3.6e35 & 4.8 &  \citet{Falanga06,Degenaar12_two_new} \\
    Swift J1922.7-1716 & 2011/7 & 2011/8 & 2.1e-10 &   2.7e21 & 2.0 & 5.8e35 & 4.8 &  \citet{Nakahira11,Degenaar12_two_new} \\
    Swift J1922.7-1716 & 2024/2 & 2024/5 & 4.2e-10 & - & 2.2 & 1.2e36 & 4.8 & \citet{Nakajima24},Swift/XRT \\
    \hline 
    MAXI J1957+032 & 2015/5 & 2015/5 & 2.0e-10 & 1.7e21 & 2 & 6.0e35 & 5 & \citet{Negoro15_1957,MataSanchez17} \\
    MAXI J1957+032 & 2015/10 & 2015/10 & 3.2e-10 & -&- & 9.5e35 & 5 &  \citet{Sugimoto15,MataSanchez17} \\
    MAXI J1957+032 & 2016/1 & 2016/1 & 4e-10 & -&- & 1.2e36 & 5 & \citet{Tanaka16,Beri19} \\
    MAXI J1957+032 & 2016/9 & 2016/9 & 6.6e-10 & -&- & 2.0e36 & 5 & \citet{Negoro16_1957,MataSanchez17} \\
    MAXI J1957+032 & 2022/6 & 2022/6 & 8e-10 & -&-  & 2.4e36 & 5 & \citet{Negoro22,Sanna22}  \\
    MAXI J1957+032 & 2025/5 & 2025/5  & 9e-10 & - &  -& 3e36 & 5 & \citet{Sun25,Negoro25,Illiano25} \\
    \hline
    XTE J2123-058 & \revise{1998/6} & \revise{1998/8} & 1.7e-9 & 6e20 & 2 & 1.5e37 & 8.5 & \citet{Levine98,Tomsick99,Tomsick04}  \\
    \hline  
    4U 2129+47 & 1970 & 1983/9 & 4.8e-10 & 2e21 & - & 1.8e35 & 1.75 & \citet{Forman78,Bozzo07}  \\
   % 4U 2129+47 & 2019/12 & & nope  \\
    \hline % reviewed to here May 2025 ***************
    \enddata
     	\label{tab:outbursts_app}
    \tablecomments{Published outbursts of verified neutron star LMXBs. Max $F_X$ is highest reported flux in each outburst, transformed to 2-10 keV, generally using the power-law spectral index \& $N_H$ noted. We assume 1 Crab is 
    $F_X$(2-10)=$2\times10^{-8}$ erg cm$^{-2}$ s$^{-1}$. 
    Peak observed intrinsic $L_X$ in 2-10 keV, for the assumed distance (listed). A * at the end date indicates that we do not have evidence that the outburst ended--the last recorded observation found it active. Reference to an instrument (e.g. MAXI) indicates we have made use of that instrument's archive to generate some information for that line.}  
    %We give one or two references that include the most information about the outbursts; details (e.g. distance estimates) were often produced in papers cited by the references, such as \citet{Galloway08}.
  \end{deluxetable}
\end{longrotatetable}
\twocolumngrid

Notes on individual sources, except the frequent outbursters (which are in Appendix B):

{\bf IGR J00291+5934}; the distance estimate is 4.2$\pm0.5$ kpc, from a photospheric radius expansion (PRE) burst \citep{deFalco17}.

{\bf MAXI J0556-332}; the distance of 43.6$^{+0.9}_{-1.6}$ kpc is from modeling the quiescent X-ray spectrum \citet{Parikh17}, and is supported by comparison of its outburst behaviour \citep{Homan14}.

{\bf EXO 0748-676}; \citet{Galloway08} derive the distance of 7.4$\pm0.9$ kpc from a PRE burst. %\revise{Swift/XRT observations indicate EXO 0748-676's 2024 outburst ended between February and April 2025.}

{\bf GS 0836-429}; also known as 4U 0836-429, 1M 0836-425, MX 0836-42. \citet{Aranzana16} place a 9.2 kpc upper limit on the distance from a burst. Note that we use the $N_H$ value of $2.7\times10^{22}$ cm$^{-2}$ quoted by \citet{Chelovekov05}, which \citet{Aranzana16} misquote as $2.7\times10^{21}$ cm$^{-2}$.

{\bf MAXI J0911-655}; also known as Swift J0911.9-6452. The 9.5 kpc distance is the estimate to the host globular cluster, NGC 2808 \citep{Watkins15}.  Its long outburst appears to have ended around October 2023 \citep{HeinkeDegenaar23}.
%MAXI no longer detects it after Oct. 2023, and a short Swift/XRT observation on Nov. 30, 2023 places a limit of $L_X$(2-10 keV)$<3\times10^{33}$ erg/s, indicating its outburst has ended.

{\bf XTE J0929-314}; \citet{Galloway02} estimate a lower limit on the distance ($>6$ kpc) from binary evolution considerations.

{\bf SAX J1324.5-6313}; the distance limit of $<6.2$ kpc is from a burst \citep{Cornelisse02}.

{\bf MAXI J1421-613}; \citet{Nobukawa20} estimate the distance as 3 kpc from analysis of dust scattering echoes.

{\bf Cen X-4} or 4U 1456-32; \citet{Kaluzienski80} estimate a distance of 1.3$\pm0.7$ kpc from the X-ray burst. We reference $N_H=9\times10^{20}$ from \citet{Chakrabarty14}. 

{\bf MAXI J1621-501}; \citet{Chenevez18} estimate 8.4$\pm2$ kpc from a PRE burst. 

{\bf MAXI J1647-227}; the 10 kpc rough distance estimate was made by \citet{Onodera14}.

{\bf MXB 1659-298} or MXB 1658-298, or MXB 1659-29, or H 1658-298; the 10 kpc distance estimate is from a PRE burst \citep{Sharma18}.

{\bf IGR J16597-3704}; the 9.1 kpc distance estimate is from the globular cluster NGC 6256 \citep{Valenti07}.

{\bf XTE J1701-407}; the distance of 6.2$\pm0.5$ kpc was estimated from a PRE burst \citep{Falanga09,Linares09}. It is not clear whether XTE J1701-407 has ever reached full quiescence since 2016.

{\bf XTE J1701-462}; \citet{Lin09b} measure a distance of 8.8$\pm1.3$ kpc via PRE bursts. Swift/XRT observations in March 2023 showed its 2023 outburst declining below $L_X<10^{34}$ erg/s.

{\bf IGR J17062-6143}; \citet{Keek17} measure a distance of   7.3$\pm0.5$ kpc from a PRE burst. We use $N_H=2.3\times10^{21}$ cm$^{-2}$ from \citet{Degenaar17_igr_j17062}.

{\bf XTE J1709-267}, or RX J1709.5-2639; \citet{Jonker03} suggest a distance of 8.5 kpc if it is associated with the nearby globular cluster NGC 6293.

{\bf 2S 1711-339} or 4U 1711-34; the distance limit of 7.5 kpc is from a burst \citep{Cornelisse02}.

{\bf 1M 1716-315}; also known as MXB 1715-32, XB 1715-321, or 1H 1715-321. The distance of 6.9 kpc is estimated from a PRE burst \citet{Jonker04}.

{\bf IGR J17191-2821}; the distance limit of 11 kpc is from a burst \citep{Markwardt07}.

{\bf XTE J1723-376}; the 8 kpc distance limit is from a burst \citep{Marshall99b}.

{\bf 4U 1730-22}; \citet{Bult22} measure the distance of 6.9$\pm0.2$ kpc from PRE bursts. \revise{The end of its second outburst is uncertain, but it was clearly in quiescence in a Swift/XRT observation in May 2025.}

 {\bf KS 1731-260}; \citet{Muno00} estimate the distance from PRE bursts. 

{\bf XB 1733-30}; this quasi-persistent transient in Terzan 1 is also known as SLX 1732-304, XB 1732-304, and 1RXS J173546.9-302859. We use the globular cluster distance estimate of 5.7$\pm0.2$ kpc from \citet{Baumgardt21}.

{\bf Swift J1734.5-3027} or IGR J17344-3023; \citet{Bozzo15} measured a distance of 7.2$\pm1.5$ kpc from a PRE burst.

{\bf IGR J17379-3747} or IGR J17380-3749; this is assumed to be located in the Galactic bulge, for a distance of roughly 8.5 kpc \citet{Chelovekov10}. 

{\bf XTE J1739-285}; \citet{Galloway08} find a distance constraint of $<$7.3 kpc from the brightest bursts, assuming a H-rich composition, which appears appropriate for the brightest bursts given their longer rise and decay times \citep{Galloway08}.

{\bf GRS 1741.9-2853}; \citet{Pike21} measure a distance of 9.0$\pm0.5$ kpc from a 2020 May PRE burst. (This is farther than the 7 kpc distance estimate inferred from prior X-ray bursts, which however did not show PRE; \citet{Barriere15}.)
The proximity of this source to the Galactic Centre prevents us from using all-sky monitors to track its recurrences.

{\bf KS 1741-293}; \citet{Chelovekov11} place an upper limit of $<$6.2 kpc from analyzing bursts. 
While this source has generally been seen not to exceed a few $10^{36}$ ergs s$^{-1}$ (see e.g. \citealt{Degenaar13_KS1741}), we notice that the 2011-2012 outburst was reported by \citet{Chenevez12} using INTEGRAL/JEM-X to reach a 3-10 keV flux of 415$\pm75$ mCrab, an order of magnitude brighter. We also identify a Swift/XRT observation on May 14, 2017 that reached 5.2 cts/s, or $L_X\sim6.4*10^{36}$ erg/s.

 {\bf 1A 1744-361}, or A 1744-361; its proximity to the Galactic center has led many workers to assume an 8 kpc distance, though the only burst constraint is $<$11 kpc \citep{Galloway08}. 
 The RXTE/PCA Bulge scans show three likely small outbursts, at MJD 52020, 53281, and 54188, all reaching around $F_X=5\times10^{-11}$ ergs \revise{cm$^{-2}$ s$^{-1}$}. Given the crowded environment, it would be useful to verify these signals. 1A 1744-361 is not clearly resolved by MAXI, so our sensitivity to its outbursts declines after 2011.

 {\bf IGR J17445-2747}; the proximity of this source to the Galactic Centre suggests an 8 kpc distance. \citet{Mereminskiy17} place limits of $>$5 and $<$12.3 kpc, from analysis of a burst. 

{\bf XMM J174457-2850.3}; \citet{Degenaar14} measure a distance of 6.5$\pm0.2$ kpc from a PRE burst. We use the long-term Swift XRT monitoring lightcurve \citep{Degenaar15} to check for additional outbursts (with signal/error $>$4,  $L_X>2\times10^{34}$ erg/s), and find only one \citep[in 2016,][]{Degenaar16_174457} between 2012 and 2022. This appears to be a substantial change from its frequent outbursting behavior before 2012 \citep{Degenaar14_XMMJ1744}.

{\bf AX J1745.6-2901}; the proximity to the Galactic Centre suggests an 8 kpc distance, which is supported by the peak fluxes of the brightest X-ray bursts \citep{Degenaar09}.  (Note this is not the same as 1A 1742-289, \citealt{Kennea96}.)
 
{\bf GRO J1744-28}; although the proximity to the Galactic Centre suggests a distance of 8 kpc, several methods of estimating the distance seem to converge on a value around 4 kpc \citep{Sanna17_GRO_J1744,Gosling07,Wang07}.
This adjusts the $L_X$ of the brightening episode in Sept. 2008 to $L_X=5\times10^{33}$ erg/s, below our outburst definition. The 2011 MAXI outburst initially identified as GRO 1744-28 \citep{Suwa11} was from SAX J1747.0-2853 \citep{Kennea11_SAX_J1747}. 

 {\bf EXO 1745-248 (or XB 1745-248), IGR J17480-2446 (or Ter 5 X-2), Swift J17480.5-244637 (or Ter 5 X-3, or MAXI J1747-249)}: These are all located in the globular cluster Terzan 5.  Additional outbursts have been seen in 1981, 1991 \citep{Johnston95}, 2002 \citep{Wijnands02_ter5}, and 2004 \citep{Markwardt04b}, but these cannot be confidently identified with one of the identified transients, which require arcsecond-scale localizations in this crowded cluster. We use the distance of 5.9 kpc from \citet{Valenti07}.

{\bf IGR J17473-2721} is also known as XTE J1747-274.  The distance estimate is from PRE bursts \citep{Chenevez11_IGR_J17473}.

\begin{figure}
   % \centering
    \includegraphics[width=3.3in]{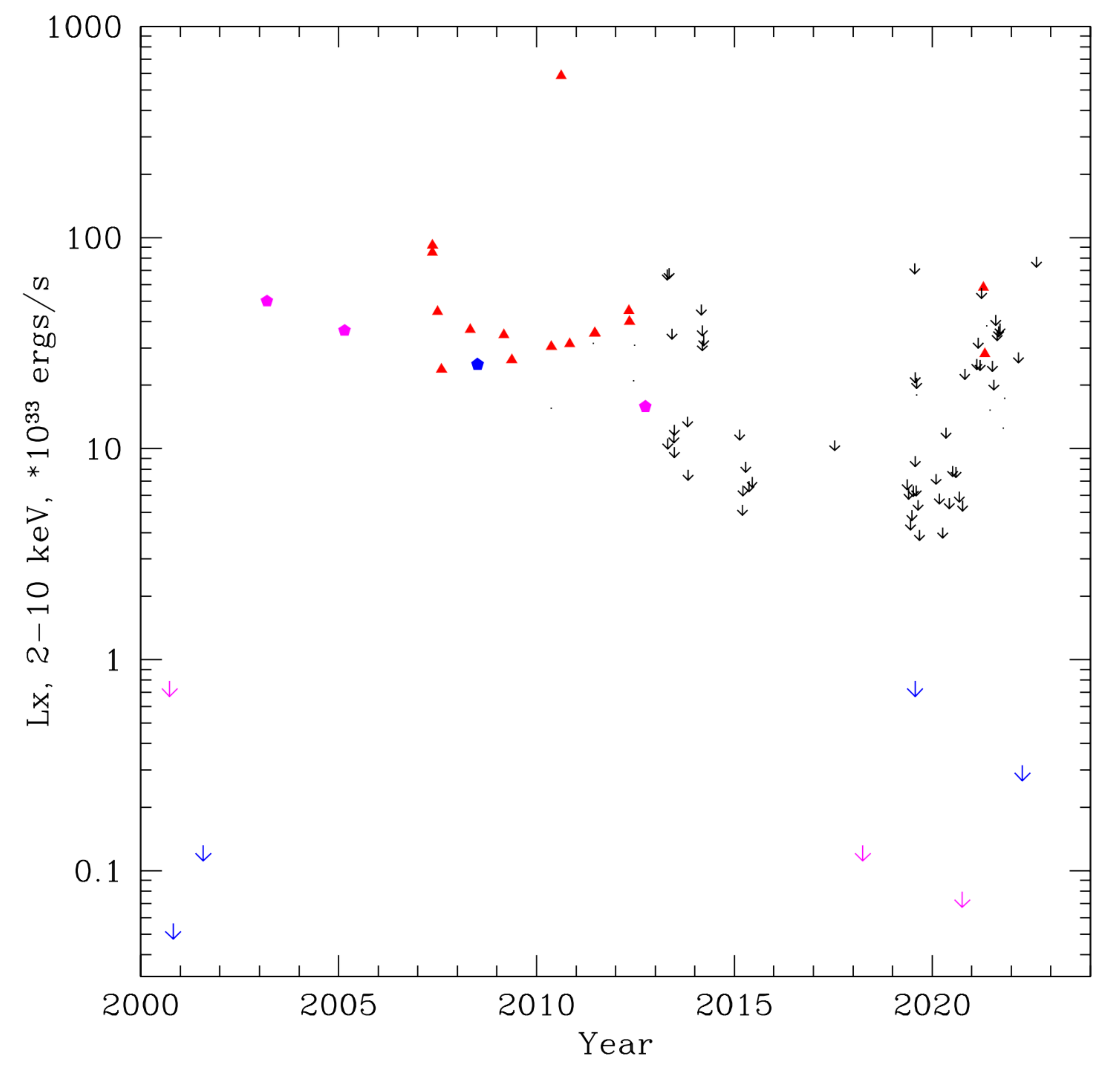}
 \caption{The Swift/XRT, Chandra, and XMM-Newton lightcurve of XMMU J174716.1-281048. Red filled triangles are Swift/XRT detections, black downward arrows are Swift/XRT upper limits. Blue filled pentagons, and downward triangles, are Chandra data, while magenta filled pentagons and downward triangles are from XMM-Newton. 
 }
    \label{fig:XMMU_J1747_lc}
\end{figure}

{\bf XMMU J174716.1-281048}; also known as IGR J17464-2811. The distance estimate is an upper limit from a burst \citep{Degenaar11_XMMU_J174716}. 
%The last reported
\citet{DelSanto15} report that XMMU J174716.1-281048 was not detected by Swift  in 2014 or 2015, but it was not clear if it had returned to full quiescence.
Extensive Swift/XRT observations, and several archival XMM-Newton and Chandra observations, allow an accurate flux history of this source. We utilize the Leicester Swift/XRT lightcurves \citep{Evans07}, archival Chandra observations from 2000, 2001, 2019, and 2022 (ObsIDs 1036, 2271, 21624, and 24222, the first 2 previously reported by \citealt{delSanto07}), and archival XMM-Newton observations from 2012, 2018, and 2020 (ObsIDs 0694641401, 0802410101, 0862471201), along with XMM flux values from 2003 and 2005 from \citet{delSanto07}, to construct a longer lightcurve for XMMU J174716.1-281048 (Fig.~\ref{fig:XMMU_J1747_lc}. This reveals evidence that XMMU J174716.1-281048 declined significantly in late 2012 and had entered deep quiescence (2-10 keV $L_X<10^{33}$ erg/s) by 2018, most likely earlier. The Swift/XRT data also reveal that XMMU J174716.1-281048 experienced a second outburst in April 2021, but limit the period brighter than $L_X=3\times10^{34}$ erg/s to between 2 weeks and 3 months. XMM and Chandra show that it was in deep quiescence in October 2020 and again by April 2022.

{\bf EXO 1747-214}; the distance upper limit is from the EXOSAT burst \citep{Tomsick05}.

{\bf SAX J1747.0-2853:} 
we use the distance of 7.5$\pm1.3$ kpc from PRE bursts of \citet{Werner04}. We find a much smaller flux ($6\times10^{-9}$ ergs \revise{cm$^{-2}$ s$^{-1}$}, 2-10 keV unabsorbed) in Feb. 2011 than \citet{Kennea11_SAX_J1747}. The MAXI page notes that the apparent flare in MAXI data in early 2014 is from a different source. %It is unclear whether SAX J1747.0-2853 has been active since May 2023.

{\bf SAX J1748.9-2021} and {\bf NGC 6440 X-2}: these two accreting millisecond X-ray pulsars are both located in the globular cluster NGC 6440 X-2, at d=8.5 kpc \citep{Harris1996}. They can be distinguished by their spatial positions (using {\it Chandra}, Swift/XRT, or radio interferometric imaging), or by detection of X-ray pulsations. SAX J1748.9-2021 has been verified to be the object in outburst in all bright detected outbursts, except the 1971 outburst, but the similarity of that outburst to others verified to be SAX J1748.9-2021 suggest that it was responsible. 

NGC 6440 X-2 showed numerous weak, short X-ray outbursts in 2009 and 2010, often with a 1-month recurrence time \citep{Heinke10_6440paper}\footnote{Note that \citet{Heinke10_6440paper} quote $N_H=0.59\times10^{21}$ cm$^{-2}$ in their Table 2, but actually used $N_H=0.59\times10^{22}$ cm$^{-2}$.}. 
The RXTE PCA bulge scan reveals at least 7 faint outbursts during 2009-2011, but does not show other similar features except one in 2007, and possibly 4 in 1999. 
It is not certain whether NGC 6440 X-2 only produced outbursts in 2007 and 2009-2011, or whether we missed others (due perhaps to changes in the outburst peak flux). We list NGC 6440 X-2's outbursts in a separate table (Table~\ref{tab:6440_x2}), to quote their MJD dates.

\onecolumngrid

\begin{table}
	\centering
	\begin{tabular}{lcccccl} %
    \hline
	NS XRB & Year/Month & MJD & Peak $F_X$  & Index & $L_X$ &   References \\
	\hline
    NGC 6440 X-2 & 2007/4 & 54205 & 1.4e-10 & 2 & 1.2e36 & \citet{Heinke10_6440paper},  RXTE/PCA \\
    NGC 6440 X-2 & 2009/3 & 54910 & 1.9e-10 & 2 & 1.6e36 & \citet{Heinke10_6440paper},  RXTE/PCA \\
    NGC 6440 X-2 & 2009/5-6 & 54980 & 1e-10 & 2 & 8.6e35 & \citet{Heinke10_6440paper},  RXTE/PCA \\
    NGC 6440 X-2 & 2009/7 & 55040 & 1.7e-10 & 2 & 1.5e36 & \citet{HeinkeBudac09,Heinke10_6440paper},  RXTE/PCA  \\
    NGC 6440 X-2 & 2009/8-9 & 55072 & 1.6e-10 & 1.79 & 1.4e36 & \citet{Heinke09_again6440,Heinke10_6440paper,Altamirano10_6440},  RXTE/PCA \\
    NGC 6440 X-2 & 2009/10 & 55105 & 5.6e-11 & 1.79 & 4.8e35 & \citet{Heinke10_6440paper} \\
    NGC 6440 X-2 & 2009/10-11 & 55132  & 8.1e-11 & 1.8 & 7e35 & \citet{Heinke10_6440paper} \\
    NGC 6440 X-2 & 2010/3 & 55274 &  1.5e-10 & 1.9 & 1.3e36 & \citet{Altamirano10_6440march}, RXTE/PCA \\
    NGC 6440 X-2 & 2010/6 & 55359 & 2e-10 & -& 1.7e36 & \citet{Patruno10_6440,Patruno13}, RXTE/PCA \\
    NGC 6440 X-2 & 2010/10 & 55473 & 1.8e-10 & -& 1.6e36 & \citet{Patruno13}, RXTE/PCA \\
    NGC 6440 X-2 & 2011/1 & 55584 & 2e-10 & -& 1.7e36 & \citet{Patruno13} \\
    NGC 6440 X-2 & 2011/3 & 55639 & 2.8e-10 & -& 2.4e36 & RXTE/PCA \\
    NGC 6440 X-2 & 2011/9 & 55819 & 2.2e-10 & -& 1.9e36 & RXTE/PCA \\
    NGC 6440 X-2 & 2011/11 & 55871 & 2e-10 & -& 1.7e36 & \citet{Patruno13} \\ 
    \hline 
	\end{tabular}
	\caption{Outbursts of NGC 6440 X-2. $F_X$ in 2-10 keV ergs \revise{cm$^{-2}$ s$^{-1}$}, $L_X$ in 2-10 keV ergs s$^{-1}$. $N_H$ is about $6\times10^{21}$ cm$^{-2}$.  D=8.5 kpc \citep{Harris1996}.  RXTE/PCA indicates detection in RXTE/PCA bulge scans.
 }
 \label{tab:6440_x2}
\end{table}

\twocolumngrid

{\bf IGR J17494-3030}; \citet{ArmasPadilla13} assume a distance of 8 kpc, due to its proximity to the Galactic Center.

{\bf Swift J1749.4-2807}; \citet{Wijnands09} calculate a distance of 6.7$\pm1.3$ kpc, but this is an upper limit as the burst is not clearly PRE.

{\bf IGR J17498-2921}; \citet{Linares11} estimate a 7.6 kpc distance from a PRE burst, but \citet{Galloway24} infer $5.7^{+0.6}_{-0.5}$ kpc from more detailed modeling, which we adopt. NICER observations of the 2023 outburst show a rapid drop to quiescence in May 2023.

{\bf SAX J1750.8-2900}; \citet{Galloway08} calculate a distance of 6.79$\pm0.14$ kpc from PRE bursts.

{\bf XTE J1751-305}; \citet{Papitto08} discuss various estimates for the distance, from 6.7 to 11.6 kpc; they and other authors use a default distance of 8.5 kpc, as it is close to the Galactic Center.

{\bf IGR J17511-3057}; the distance upper limit is 6.9 kpc, from bursts \citep{Altamirano10_IGR_J17511}.

{\bf SAX J1753.5-2349}; the distance is assumed to be 8 kpc. We include an INTEGRAL-detected outburst in 2006, designated IGR J17536-2339, which may or may not be this object (and thus give it a ? in the table).

{\bf AX J1754.2-2754}; The distance estimate of 6.6$\pm0.4$ kpc, from an INTEGRAL burst reported by \citet{Chenevez17}, assumes a helium composition for the atmosphere.

\begin{figure}
   % \centering
    \includegraphics[width=3.3in]{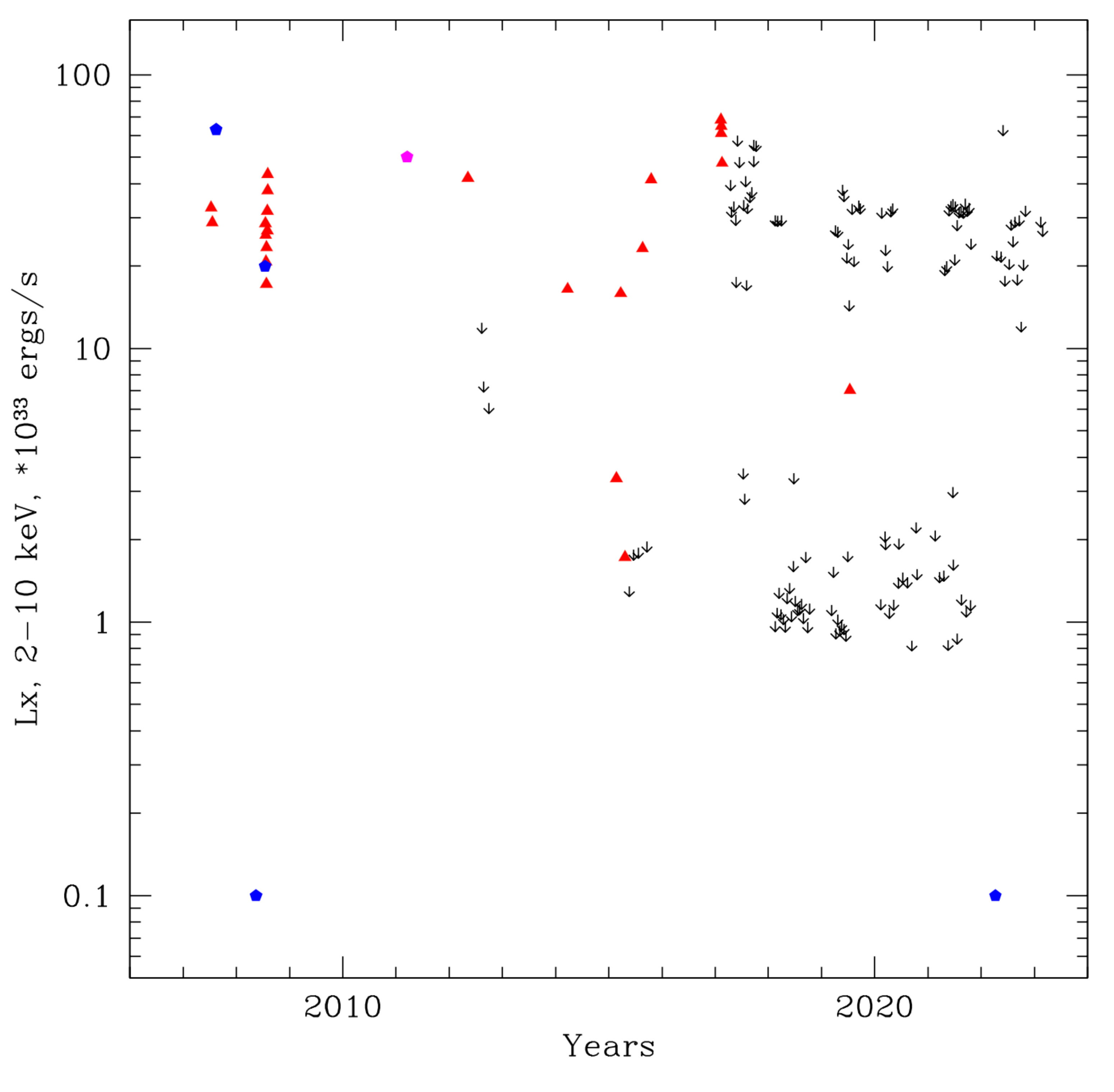}
 \caption{The Swift/XRT, Chandra, and XMM-Newton lightcurve of AX J1754.2-2754. Red filled triangles are Swift/XRT detections, black downward arrows are Swift/XRT upper limits. Blue filled pentagons are Chandra data, while the magenta filled pentagon is from XMM-Newton. 
 }
    \label{fig:AX_J1754.2-2754}
\end{figure}

After its initial ASCA detection in 1999, one short interval of quiescence has been reported (in 2007/2008, \citealt{Bassa08_AXJ1754}). 
However, the Swift/XRT lightcurve presented by \citet{Shaw17} suggests multiple additional quiescent periods.
We use Swift/XRT observations to identify additional quiescent episodes (with $L_X$(2-10 keV)$<10^{34}$ erg/s, or Swift/XRT count rates $<$0.04 cts/s), 
in August 2012, May-June 2015, Sept. 2015. Since Feb. 2017, it has been routinely observed by the Swift Bulge Survey \citep{Bahramian21}, with only one detection in July 2019 at $L_X$(2-10 keV)$=7\times10^{33}$ erg/s.
We used a recent 4.2 ks {\it Chandra} observation (May 5, 2022, ObsID 24082; PI G. Ponti) to place a strong upper limit of $L_X$(2-10 keV)$<1.0\times10^{32}$ erg/s on AX J1754.2-2754. We show the Swift/XRT data, the new and previously published {\it Chandra} data, and the {\it XMM-Newton} point of \citet{ArmasPadilla13} in Fig.~\ref{fig:AX_J1754.2-2754}.
We conclude that AX J1754.2-2754  spends much of its time 
at luminosities at least 10-100 times lower than its peak.
%in quiescence.

{\bf Swift J1756.9-2508}; the 8 kpc distance estimate is based on its proximity to the Galactic Center. \citet{Li21_Swift_J1756.9} place a lower limit on the distance of 4 kpc.

{\bf IGR J17591-2342}; \citet{Russell18} argue that the $N_H$ implies a distance $>$6 kpc; since it is near the Galactic Centre, it is typically taken to be 8 kpc away.

\revise{{\bf IGR J17597-2201}, or XTE J1759-220; the 8 kpc distance estimate assumes it is near the Galactic Center, which is consistent with the (unconstraining, $<16$ kpc) X-ray burst constraint \citep{Galloway08}.}

{\bf 2S 1803-245} is also known as XTE J1806-246, and SAX J1806.8-2435. \citet{Cornelisse07} place an upper limit of $<$7.3 kpc, from a burst in 1998.

{\bf 1RXS J180408.9-342058};
the 10.0$\pm1.4$ kpc distance estimate is from PRE bursts  \citep{Marino19}.

{\bf SAX J1806.5-2215} showed a long outburst, which started in Feb. 2011 \citep{Altamirano11_SAX_J1806}. This was reported in Atels until 2015 \citep{Sguera15}, but Swift/XRT observations indicate it continued until July 2017, and was no longer visible to Swift/XRT by January 2018.
The $<$8 kpc upper limit comes from an analysis of bursts by \citep{Cornelisse02_chandra}.

{\bf MAXI J1807+132} has an upper limit on its distance of 12.4 kpc, from bursts  \citep{Albayati21}.  Its 2023 outburst experienced two re-flares through September and October, visible in the Swift/XRT lightcurve; while it might rebrighten yet again, it appears to be in quiescence in Nov. 2023.

{\bf SAX J1808.4-3658}; \citet{Galloway06} estimated a distance of 3.5$\pm$0.1 kpc, using PRE bursts; %and the expected mass transfer rate from gravitational radiation. 
this was revised to $2.7\pm0.3$ kpc by \citet{Galloway24}.

{\bf XTE J1807-294}; the distance is assumed to be 8 kpc (based on its proximity to the Galactic Centre).

{\bf XTE J1810-189}; \citet{Weng15} estimate a distance of 5.5$\pm1.7$ kpc from a PRE burst. 

{\bf SAX J1810.8-2609}: the distance was constrained by an X-ray burst to $<$5.7 kpc by \citet{Fiocchi09}. 
We note several likely weak, single-datapoint outbursts in the RXTE/PCA lightcurve, that are not remarked on in the literature. At MJD 52384 we see a 10-sigma detection; also there is a 5-sigma RXTE/ASM detection at 52373. We include this outburst in our list. At MJD 53207, and 54989, we see 3-sigma possible outbursts, without RXTE/ASM detections; we leave these off our list. 

{\bf XTE J1812-182}: also known as 
XMMU J181227.8-181234. This may be the same as 1H1812-182, detected by HEAO-1 with $F_X$(2-10 keV)$=7\times10^{-10}$ erg/cm$^2$/s in 1977/78 \citep{Wood84}, but the errorbox is large. \citet{Goodwin19} estimated the distance as 14$\pm$2 kpc, from analysis of a train of several (non-PRE) bursts. 

{\bf XTE J1814-338}: 
\citet{Wijnands03_1814} argued this is the same as EXMS B1810-337. The distance estimate of 8 kpc from X-ray bursts is from \citet{Strohmayer03}.

{\bf MAXI J1816-195}; we use the $<$6.3 kpc distance constraint from bursts from 
\citet{Chen22_MAXI_J1816}.

{\bf Swift J181723.1-164300}; the 7$\pm$1 kpc distance comes from a burst \citep{Parikh17}.

{\bf IGR J18245-2452}; this accreting, bursting transitional millisecond pulsar is in the globular cluster M28, at a distance of 5.5 kpc \citep[][2010 revision]{Harris1996}. 
Although no other outbursts have been recorded from this cluster,
a burst during quiescence has been seen (see Appendix C for AX J1824.5-2451).

{\bf SAX J1828.5-1037}; the distance upper limit of $<$6.2 kpc is from a burst \citep{Cornelisse02}. The  flux quoted for the 2008 outburst by \citet{Campana09} appears to be in error by a factor of ten. 

{\bf Swift J185003.2-005627}; the distance estimate upper limit of 3.7 kpc is from the single observed burst \citep{Degenaar12_two_new}.

{\bf Swift J1858.6–0814}; \citet{Buisson20} use PRE bursts to estimate a distance of 12.8$^{+0.8}_{-0.6}$ kpc, though the large and variable $N_H$ ($1-4\times10^{23}$ cm$^{-2}$, \citealt{Hare20}) implies systematic uncertainties in this measurement.

{\bf HETE J1900.1-2455}; \citet{Galloway08} compute a distance of 4.7$\pm$0.6 kpc from a PRE burst.

{\bf 1H 1905+000} or 4U 1857+01, or 4U 1905+000; \citet{Jonker04} suggest either 7.5 or 10 kpc based on a PRE burst detected by \citet{Chevalier90}, depending if the burst was H-rich or He-rich. As \citet{Chevalier90} identify a moderately long rise time (2.5 s) and decay time ($\tau$=19 s), we suggest the burst was H-rich, and prefer the shorter 7.5 kpc distance.

{\bf Swift J1922.7-1716}; the distance estimate of 4.8 kpc is from two PRE bursts \citep{Degenaar12_two_new}.

{\bf MAXI 1957+032}, or IGR J19566+0326; the distance estimate of 5$\pm$2 kpc is from the hypothesized optical counterpart (which would require a triple system, \citealt{Ravi17}). 

{\bf XTE J2123-058}; the distance estimate of 8.5$\pm2.5$ kpc combines several constraints from bursts and studies of the optical counterpart in quiescence \citep{Tomsick04}.

{\bf 4U 2129+47} or XB 2129+47; \citet{Arnason21} show that the Gaia distance to the star generally believed to be the counterpart is only 1.75 kpc. This is an accretion-disk corona source, so the $L_X$ estimate may be underestimated.

MAXI reported an outburst in December 2019 at 14$\pm$3 mCrab, or $L_X=$1e35 erg/s  \citep{Kurihara19}, but it is unclear that this was a real detection; Swift observed within 2 days and only placed (unabsorbed) flux limits $F_X<7\times10^{-13}$ erg/s/cm$^{2}$, or $L_X<2.6\times10^{32}$ erg/s. We leave this possible outburst off our list.

\onecolumngrid

\section{Frequently outbursting sources}
\label{frequent}

{\bf 4U 1608-52}, or 4U 1608-522, X1608-52, QX Nor, \revise{in Table~\ref{tab:1608-52}.}
We assume a distance of 3.2$\pm$0.3 kpc \citep{Galloway08}. \revise{This is reasonably consistent with the estimate of \citet{Ozel16} of $3.6\pm0.3$ kpc.}
 %We note that \citet{Guver10} find a distance of 5.8$^{+2.0}_{-1.8}$ kpc, which is inconsistent with this value. A problem with the analysis of \citet{Guver10} is that they used the abundances of \citet{Wilms00} to estimate $N_H$, but inconsistently used the relation of \citet{GuverOzel09} to convert $N_H$ to optical, then infrared extinction. The \citet{GuverOzel09} relation assumed a different \citep{Anders89} set of abundances; the updated $N_H$/extinction relations of \citet{Bahramian15} or \citet{Foight16} which use \citet{Wilms00} abundances give an extinction only 79\% as large for the same $N_H$. From Fig. 6 of \citet{Guver10}, this suggests a revised distance estimate of 4.1$^{+1.0}_{-0.4}$ kpc, close to consistency with the \citet{Galloway08} estimate. 
 4U 1608-52 has remained continually in low-level outburst for several years at several points in its history, most notably 
 1999-2000 \citep{Wachter02} and 
 2007-2009 \citep{Simon20}, but also probably in the 1970s \citep{Simon20}. \revise{We see 4U 1608-52 to be in a continuous outburst from July 2022 through July 2023, with a normal,  bright, short outburst superimposed in November 2022.}

\begin{table}
	\centering
	\begin{tabular}{lcccl} %
		\hline
		NS XRB & Year/Month & Peak $F_X$  & $L_X$ &   References \\
        & & 2-10 keV &  2-10 keV &   \\
		\hline
	4U 1608-52 & 1970/4 & 2.0e-8 & 2.4e37 &    \citet{Lochner94} \\
 	4U 1608-52 & 1970/9 & 1.7e-8 &   2.1e37 &      \citet{Lochner94} \\
  	4U 1608-52 & 1971/6 & 1.5e-8 &   1.8e37 &      \citet{Matilsky72,Lochner94} \\
   	4U 1608-52 & 1971/9 & 1.7e-8 &   2.1e37 &      \citet{Lochner94,Tananbaum76} \\
   % \revise{4U 1608-52} & 1972 & e-8 & e37 & \citet{Tananbaum} \\
    4U 1608-52 & 1975/11 & 1.2e-8 &   1.5e37 &      \citet{Kaluzienski75} \\
    4U 1608-52 & 1977/7 & 2.4e-8 &   2.9e37 &      \citet{ClarkLi77} \\
    4U 1608-52 & 1979/1 & 6.5e-9 &   7.9e36 &      \citet{KaluzienskiHolt79} \\
    4U 1608-52 & 1979/4 & 1.3e-8 &   1.6e37 &      \citet{Oda79} \\
    4U 1608-52 & 1980/4 & $<$2.1e-10 &   $<$2.6e35 &      \citet{Oda80} \\
    4U 1608-52 & 1983/4 & 1.6e-8 &  2.0e37 &      \citet{Mitsuda84} \\
    4U 1608-52 & 1984/7 & 1.0e-9 &   1.2e36 &      \citet{Gottwald87} \\
    4U 1608-52 & 1986/3 & 5.8e-10 &   7.1e35 &      \citet{Penninx89} \\
    4U 1608-52 & 1991/6-12 & 9.7e-10 &    1.2e36 &      \citet{Zhang96} \\
    4U 1608-52 & 1996/3 & 7.1e-9 &   8.7e36 &      \citet{Marshall96_1608,Mendez98} \\
    4U 1608-52 & 1998/2-4 & 1.2e-8 &   1.5e37 &      \citet{Cui98_1608,Rutledge01atel} \\
    4U 1608-52 & 2000/1 & 4.4e-9 &   5.4e36 &      \citet{Simon04} \\
    4U 1608-52 & 2000/12 & 7.4e-9 &   9.0e36 &      \citet{Simon04} \\
    4U 1608-52 & 2001/11 & 1.5e-8 &   1.8e37 &      \citet{Rutledge01atel} \\
    4U 1608-52 & 2002/7 & 1.7e-8 &   2.1e37 &      \citet{Simon04,Keek08} \\
    4U 1608-52 & 2003/3 & 6.3e-9 &   7.7e36 &      \citet{Miller03atel} \\
    4U 1608-52 & 2004/3 & 5.4e-9 &   6.6e36 &     \citet{Remillard04atel} \\
    4U 1608-52 & 2004/7 & 2.9e-9 &  3.5e36 &   \citet{YanYu15} \\
    4U 1608-52 & 2005/3 & 2e-8 &   2.4e37 &      \citet{Keek08} \\
    4U 1608-52 & 2005/9 & 6.1e-9 &   7.4e36 &      \citet{Asai12} \\
    4U 1608-52 & 2006/7 & 1.9e-9 &   2.3e36 &     \citet{Markwardt06,Greco06} \\
    4U 1608-52 & 2007/6-2009/10 & 7.8e-8 &   9.5e37 &      \citet{Palmer07,Russell09atel} \\
    4U 1608-52 & 2010/2 & 6.3e-9 &   7.7e36 &      \citet{Morii10,ArmasPadilla17} \\
    4U 1608-52 & 2011/3 & 1.7e-8 &   2.1e37 &      \citet{Suguzaki11,Asai13} \\
    4U 1608-52 & 2012/10 & 1.2e-8 &  1.5e37 &   \citet{Sugimoto12} \\
    4U 1608-52 & 2013/5 & 3e-9 &   3.7e36 &      \citet{ZhangYu13} \\
    4U 1608-52 & 2014/1 & 4.1e-9 &   5.0e36 &      \citet{ZhangYu14} \\
    4U 1608-52 & 2014/10 & 6.4e-9 &   7.8e36 &      \citet{Negoro14} \\
    4U 1608-52 & 2016/8 & 5.5e-9 &   6.7e36 &      \citet{Ono16} \\
    4U 1608-52 & 2017/6 & 1.1e-8 &  1.3e37 &      \citet{Jaisawal19} \\
    4U 1608-52 & 2018/6 & 8.5e-9 &  1.0e37 &      \citet{vandenEijnden20} \\
    4U 1608-52 & 2019/3 & 5.7e-9 &   7.0e36 &     MAXI  \\
    4U 1608-52 & 2020/5 & 1.5e-8 &  1.8e37 &      \citet{Guver21} \\
    4U 1608-52 & 2021/8 & 2.8e-9 &   3.4e36 &       MAXI \\
    4U 1608-52 & 2022/7-2023/7 & 2.1e-8 &   2.6e37 &    \citet{Chen23_4U1608,Chen24}, MAXI \\
    4U 1608-52 & 2024/2 & 4e-9 & 5e36 & MAXI  \\
    4U 1608-52 & 2025/3 &  1.8e-8 & 2.2e37 & \citet{Wang25},MAXI \\
	\end{tabular}
	\caption{Outbursts of 4U 1608-52. The distance is assumed to be 3.2 kpc. $F_X$ in ergs \revise{cm$^{-2}$ s$^{-1}$}, $L_X$ in ergs s$^{-1}$.  Spectral fits assume $N_H=2\times10^{22}$ cm$^{-2}$, $\Gamma$=1.7 \citep{Mitsuda84} and $N_H=1.5\times10^{22}$, $\Gamma$=1.9 \citep{Gottwald87}. Reference to an instrument (e.g. MAXI) indicates we have made use of that instrument's archive to generate some information for that line. 
    }
    \label{tab:1608-52}
\end{table}

{\bf MXB 1730-335, aka Rapid Burster}, 4U 1730-335, or MXB 1730-33. We use a distance of 8.0$\pm$0.3 kpc, as measured to the host globular cluster, Liller 1 \citep{Baumgardt21}. 
We include a table for its frequent outbursts (Table~\ref{tab:RapidBurster}). 
We note that some of these outbursts may be produced by the second known transient in this cluster, CXOU J173324.6-332321, which was discovered in low-level outburst ($L_X$[0.5-10]$=5.5\times10^{35}$ erg/s) in April 2018 between two larger outbursts \citep{Homan18}, but \revise{likely} does not show Type II bursts as reported from many outbursts \citep[e.g.][]{Bagnoli15}. 
We quote MJD values for the highest observed countrate of each outburst. Where only intervals are given (e.g. in 1971), we estimate the midpoint of the interval.  
For spectral fitting, we assume a 1.6 keV blackbody, with $N_H=1.8\times10^{22}$ cm$^{-2}$ \citep{Marshall01,Mahasena03}.
We use RXTE/PCA Bulge Scan values where available, then RXTE/ASM, MAXI, or Swift/XRT. For RXTE/ASM we multiply countrates by 3.4$\times10^{-10}$ 
to get 2-10 keV fluxes, 
for RXTE/PCA by 2.3$\times10^{-12}$. 
For MAXI we subtract 0.07 cts/cm$^{-2}$/s (to account for confusion) from the 2-20 keV fluxes, and multiply by 1.1$\times10^{-8}$ to get 2-10 keV fluxes. 
For Swift/XRT data we multiply by 7.5$\times10^{-11}$ to get 2-10 keV fluxes, and for INTEGRAL/JEM-X 3-10 keV data we multiply by $1.7\times10^{-10}$. 
These different flux estimates are generally in reasonable agreement where they overlap, though the RXTE/ASM \revise{data} has some untrustworthy points when the Sun is very close. We identify ******* 52 outbursts that we do not see specifically identified elsewhere in the literature. 

\citet{Masetti02} identified a change in the Rapid Burster's outburst interval, from 200 to 100 days, and a factor of two decrease in peak luminosity, in the year 2000. We confirm that the outburst interval has remained generally between 100 and 170 days from 2000 to 2023, while the peak flux has generally averaged $L_X=2\times10^{37}$ erg/s in this time frame.  However, in 2023, the average peak flux decreased to $L_X=1\times10^{37}$ erg/s.  We see two intervals in the full record that are twice as long as nearby intervals, suggesting the RXTE instruments may have missed outbursts in Dec. 2001 and Dec. 2003 (due to proximity of the Sun).  We also see two unusually short intervals, of 28 and 37 days, suggesting the outbursts on MJD 56491 and 57131 (each observed only by MAXI) appear out of sequence, and thus might be due to other transients.

\begin{figure}
\centering
\includegraphics[width=0.48\textwidth]{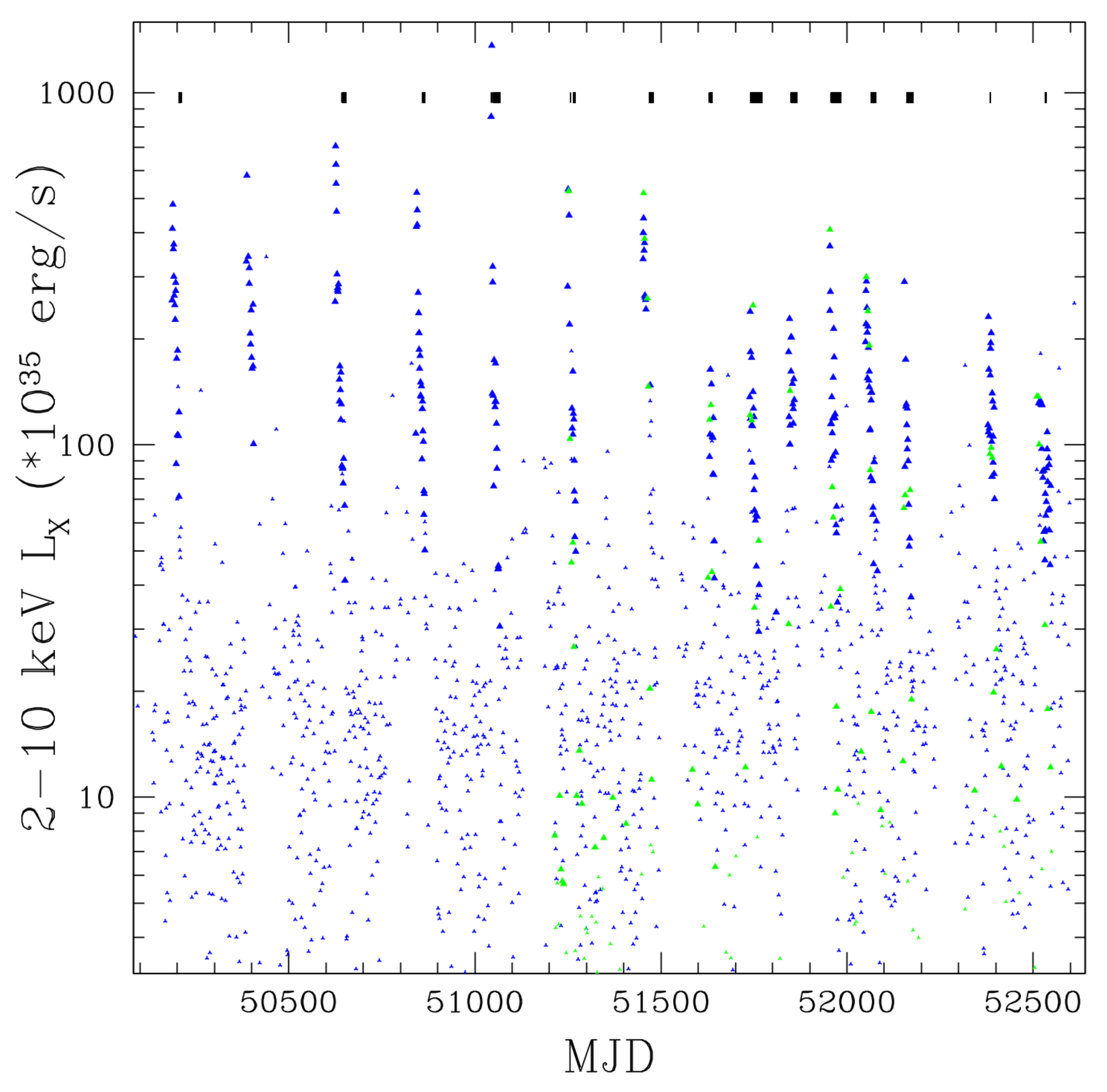} 
\includegraphics[width=0.485\textwidth]{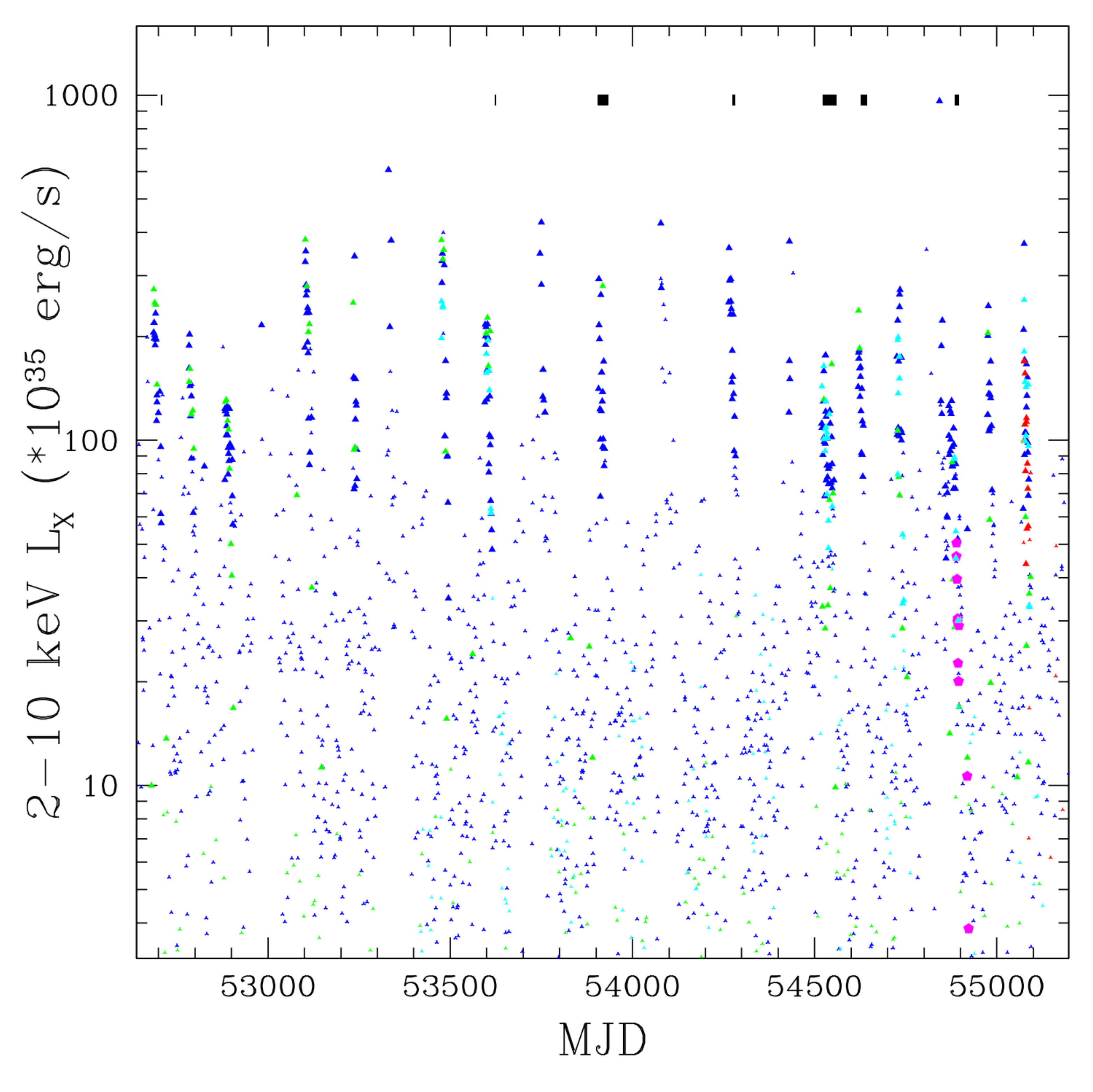} 
\includegraphics[width=0.48\textwidth]{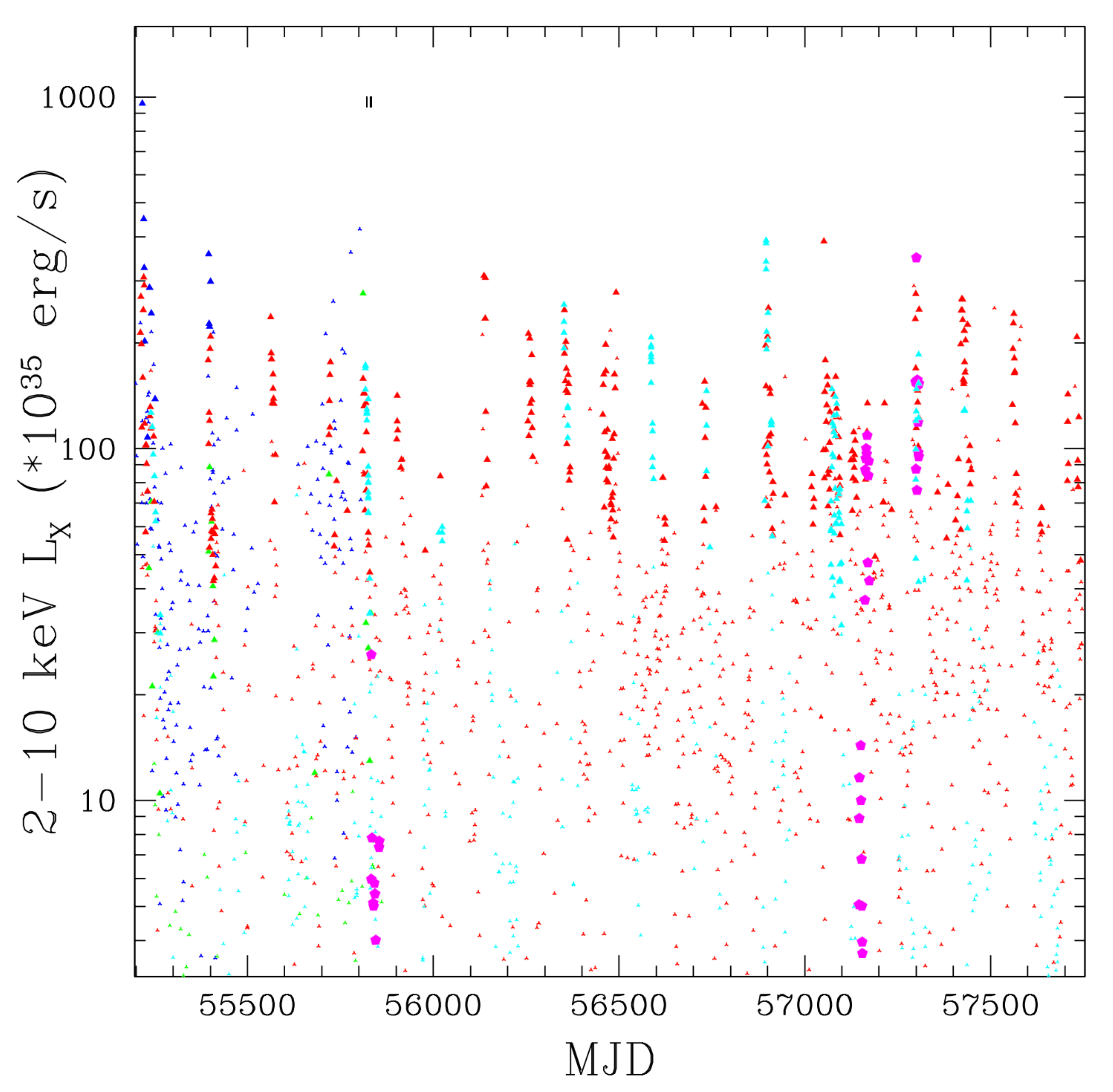} 
\includegraphics[width=0.485\textwidth]{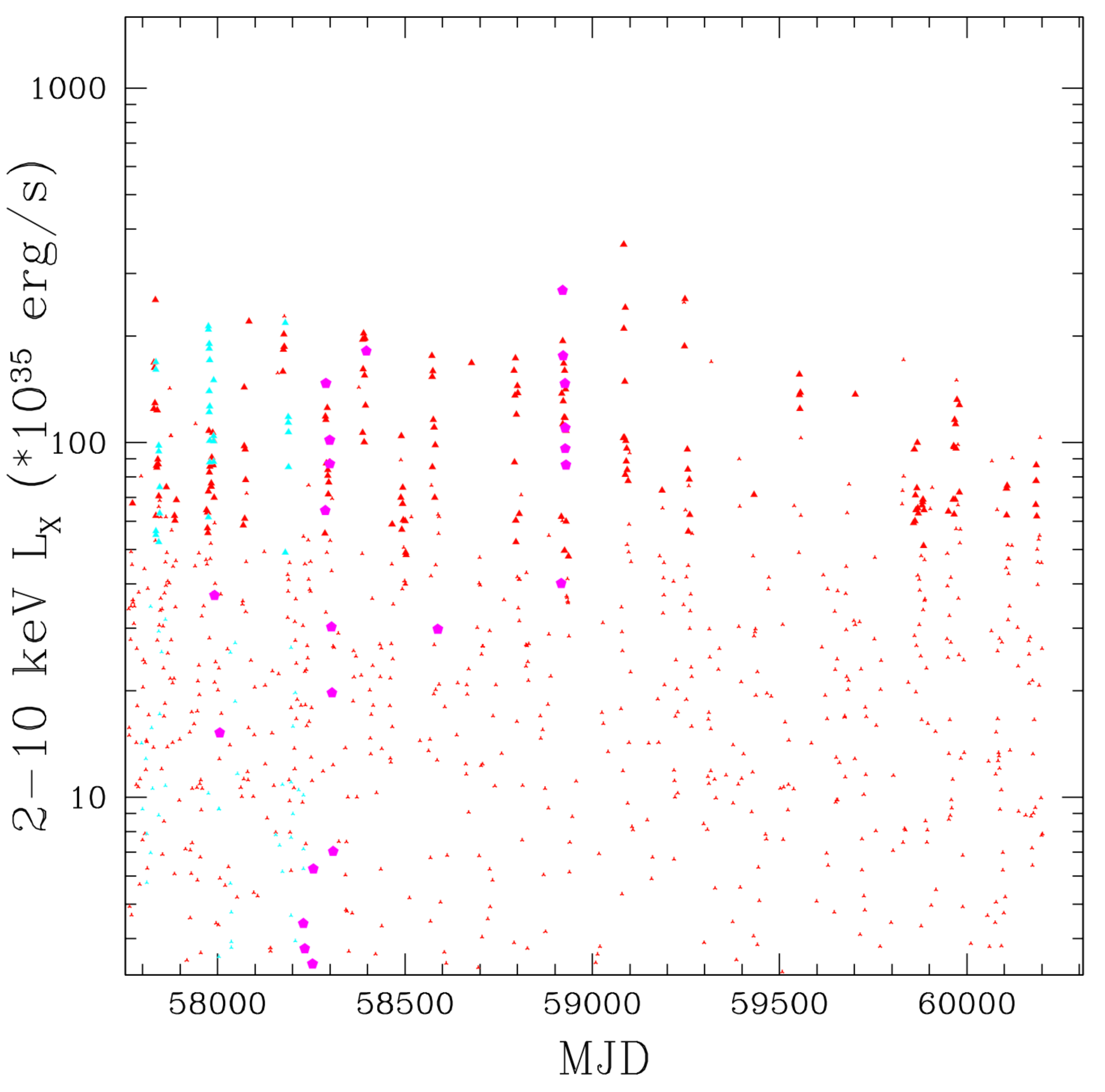} 
\caption{RXTE/PCA (green), RXTE/ASM (blue), MAXI/GSC (red), Swift/XRT (magenta), and INTEGRAL/JEM-X (cyan) lightcurves of MXB 1730-335 (the Rapid Burster). We also include the times of Type II bursts, as observed with the RXTE/PCA \citep{Bagnoli15}, indicated by short black lines at the top. Upper left, 1996-2002; upper right, 2003-2009; lower left, 2010-2016; lower right, 2017-2023.
 }
 \label{fig:RapidBurster}
\end{figure}

\clearpage

\startlongtable
\centerwidetable
\begin{deluxetable*}{lccccl}
\tablehead{		\colhead{NS XRB}  & \colhead{Year/Month} & \colhead{MJD} & \colhead{Peak $F_X$}  & \colhead{$L_X$} &   \colhead{References} \\
       \colhead{} & \colhead{} & \colhead{} & \colhead{2-10 keV} &  \colhead{2-10 keV} & }
  \startdata
     MXB 1730-335 & 1971/2 & 41028 & &    & \citet{GrindlayGursky77}\\
     MXB 1730-335 & 1972/4 & 41448 &  &   &\citet{GrindlayGursky77}\\
     MXB 1730-335 & 1976/2 & 42840 & 1.7e-9 &  1.3e37  & \citet{Lewin76} \\
     MXB 1730-335 & 1977/3-5 & 43266 & 2.2e-9 &   1.7e37  &\citet{White78}\\
     MXB 1730-335 & 1977/8 & 43408 & &    & \citet{GrindlayGursky77}\\
     MXB 1730-335 & 1978/3 & 43598 & &    & \citet{Hoffman78} \\
     MXB 1730-335 & 1978/9 & 43796 & &    & \citet{Lewin81} \\
     MXB 1730-335 & 1979/2-3 & 43936 & &    & \citet{Inoue80,Lewin81} \\
     MXB 1730-335 & 1979/8-9 & 44100 & &    & \citet{Inoue80}   \\
     MXB 1730-335 & 1983/4 & 45437 & &    & \citet{Kaminker90} \\
     MXB 1730-335 & 1983/8 & 45564 & 1.2e-9 &   9e36 & \citet{Tanaka83,Kunieda84,Barr87} \\
     MXB 1730-335 & 1984/7 & 45886 & 1.2e-9 &   9e36 & \citet{Makino84,Kawai90} \\
     MXB 1730-335 & 1985/8-9 & 46305 & 2.6e-9 &   2e37  & \citet{Stella85,Stella88} \\
     MXB 1730-335 & 1988/8 & 47385 & &    & \citet{Dotani90} \\
     MXB 1730-335 & 1989/1 & 47541 & &    & \citet{Lewin93} \\
     MXB 1730-335 & 1996/4 & 50189 & 6.3e-9 &  4.8e37 & \citet{Levine96_iauc,Guerriero99},RXTE/ASM  \\
     MXB 1730-335 & 1996/10 & 50387 & 7.7e-9 &  5.8e37 & \citet{Remillard96,Guerriero99},RXTE/ASM  \\
     MXB 1730-335 & 1997/6 & 50625 & 9.3e-9 &  7.0e37 & \citet{Guerriero97,Guerriero99},RXTE/ASM  \\
     MXB 1730-335 & 1998/1 & 50843 & 6.8e-9 &  5.2e37 & \citet{Guerriero98,Guerriero99,Masetti00},RXTE/ASM  \\
     MXB 1730-335 & 1998/8 & 51045 & 1.8e-8 &  1.4e38 & \citet{Fox98,Masetti02,Mahasena03},RXTE/ASM  \\
     MXB 1730-335 & 1999/3 & 51252 & 6.9e-9 & 5.3e37 & \citet{Masetti02,Mahasena03},RXTE/PCA, RXTE/ASM  \\
     MXB 1730-335 & 1999/10 & 51452 & 6.8e-9 &  5.2e37 & \citet{Fox99,Masetti02,Marshall01},RXTE/PCA,ASM  \\
     MXB 1730-335 & 2000/3 & 51633 & 2.2e-9 &  1.6e37 & \citet{Fox00,Masetti02},RXTE/ASM, RXTE/PCA   \\
     MXB 1730-335 & 2000/7 & 51747 & 3.3e-9 &  2.5e37 & \citet{Masetti02,Homer01},RXTE/PCA, RXTE/ASM  \\
     MXB 1730-335 & 2000/10 & 51849 & 2.7e-9 &  2.0e37 & \citet{Masetti02},RXTE/ASM, RXTE/PCA   \\
     MXB 1730-335 & 2001/2 & 51954 & 5.4e-9 &  4.1e37 & \citet{Masetti02},RXTE/PCA, RXTE/ASM  \\
     MXB 1730-335 & 2001/5 & 52052 & 3.9e-9 &  3.0e37 & \citet{Masetti02},RXTE/PCA, RXTE/ASM  \\
     MXB 1730-335 & 2001/9 & 52154 & 3.8e-9 &  2.9e37 & \citet{Masetti02},RXTE/PCA, RXTE/ASM  \\     
     MXB 1730-335 & 2002/4 & 52387 & 2.7e-9 &  2.1e37 & \citet{Simon08},RXTE/PCA, RXTE/ASM  \\
     MXB 1730-335 & 2002/8 & 52510 & 1.8e-9 &  1.4e37 & \citet{Simon08},RXTE/PCA, RXTE/ASM  \\
     MXB 1730-335 & 2003/2 & 52686 & 3.6e-9 &  2.7e37 & \citet{Falanga04,Simon08},RXTE/PCA, RXTE/ASM  \\
     MXB 1730-335 & 2003/5 & 52785 & 2.7e-9 &  2.0e37 & \citet{Simon08},RXTE/ASM, RXTE/PCA   \\
     MXB 1730-335 & 2003/9 & 52886 & 1.7e-9 &  1.3e37 & \citet{Simon08},RXTE/PCA, RXTE/ASM  \\
     MXB 1730-335 & 2004/4 & 53102 & 5.0e-9 &  3.8e37 & \citet{Simon08},RXTE/PCA, RXTE/ASM  \\
     MXB 1730-335 & 2004/8 & 53237 & 4.5e-9 &  3.4e37 & RXTE/PCA, RXTE/ASM  \\
     MXB 1730-335 & 2004/11 & 53330 & 8.0e-9 &  6.1e37 & RXTE/ASM  \\
     MXB 1730-335 & 2005/4 & 53475 & 5.0e-9 &  3.8e37 & RXTE/PCA, RXTE/ASM, IGR/JEM-X  \\
     MXB 1730-335 & 2005/8 & 53602 & 3.0e-9 &  2.3e37 & \citet{Kretschmar05}, RXTE/PCA, RXTE/ASM, IGR/JEM-X  \\
     MXB 1730-335 & 2006/1 & 53749 & 5.6e-9 &  4.3e37 & RXTE/ASM  \\
     MXB 1730-335 & 2006/7 & 53918 & 3.7e-9 &  2.8e37 & \citet{Bagnoli15},RXTE/PCA, RXTE/ASM  \\
     MXB 1730-335 & 2006/12 & 54078 & 5.6e-9 &  4.2e37 & RXTE/ASM  \\
     MXB 1730-335 & 2007/6 & 54265 & 4.8e-9 &  3.6e37 & RXTE/ASM  \\
     MXB 1730-335 & 2007/11 & 54430 & 5.0e-9 &  3.8e37 & RXTE/ASM  \\
     MXB 1730-335 & 2008/2 & 54546 & 2.2e-9 &  1.7e37 & \citet{Kuulkers08}, RXTE/PCA, RXTE/ASM, IGR/JEM-X  \\
     MXB 1730-335 & 2008/6 & 54620 & 3.1e-9 &  2.4e37 & RXTE/PCA, RXTE/ASM  \\
     MXB 1730-335 & 2008/9 & 54727 & 2.9e-9 &  2.2e37 & RXTE/ASM, RXTE/PCA, IGR/JEM-X  \\
     MXB 1730-335 & 2009/2 & 54876 & 1.6e-9 &  1.1e37 & \citet{Sala12}, RXTE/ASM, RXTE/PCA, Swift/XRT, JEM-X \\
     MXB 1730-335 & 2009/5 & 54976 & 3.2e-9 &  2.4e37 & RXTE/ASM, RXTE/PCA   \\
     MXB 1730-335 & 2009/8 & 55074 & 4.9e-9 &  3.7e37 & RXTE/ASM, RXTE/PCA, MAXI, IGR/JEM-X  \\
     MXB 1730-335 & 2010/2 & 55241 & 4.3e-9 &  3.3e37 & RXTE/PCA, RXTE/ASM, MAXI, IGR/JEM-X  \\
     MXB 1730-335 & 2010/7 & 55395 & 4.7e-9 &  3.6e37 &  RXTE/ASM, MAXI, RXTE/PCA  \\
     MXB 1730-335 & 2011/1 & 55562 & 3.1e-9 &  2.4e37 & MAXI \\
     MXB 1730-335 & 2011/6 & 55723 & 2.3e-9 &  1.8e37 & MAXI, RXTE/PCA  \\
     MXB 1730-335 & 2011/9 & 55811 & 3.6e-9 &  2.8e37 & \citet{Chenevez11}, RXTE/PCA, MAXI, Swift/XRT, IGR/JEM-X  \\
     MXB 1730-335 & 2011/12 & 55902 & 1.9e-9 & 1.4e37 & MAXI \\
     MXB 1730-335 & 2012/4 & 56019 & 1.1e-9 &  8.4e36 & MAXI, IGR/JEM-X \\
     MXB 1730-335 & 2012/7 & 56136 & 4.1e-9 &  3.1e37 & MAXI \\
     MXB 1730-335 & 2012/11 & 56256 & 2.8e-9 &  2.1e37 & MAXI \\
     MXB 1730-335 & 2013/3 & 56352 & 3.3e-9 &  2.5e37 & \citet{Chenevez13a_RapidBurster}, MAXI, IGR/JEM-X \\
     MXB 1730-335 & 2013/6 & 56463 & 2.6e-9 &  2.0e37 & MAXI \\
     MXB 1730-335 & 2013/7 & 56491 & 3.7e-9 &  2.8e37 & MAXI \\
     MXB 1730-335 & 2013/10 & 56586 & 2.6e-9 &  2.0e37 &  \citet{Chenevez13b}, IGR/JEM-X, MAXI \\
     MXB 1730-335 & 2014/3 & 56730 & 2.0e-9 &  1.5e37 & MAXI, IGR/JEM-X \\
     MXB 1730-335 & 2014/9 & 56896 & 5.2e-9 &  3.9e37 &  IGR/JEM-X, MAXI \\
     MXB 1730-335 & 2015/1 & 57051 & 5.1e-9 &  3.9e37 & MAXI, IGR/JEM-X \\
     MXB 1730-335 & 2015/4 & 57131 & 1.5e-9 & 1.1e37 & MAXI \\
     MXB 1730-335 & 2015/5 & 57168 & 1.8e-9 &  1.3e37 & Swift/XRT, MAXI \\
     MXB 1730-335 & 2015/10 & 57298 & 3.6e-9 &  2.8e37 & \citet{vandenEijnden17},MAXI, IGR/JEM-X, Swift/XRT \\
     MXB 1730-335 & 2016/2 & 57422 & 3.5e-9 &  2.7e37 & MAXI, IGR/JEM-X \\
     MXB 1730-335 & 2016/6 & 57562 & 3.2e-9 &  2.4e37 & MAXI \\
     MXB 1730-335 & 2016/12 & 57732 & 2.7e-9 &  2.1e37 & MAXI \\
     MXB 1730-335 & 2017/3 & 57833 & 3.3e-9 &  2.5e37 & MAXI, IGR/JEM-X \\
     MXB 1730-335 & 2017/8 & 57975 & 2.8e-9 &  2.1e37 & \citet{Chen21}, IGR/JEM-X, MAXI, Swift/XRT \\
     MXB 1730-335 & 2017/11 & 58071 & 1.9e-9 &  1.4e37 & MAXI \\
     MXB 1730-335? & 2018/3 & 58178 & 3.0e-9 &  2.2e37 & \citet{Homan18}, MAXI, IGR/JEM-X \\
     MXB 1730-335 & 2018/6 & 58292 & 1.7e-9 &  1.3e37 & \citet{Bahramian18}, MAXI,Swift/XRT \\
     MXB 1730-335 & 2018/9 & 58389 & 2.7e-9 &  2.0e37 & MAXI \\
     MXB 1730-335 & 2019/1 & 58490 & 1.4e-9 &  1.0e37 & MAXI \\
     MXB 1730-335 & 2019/3 & 58571 & 2.3e-9 &  1.8e37 & MAXI, Swift/XRT \\
     MXB 1730-335 & 2019/7 & 58677 & 2.2e-9 & 1.7e37 & MAXI \\
     MXB 1730-335 & 2019/11 & 58794 & 2.3e-9 &  1.7e37 & MAXI \\
     MXB 1730-335 & 2020/3 & 58920 & 3.5e-9 &  2.7e37 & \citet{vandenEijnden24},Swift/XRT, MAXI \\
     MXB 1730-335 & 2020/8 & 59083 & 4.8e-9 &  3.6e37 & MAXI \\
     MXB 1730-335 & 2021/2 & 59247 & 3.3e-9 &  2.5e37 & MAXI \\
     MXB 1730-335 & 2021/8 & 59429 & 6.3e-10 & 4.8e36 & MAXI \\
     MXB 1730-335 & 2021/12 & 59553 & 2.0e-9  & 1.6e37 & MAXI \\
     MXB 1730-335 & 2022/5 & 59702 & 1.8e-9 & 1.4e37 & MAXI \\
     MXB 1730-335 & 2022/10 & 59867 & 1.3e-9  & 1.0e37 & MAXI \\
     MXB 1730-335 & 2023/1 & 59973 & 1.8e-9  & 1.4e37 & MAXI \\
     MXB 1730-335 & 2023/6 & 60105 & 9.9e-10  & 7.5e36 & MAXI \\
    MXB 1730-335 & 2023/8 & 60185 & 1.1e-9 & 8.6e36 & MAXI \\
    MXB 1730-335 & 2024/1 & 60329 & 1.9e-9 & 1.4e37 & MAXI  \\
    MXB 1730-335 & 2024/6 & 60490 & 2.6e-9 & 2.0e37 & MAXI \\
    MXB 1730-335 & 2024/10 & 60594 & 2.1e-9 & 1.6e37 & MAXI \\
    MXB 1730-335 & 2025/1 & 60689 & 1.3e-9 & 9.5e36 & MAXI  \\
    MXB 1730-335 & 2025/4 & 60791 & 1.0e-9 & 7.5e36 & MAXI  
	\enddata
	\tablecomments{Outbursts of MXB 1730-335, the Rapid Burster in the globular cluster Liller 1. $F_X$ in ergs \revise{cm$^{-2}$ s$^{-1}$}, $L_X$ in ergs s$^{-1}$. Distance of 8 kpc assumed. Reference to an instrument (e.g. MAXI) indicates we have made use of that instrument's archive to generate some information for that line. 
    }
\label{tab:RapidBurster}
\end{deluxetable*}

\begin{figure}
\centering
\includegraphics[width=0.485\textwidth]{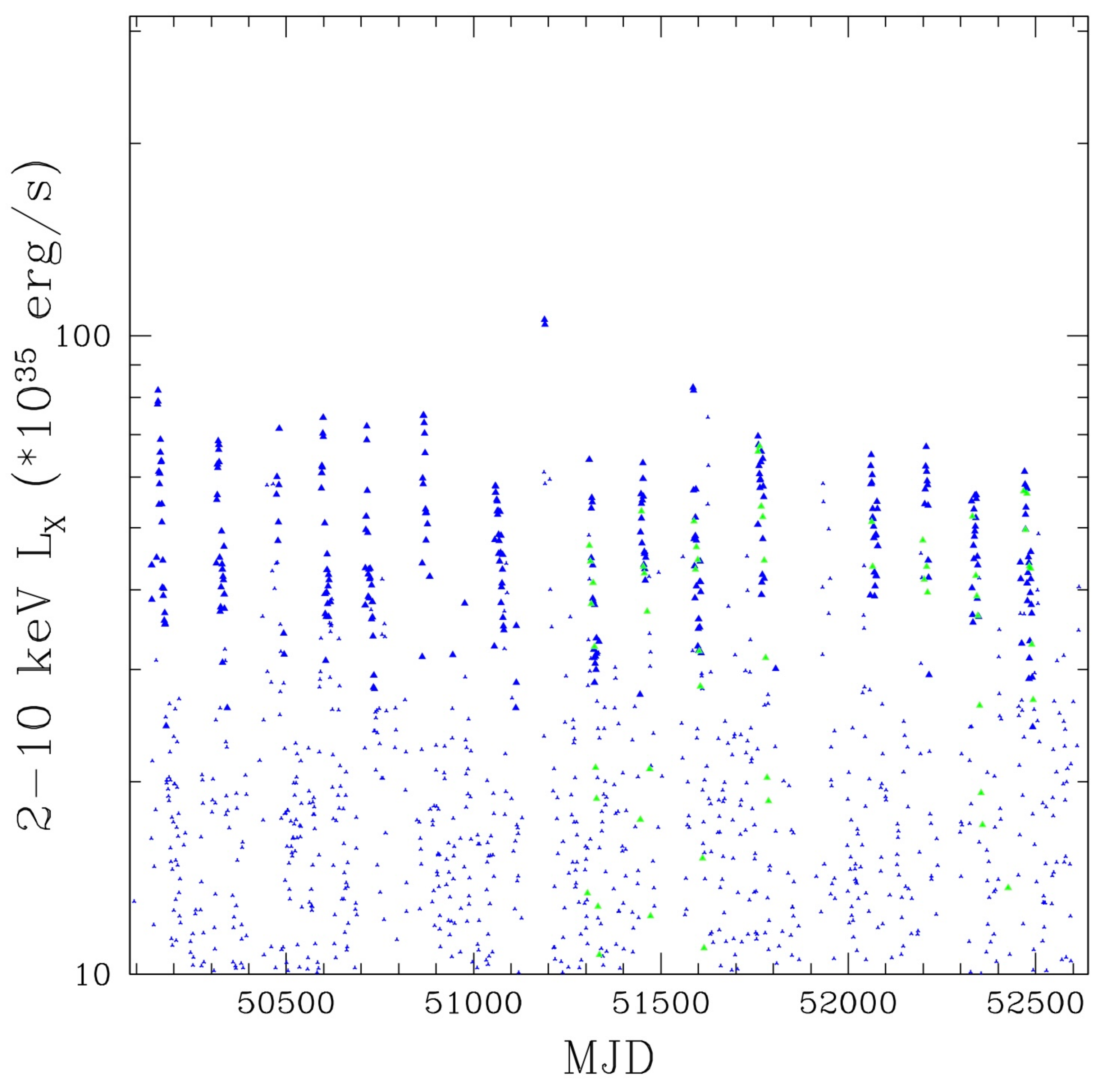} 
\includegraphics[width=0.48\textwidth]{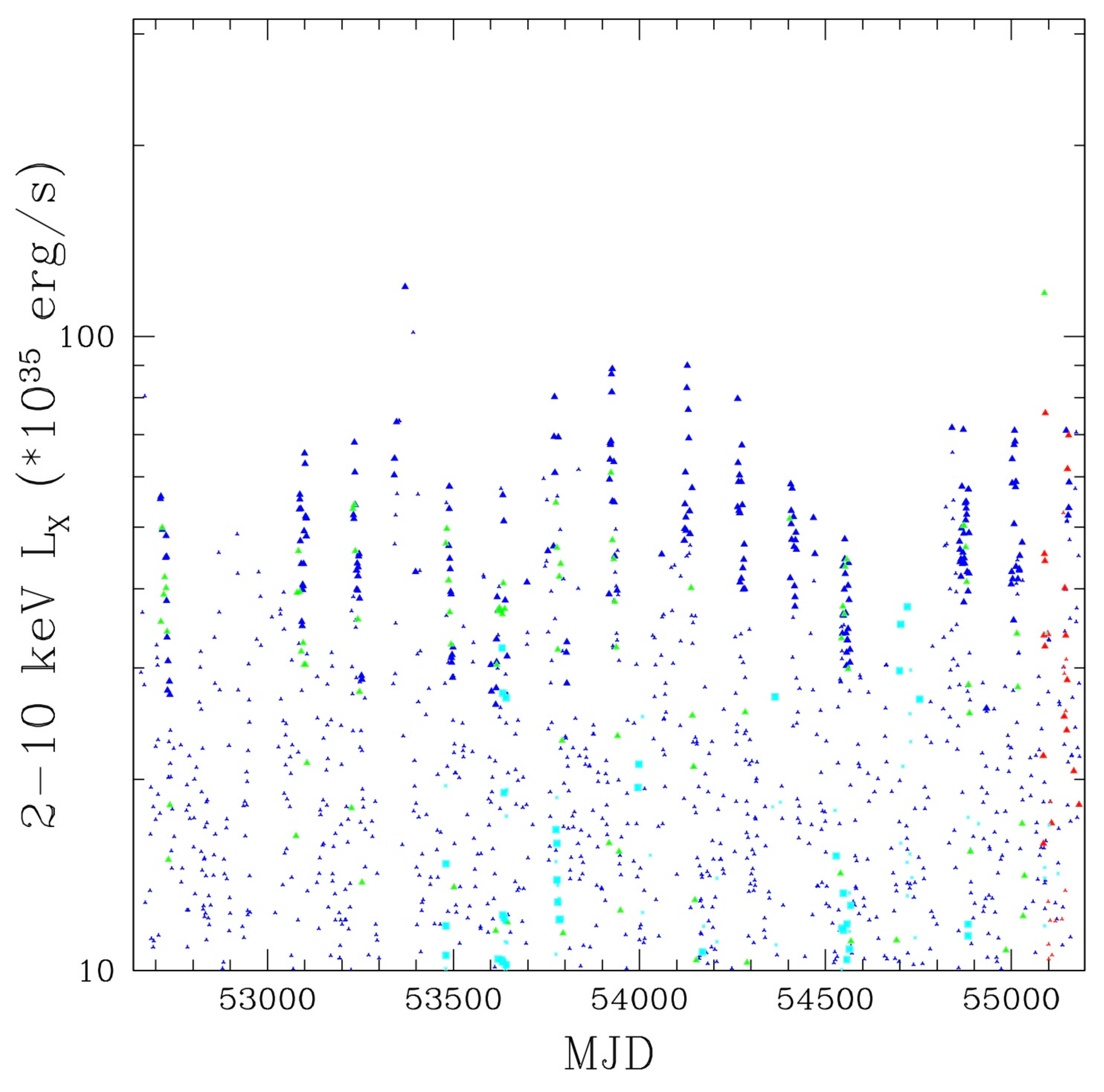} 
\includegraphics[width=0.48\textwidth]{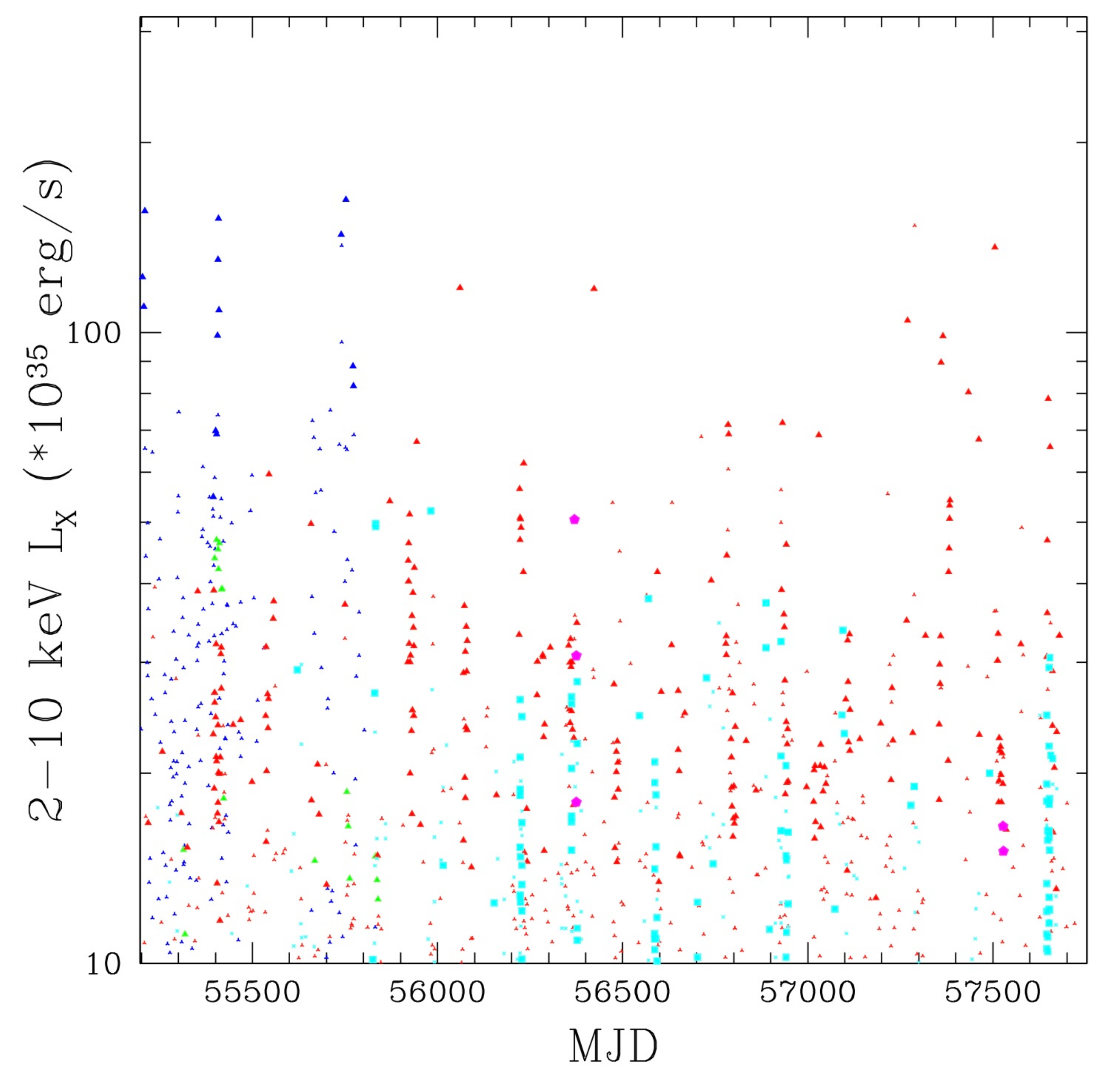} 
\includegraphics[width=0.485\textwidth]{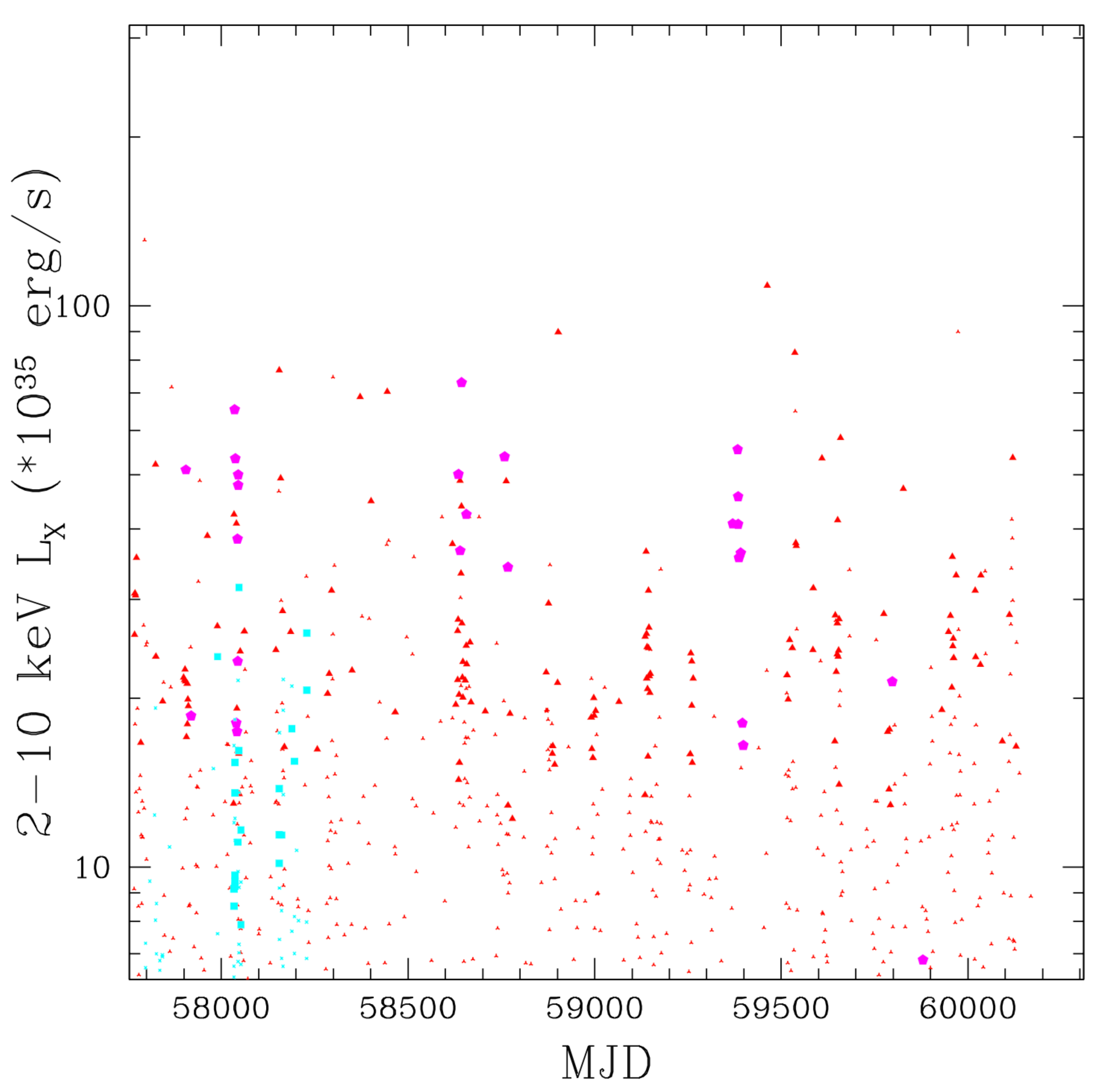} 
\caption{RXTE/PCA (green), RXTE/ASM (blue), MAXI/GSC (red), Swift/XRT (magenta), and INTEGRAL/JEM-X (cyan) lightcurves of GRS 1747-312 (in Terzan 6).  Upper left, 1996-2002; upper right, 2003-2009; lower left, 2010-2016; lower right, 2017-2023 (slightly different scale, to show one Swift data point).
 }
 \label{fig:Ter6}
\end{figure}

{\bf GRS 1747-312}; We use the distance estimate for the globular cluster Terzan 6 of 7.3$\pm0.4$ kpc from \citet{Baumgardt21}.  GRS 1747-312's outbursts have not been cataloged since 2008, so we turn to all-sky monitors to produce a full list (Table~\ref{tab:Terzan6}). We assume $N_H=1.4\times10^{22}$ cm$^{-2}$ and a photon index of 1.0 \citep{intZand00} to transform countrates to fluxes. 
 %We use RXTE/PCA Bulge Scan values where available, then RXTE/ASM, MAXI, or Swift/XRT. 
 For RXTE/ASM we multiply countrates by 3.6$\times10^{-10}$ to get 2-10 keV fluxes, for RXTE/PCA by 2.1$\times10^{-12}$.
For Swift/XRT data we multiply by 9.4$\times10^{-11}$ to get 2-10 keV fluxes, for MAXI by $8.7\times10^{-9}$, and for INTEGRAL/JEM-X by $7.1\times10^{-11}$.
\revise{We average RXTE/ASM data over a 3-day window (weighting by the error), to reduce the noise. }
We identify outbursts in the RXTE/ASM, RXTE/PCA, INTEGRAL/JEM-X, and MAXI lightcurves when we see at least two detections (among all the lightcurves) with the countrate exceeding 3 times the error \revise{above the average quiescent rate of 0.5 ct/s} (see Fig.~\ref{fig:Ter6}). This generally agrees with the identifications in \citet{intZand03} where they overlap, though the ASM data is noisy in this crowded region. 
%except that the June 1998 outburst (epoch 6 according to \citealt{Simon09_GRS}; VII by \citealt{intZand03}) is not significant by our metric (it is also unusually close in time to the outbursts before and after), so we are not certain of its reality or, if real, attribution to GRS 1747-312 (we include it with a ?). 
We are also in \revise{good} agreement with the identifications of \citet{Simon09_GRS}, \revise{which use a more complicated smoothing algorithm to construct their lightcurves}.
\revise{We note that the last two outbursts detected by the RXTE/ASM (late 2010 to early 2011) are especially noisy; we do not use these data, though we plot them in Fig.~\ref{fig:Ter6}.}
%but are not completely confident of the reality of their epochs 13, 20, and 21 (not included in our list), as the signal-to-noise of the individual detections is not above 3 sigma per day, though the timing of the outbursts appears plausible. 
The outbursts average every 138 days, with the average varying between 120 and 150 days over time, consistent with previous determinations \citep{intZand03,Simon09_GRS}.
Peak fluxes average $L_X=6\times10^{36}$ erg/s. 

GRS 1747-312 is known to show regular eclipses \citep{intZand00}, which makes it possible to verify that an outburst from Terzan 6 is produced by this transient, given an accurate orbital ephemeris. \citet{Saji16} noted a burst, and an absence of a predicted X-ray eclipse,  
during a Suzaku observation in Sept. 2009, when GRS 1747-312 was not detected in RXTE/PCA monitoring.\footnote{\citet{vandenBerg24} point out that a spike is visible in RXTE/PCA scans on MJD 55088, two days earlier, but that pulsations in the scan reveal that this spike was produced by the known accreting millisecond pulsar IGR J17511–3057, then in outburst.} 
More recently, \citet{vandenBerg24} have shown conclusive evidence that a second eclipsing, bursting X-ray binary, Terzan 6 X2, is located 0.7" from GRS 1747-312. 
We label the 2009 detection by Suzaku as "Ter 6 X2" in Table~\ref{tab:Terzan6}. %and Fig.~\ref{fig:Ter6}.
%in the Sept. 2009 outburst. %We note that this outburst was unusual in other ways as well. %The RXTE/PCA bulge scan reported a value twice as high as previously observed from this source, but only for one scan, while most other outbursts showed $\sim$5 scan detections, implying a more rapid decline.  Interestingly, this outburst occurred "out of sequence". 
%There are three other outbursts in 2009 with intervals between them of 139 and 137 days, but the Sept. 2009 outburst lies in the middle of (80 days into, 57 days from the end of) one such interval. This supports the suggestion by \citet{Saji16} that this outburst was produced by a different X-ray transient in Terzan 6 (which we label "Ter 6 X-2?" in Table~\ref{tab:Terzan6}, and Fig.~\ref{fig:Ter6}).

\clearpage

\startlongtable
\centerwidetable
\begin{deluxetable*}{lcccccl}
\tablehead{		\colhead{NS XRB}  & \colhead{Year/Month} & \colhead{E} & \colhead{MJD} & \colhead{Peak $F_X$}  & \colhead{$L_X$} &   \colhead{References} \\
       \colhead{} & \colhead{} & \colhead{} & \colhead{} & \colhead{2-10 keV} &  \colhead{2-10 keV} & }
  \startdata
    GRS 1747-312 & 1990/9 & & 48143 & 8.8e-10 &  5.5e36 & \citet{Pavlinsky92,intZand00}\\
    GRS 1747-312 & 1996/3 & 0 & 50157 & 1.3e-9 &  8.2e36 &  \citet{intZand03,Simon09_GRS}, RXTE/ASM \\
    GRS 1747-312 & 1996/8 & 1 & 50318 &  1.1e-9 &  6.8e36 &  \citet{intZand00,intZand03,Simon09_GRS}, RXTE/ASM \\
    GRS 1747-312 & 1997/1 & 2 & 50475 &  9.4e-10 &  6.0e36 &  \citet{intZand03,Simon09_GRS}, RXTE/ASM \\
    GRS 1747-312 & 1997/5 & 3 & 50598 &  1.2e-9 &  7.4e36 &  \citet{intZand03,Simon09_GRS}, RXTE/ASM \\
    GRS 1747-312 & 1997/10 & 4 & 50714 &  1.1e-9 &  7.2e36 & \citet{intZand00,intZand03,Simon09_GRS}, RXTE/ASM \\
    GRS 1747-312 & 1998/2 & 5 & 50867 &  1.2e-9 &  7.5e36 & \citet{intZand00,intZand03,Simon09_GRS}, RXTE/ASM \\
    GRS 1747-312 & 1998/6 & 6 & 50986 & 3.9e-10 &  2.5e36 & \citet{intZand03,Simon09_GRS}, RXTE/ASM \\
    GRS 1747-312* & 1998/9 & 7 & 51057 &  9.1e-10 &  5.8e36 & \citet{intZand00,intZand03,Simon09_GRS}, RXTE/ASM \\
    GRS 1747-312 & 1999/1 & (8) & 51189 & 1.7e-9 & 1.1e37 & \citet{Simon09_GRS}, RXTE/ASM \\
    GRS 1747-312 & 1999/5 & 9 & 51308 & 1.0e-9   & 6.3e36 & \citet{intZand00,intZand03,Simon09_GRS}, PCA, RXTE/ASM  \\
    GRS 1747-312* & 1999/10 & 10 & 51451 & 9.9e-10   & 6.3e36 & \citet{intZand03,Simon09_GRS}, RXTE/ASM, PCA \\
    GRS 1747-312* & 2000/2 & 11 & 51585 & 1.3e-9   & 8.2e36 & \citet{intZand03,Simon09_GRS}, RXTE/ASM, PCA \\
    GRS 1747-312* & 2000/8 & 12 & 51759 & 1.1e-9 &   6.9e36 & \citet{intZand03,Simon09_GRS}, RXTE/ASM, PCA \\
    % outburst 13 is not significant
    GRS 1747-312* & 2001/5 & 14 & 52063 & 9.5e-10 &   6.0e36 & \citet{intZand03,Simon09_GRS}, RXTE/ASM, PCA \\
    GRS 1747-312* & 2001/10 & 15 & 52207 & 1.0e-9 &  6.7e36 & \citet{intZand03,Simon09_GRS}, RXTE/ASM, PCA \\
    GRS 1747-312* & 2002/2 & 16 & 52337 & 8.8e-10 &  5.6e36 & \citet{intZand03,Simon09_GRS}, RXTE/ASM, PCA \\
    GRS 1747-312* & 2002/7 & 17 & 52470 & 9.6e-10 &  6.1e36 & \citet{intZand03,Simon09_GRS}, RXTE/ASM, PCA \\
    %GRS 1747-312 & 2002/10 & 18 & 52574 & 8.8e-10 & 5.6e36 & \citet{Simon09_GRS}, RXTE/ASM \\  no longer significant
    GRS 1747-312 & 2003/3 & 19 & 52713 & 8.7e-10 &   5.5e36 & \citet{Simon09_GRS}, RXTE/ASM, PCA \\
    GRS 1747-312 & 2004/3 & 22 & 53100 & 1.0e-9 &  6.5e36 & \citet{Markwardt04b,Simon09_GRS}, RXTE/ASM, PCA \\
    GRS 1747-312 & 2004/8 & 23 & 53233 & 1.1e-9 & 6.7e36 & \citet{Simon09_GRS}, RXTE/ASM, PCA \\
    GRS 1747-312 & 2004/11 & 24 & 53341 & 1.0e-9 &  6.4e36 & \citet{Simon09_GRS}, RXTE/ASM \\
    GRS 1747-312 & 2005/4 & 25 & 53489 & 9.1e-10 &  5.8e36 & \citet{Simon09_GRS},  RXTE/ASM, PCA\\
    GRS 1747-312 & 2005/9 & 26 & 53633 & 8.8e-10 & 5.6e36 & \citet{Simon09_GRS},  RXTE/ASM, PCA, IGR/JEM-X \\
    GRS 1747-312 & 2006/2 & 27 & 53782 & 1.1e-9 &  6.9e36 & \citet{Chenevez06,Simon09_GRS}, RXTE/ASM, PCA, IGR/JEM-X\\
    GRS 1747-312 & 2006/7 & 28 & 53926 & 1.4e-9 & 8.8e36 & \citet{Simon09_GRS}  RXTE/ASM, PCA\\
    GRS 1747-312 & 2007/1 & 29 & 54128 & 1.4e-9 &  8.9e36 & \citet{Simon09_GRS},  RXTE/ASM, PCA \\
    GRS 1747-312 & 2007/6 & 30 & 54264 & 1.2e-9 & 7.9e36 & \citet{Simon09_GRS},  RXTE/ASM, PCA \\
    GRS 1747-312 & 2007/11 & 31 & 54406 & 9.1e-10 & 5.8e36 & \citet{Simon09_GRS}, RXTE/ASM, PCA  \\
    GRS 1747-312 & 2008/3 & 32 & 54552 & 7.5e-10 &  4.8e36 & \citet{Simon09_GRS}, RXTE/ASM, PCA, IGR/JEM-X  \\
    GRS 1747-312 & 2008/9 & & 54726 & 5.9e-10 & 3.7e36 &  IGR/JEM-X \\ % RXTE/ASM,
    GRS 1747-312 & 2009/2 & & 54870 & 1.1e-9 &  7.1e36 & RXTE/ASM, PCA \\
    GRS 1747-312 & 2009/6 & & 55008 & 1.1e-9 &  7.0e36 & RXTE/ASM, PCA \\
    Ter 6 X2 & 2009/9 & & 55090 & 1.0e-11 &  6e34 &  \citet{Saji16,vandenBerg24} %RXTE/PCA,MAXI 
    \\
    GRS 1747-312 & 2009/11 & & 55147 & 1.1e-9 &  7.0e36 & RXTE/ASM, MAXI \\
    GRS 1747-312 & 2010/7 & & 55402 & 7.3e-10 & 4.6e36 & MAXI, RXTE/PCA, ASM \\ %ignore ASM points
    GRS 1747-312 & 2010/12 & & 55544 & 1.0e-9 & 6.5e36 & MAXI \\
    GRS 1747-312 & 2011/4 & & 55658 & 8.6e-10 & 5.5e36 & MAXI, RXTE/PCA \\ %ignore ASM points
    GRS 1747-312 & 2011/7 & & 55749 & 6.6e-10 & 4.2e36 & MAXI, RXTE/PCA \\ 
    GRS 1747-312 & 2011/9 & & 55832 & 7.7e-10 & 4.9e36 & IGR/JEM-X, RXTE/PCA, MAXI \\
    GRS 1747-312 & 2011/12  & & 55924 & 8.9e-10 &  5.6e36 & MAXI \\
    GRS 1747-312 & 2012/5 & & 56072 & 6.6e-10 & 4.2e36 & MAXI \\
    GRS 1747-312 & 2012/10 & & 56221 & 9.7e-10 & 6.1e36 & MAXI, IGR/JEM-X \\
    GRS 1747-312* & 2013/3 & & 56369 & 7.9e-10 &  5.0e36 & \citet{Chenevez13_GRS_1747,Bahramian13_GRS_1747}, Swift/XRT, MAXI, JEM-X  \\
    GRS 1747-312 & 2013/7 & & 56476 & 5.2e-10 & 3.3e36 & MAXI \\
    GRS 1747-312 & 2013/10 & & 56594 & 7.0e-10 & 4.4e36 &  \citet{Chenevez13b},MAXI, IGR/JEM-X \\
    GRS 1747-312 & 2014/5 & & 56786 & 1.2e-9 & 7.4e36 & MAXI \\
    GRS 1747-312 & 2014/10 & & 56931 & 1.2e-9 & 7.7e36 & MAXI, IGR/JEM-X  \\
    GRS 1747-312 & 2015/1 & & 57030 & 1.2e-9 & 7.4e36 & MAXI \\
    GRS 1747-312 & 2015/3 & & 57107 & 6.1e-10 &  3.9e36 & MAXI, IGR/JEM-X \\
    GRS 1747-312 & 2015/7 & & 57227 & 5.1e-10 & 3.3e36 & MAXI \\
    GRS 1747-312 & 2015/12 & & 57360 & 1.5e-9 & 9.4e36 & MAXI \\%--too near Sun, untrustworthy?
    GRS 1747-312* & 2016/4 & & 57505 & 2.2e-9 &  1.4e37 & \citet{Bahramian16_GRS_1747}, MAXI, Swift/XRT \\
    GRS 1747-312 & 2016/9 & & 57649 & 1.3e-9 & 8.3e36 & MAXI, IGR/JEM-X \\
    GRS 1747-312 & 2017/1 & & 57768 & 5.7e-10 & 3.6e36 & MAXI \\
    GRS 1747-312 & 2017/5 & & 57905 & 8.0e-10 & 5.1e36 & Swift/XRT, MAXI  \\
    GRS 1747-312 & 2017/10 & & 58035 & 1.0e-9 & 6.5e36 &  \citet{DiGesu17,Bozzo17_GRS_1747}, MAXI, Swift/XRT, IGR/JEM-X \\
    GRS 1747-312 & 2018/2 & & 58155 & 1.3e-9 & 8.1e36 & MAXI, IGR/JEM-X   \\
    GRS 1747-312 & 2018/6 & & 58295 & 5.7e-10 & 3.6e36 & MAXI  \\ %poor detection?
    GRS 1747-312 & 2019/6 & &  58643 & 1.1e-9 & 7.2e36 & \citet{vandenBerg24}, Swift/XRT, MAXI \\
    GRS 1747-312* & 2019/10 & & 58758 & 8.5e-10 & 5.4e36 & \citet{Grebenev19_GRS_1747,Bahramian19_GRS_1747}, Swift/XRT, MAXI  \\
    GRS 1747-312 & 2020/2 & & 58876 & 5.5e-10 &  3.5e36 & MAXI \\
    GRS 1747-312 & 2020/5 & & 58997 & 4.0e-10 &  2.5e36 & MAXI \\
    GRS 1747-312 & 2020/10 & & 59137 & 6.6e-10 & 4.2e36 & MAXI \\
    GRS 1747-312 & 2021/2 & & 59257 & 4.6e-10 &  2.9e36 & MAXI  \\
    GRS 1747-312 & 2021/6 & & 59383 &  8.6e-10 & 5.5e36 & \citet{vandenBerg24}, Swift/XRT \\
    GRS 1747-312 & 2021/11 & & 59536 & 1.4e-9 & 8.7e36 & MAXI \\ 
    GRS 1747-312 & 2022/3 & & 59651 & 7.4e-10 & 4.7e36 & MAXI \\
    GRS 1747-312 & 2022/7 & & 59774 & 5.3e-10 & 3.4e36 & MAXI, Swift/XRT \\
    GRS 1747-312 & 2022/10 & & 59879 & 1.0e-11 & 6.6e35 & Swift/XRT \\
    GRS 1747-312 & 2023/1 & & 59958 & 6.4e-10 & 4.1e36 & MAXI, Swift/XRT  \\
    GRS 1747-312 & 2023/3 & & 60034 & 6.1e-10 & 3.9e36 & MAXI \\
    GRS 1747-312 & 2023/6 & & 60120 & 8.4e-10 & 5.3e36 &  \citet{Yang23}, MAXI \\
    GRS 1747-312 & 2024/7 & & 60508 & 8.4e-10 & 5.3e36 & MAXI \\
    GRS 1747-312 & 2025/3 & & 60747 & 5.0e-10 & 3.2e36 & MAXI \\
	\enddata
	\tablecomments{}{Outbursts of GRS 1747-312, the eclipsing burster in the globular cluster Terzan 6. $F_X$ in 2-10 keV ergs \revise{cm$^{-2}$ s$^{-1}$}, $L_X$ in 2-10 keV ergs s$^{-1}$. $N_H$ is about $1.4\times10^{22}$ cm$^{-2}$.  D=7.3 kpc \citep{Baumgardt21}. E is epoch number of \citet{Simon09_GRS}. \citet{Saji16} identify an outburst as likely produced by a different X-ray transient, labeled "Ter 6 X-2?". * indicates a reported eclipse during a pointed X-ray observation.)
    Reference to an instrument (e.g. MAXI) indicates we have made use of that instrument's archive to generate some information for that line.}
\label{tab:Terzan6}
\end{deluxetable*}

{\bf Aquila X-1}, 3U 1908+00 or 4U 1908+005. The distance is estimated at 4.5 kpc \citep{Galloway08}. 
$N_H$ is typically estimated between 3e21 and 6e21 (e.g. \citealt{Rutledge02_aql}, \citealt{Maccarone03}, \citealt{Sakurai12}, \citealt{Campana14}); we take $N_H$=4e21 cm$^{-2}$. We assume a power-law photon index of 2 for energy range conversions.
We find conversions of 5.6e-9 from MAXI 2-20 keV countrate to 2-10 keV flux, %1.5e-10 from INTEGRAL/JEM-X, 
2.7e-10 from RXTE/ASM, 2.3e-12 from RXTE/PCA (for 5 units), 2.8e-11 for Swift/XRT. 
Aquila X-1's outbursts have been studied in great detail in the literature, with significant compilations of outburst details by \citet{Kitamoto93,Lin07,Campana13,Gungor14,Ootes18} among others. We list all known X-ray outbursts from Aquila X-1 in Table~\ref{tab:aquila}.
We identify one recent X-ray outburst from Aquila X-1 (in May 2020) that appears not to have been reported in the literature.
%outburst compilations by Kitamoto93 (Ginga 1987-1992, Vela5B 1969-1979) Lin07 (1996-2006 ASM/PCA), Campana13 (1996-2011 RXTE/ASM, plus 2009-2011 MAXI), Asai13 (MAXI 2009-2012...not interesting), Gungor14 (1997-2011 RXTE/ASM, 2011, 2013 MAXI)

\startlongtable
%\centerwidetable
\begin{longrotatetable}
\begin{deluxetable}{lccccl}
\tablehead{		\colhead{NS XRB}  & \colhead{Begin} & \colhead{End} &  \colhead{Peak $F_X$}  & \colhead{$L_X$} &   \colhead{References} \\
       \colhead{} & \colhead{} & \colhead{} &  \colhead{2-10 keV} &  \colhead{2-10 keV} & }
    \startdata		
%		NS XRB & Begin & End & Peak $F_X$  &  $L_X$ &    References \\
%        & & 2-10 keV & 2-10 keV &  \\
%		\hline
        Aql X-1 & 1965/4 & 1965/4 & 1.2e-9 & 2.9e36 & \citet{Friedman67} (rocket) \\
        Aql X-1 &  1969/7 & 1969/8 &  2e-9 & 5e36 & \citet{Kitamoto93} (Vela5b) \\
        Aql X-1 &  1969/12 & 1970/1 & 1.0e-8 & 2.4e37 & \citet{Kitamoto93} (Vela5b) \\
        Aql X-1 & 1970/8 & 1970/10 & 6.4e-9 & 1.5e37 & \citet{Price72,Kitamoto93} (Vela5b,rocket) \\
        Aql X-1 & 1971/3 & 1971/4 & 5.2e-9 & 1.3e37 & \citet{Cominsky78,Kitamoto93} (Uhuru,Vela5b) \\
        Aql X-1 & 1971/9 & 1971/11 & 1.1e-8 & 2.7e37 & \citet{Kitamoto93} (Vela5b, OSO-7)   \\
        Aql X-1 & 1972/4 & 1972/5 & 6.4e-9 & 1.5e37 &  \citet{Kitamoto93} (Vela5b)  \\
        Aql X-1 & 1972/12 & 1973/3 & 1.2e-8 & 3e37 & \citet{Kitamoto93} (Vela5b, OSO-7)   \\
        Aql X-1 & 1974/4 & 1974/6 & 2e-9 & 5e36 &  \citet{Kitamoto93} (Vela5b)    \\
        Aql X-1 & 1975/6 & 1975/7 & 2.0e-8 & 5e37 &   \citet{Buff77,Kaluzienski77,Kitamoto93} (Vela5b,SAS-3,Ariel-5)\\ 
        Aql X-1 & 1976/6 & 1976/7 & 2.0e-8 & 5e37 & \citet{Wilson76,Kaluzienski77} (Ariel-5) \\
        Aql X-1 & 1977/1 & 1977/1 & 6e-9 & 1.4e37 & \citet{Holt77} (Ariel-5) \\
        Aql X-1 & 1978/5 & 1978/8 & 2.6e-8 & 6.3e37  & \citet{Charles80,Kitamoto93,Chen97} (Vela5b,Ariel-5)  \\
        Aql X-1 & 1979/3 & 1979/4 & 1e-8 & 2.4e37 & \citet{KaluzienskiHolt79_aql,Czerny87,Kitamoto93}  (Vela-5b,Ariel-5,Einstein) \\
        Aql X-1 & 1980/5 & 1980/6 & 2.4e-9 & 6e36 & \citet{Koyama81,OdaMargon80} (Hakucho) \\
        Aql X-1 & 1981/8 & 1981/8 & 1.4e-8 & 3.4e37 & \citet{Paradijs81}  (Hakucho) \\
        Aql X-1 & 1987/3 & 1987/3 & 3.1e-8 & 7.5e37 & \citet{Kitamoto93} (Ginga) \\
        Aql X-1 & 1988/11 & 1988/11 & 1.1e-8 & 2.6e37 & \citet{Makino88} (Ginga) \\
        Aql X-1 & 1989/9 & 1989/10 & 1.2e-8 & 2.9e37 & \citet{Kitamoto93} (Ginga) \\
        Aql X-1 & 1990/6 & 1990/10 & - & - & \citet{Chevalier91,Hjellming90} (optical, radio) \\
        Aql X-1 & 1991/5 & 1991/10 & - & - & \citet{Ilovaisky91,Harmon96} (optical,BATSE) \\
        Aql X-1 & 1992/4 & 1992/5 & - & - & \citet{Ilovaisky92,Harmon96} (optical,BATSE) \\
        Aql X-1 & 1993/7 & 1993/8 & - & - & \citet{Ilovaisky93,Harmon96} (optical,BATSE) \\
        Aql X-1 & 1994/4 & 1994/7 & 5e-9 & 1.2e37 & \citet{Garcia94,Rubin00} (ASCA,BATSE) \\
        Aql X-1 & 1995/7 & 1995/9 & - & - & \citet{Ilovaisky95,Ilovaisky96} (optical) \\
        Aql X-1 & 1996/2 & 1996/2 & 1.7e-9 & 4e36 & \citet{Campana13} (RXTE/ASM) \\
        Aql X-1 & 1996/5 & 1996/8 & 7.6e-10 & 1.8e36 &  \citet{Ilovaisky96,Campana13} (RXTE/ASM) \\
        Aql X-1 & 1997/1 & 1997/4 & 7.6e-9 & 1.8e37 & \citet{Levine97,Campana98} (RXTE/ASM, PCA) \\
        Aql X-1 & 1997/8 & 1997/9 & 4.4e-9 & 1.1e37 & \citet{Charles97,Campana13} (RXTE/ASM, PCA) \\
        Aql X-1 & 1998/2 & 1998/5 & 9.6e-9 & 2.3e37 & \citet{Swank98,Cui98,Campana13} (RXTE/ASM, PCA) \\
        Aql X-1 & 1999/5 & 1999/11 & 7.5e-9 & 1.8e37 & \citet{Jain99,Yu03,Campana13} (RXTE/ASM, PCA) \\
        %Aql X-1 & 2000/5 &  &  & not full outburst  \\
        Aql X-1 & 2000/9 & 2000/11 & 1.3e-8 & 3.0e37 & \citet{Rutledge00_atel,Yu03,Lin07,Campana13} (RXTE/ASM, PCA) \\
        Aql X-1 & 2001/6 & 2001/8 & 1.3e-9 & 3.1e36 & \citet{Bailyn01,Yu07,Campana13} (RXTE/ASM, PCA) \\
        Aql X-1 & 2002/2 & 2002/4 & 4.2e-9 & 1.0e37 & \citet{Swank02,Tudose09,Campana13} (RXTE/ASM, PCA) \\
        Aql X-1 & 2003/2 & 2003/4 & 1.3e-8 & 3.1e37 & \citet{Swank03,Campana13} (RXTE/ASM, PCA) \\
        Aql X-1 & 2004/2 & 2004/6 & 5.2e-9 & 1.2e37 & \citet{Molkov04,Yu07,Tudose09,Campana13} (RXTE/ASM, PCA) \\
        Aql X-1 & 2005/3 & 2005/5 & 1.3e-9 & 3.2e36 & \citet{Grebenev05_AqlX1,Rodriguez06,Tudose09,Campana13} (RXTE/ASM, PCA, IGR) \\
        Aql X-1 & 2005/11 & 2005/11 & 1.4e-9 & 3.3e36 & \citet{Rodriguez05_AqlX1,Campana13} (RXTE/ASM, PCA) \\
        Aql X-1 & 2006/7 & 2006/8 & 1.1e-9 & 2.7e36 & \citet{Wijnands06_AqlX1,Campana13} (RXTE/ASM, Swift/XRT) \\
        Aql X-1 & 2007/5 & 2007/6 & 1.3e-9 & 3.0e36 & \citet{Rodriguez07_AqlX1,Campana13} (RXTE/ASM, Swift/XRT) \\
        Aql X-1 & 2007/9 & 2007/10 & 5e-9 & 1.2e37 & \citet{Corbet07,Sakurai12,Campana13} (RXTE/ASM, Suzaku) \\
        Aql X-1 & 2008/5 & 2008/6 & 1.1e-9 & 2.6e36 & \citet{Palmer08,Campana13} (RXTE/ASM) \\
        Aql X-1 & 2009/3 & 2009/5 & 8.8e-10 & 2.1e36 & \citet{Russell09_AqlX1,Campana13} (RXTE/ASM) \\
        Aql X-1 & 2009/11 & 2009/12 & 6.2e-9 & 1.4e37 & \citet{Linares09_AqlX1,Miller-Jones10,Asai12} (RXTE/ASM,PCA,MAXI,Swift/XRT) \\
        Aql X-1 & 2010/7 & 2010/10 & 7.0e-9 & 1.7e37 & \citet{Gultekin10,Asai12,Campana14} (MAXI,RXTE/ASM,PCA,Swift/XRT) \\
        Aql X-1 & 2011/10 & 2011/10 & 1.6e-8 & 3.8e37 & \citet{Yamaoka11,Asai12,Gungor14,Waterhouse16} (MAXI,PCA) \\
        Aql X-1 & 2013/6 & 2013/6 & 2.0e-8 & 4.8e37 & \citet{Degenaar13_AqlX1,Waterhouse16,Meshcheryakov18} (MAXI) \\
        Aql X-1 & 2014/6 & 2014/7  & 6.4e-9 & 1.5e37 & \citet{Meshcheryakov14,Gandhi14,King16} (MAXI, Swift/XRT) \\
        Aql X-1 & 2015/2 & 2015/4 & 1.8e-9 & 4.3e36 & \citet{Ueno15,Waterhouse16} (MAXI) \\
      %  Aql X-1 & 2015/9 ?? & \\
        Aql X-1 & 2016/7 & 2016/8 & 1.8e-8 & 4.3e37 & \citet{Sanna16_AqlX1,DiazTrigo18,Degenaar19_Aql,Niwano23} (MAXI,Swift/XRT) \\
        Aql X-1 & 2017/5 & 2017/6 & 1.4e-9 & 3.4e36 & \citet{Dincer17,Niwano23} (MAXI) \\
        Aql X-1 & 2018/2 & 2018/3 & 8.2e-9 & 2.0e37 & \citet{Spiridonova18,Niwano23} (MAXI) \\
        Aql X-1 & 2018/11 & 2018/12 & 1e-9 & 2e36 & \citet{Lin18,Niwano23} (MAXI, Swift/XRT) \\
        Aql X-1 & 2019/8 & 2019/9 & 1.1e-8 & 2.7e37 & \citet{Motta19,Niwano23} (MAXI) \\
        Aql X-1 & 2020/5 & 2020/5 & 3.1e-9 & 7.5e36 & MAXI \\
        Aql X-1 & 2020/8 & 2020/10 & 1.0e-8 & 2.4e37 & \citet{Saikia20,Bahramian20_Aql} (MAXI,Swift/XRT)\\
        Aql X-1 & 2021/3 & 2021/4 & 1.7e-9 & 4.0e36 & \citet{Kong21_Aql}, MAXI, Swift/XRT \\
        Aql X-1 & 2021/11 & 2022/1 & 1.2e-8 & 2.8e37 & \citet{Niwano21}, MAXI \\
        Aql X-1 & 2022/5 & 2022/6 & 6.5e-9 & 1.6e37 & \citet{Pawar22},MAXI \\
        %Aql X-1 & 2022/11 --Sun
        Aql X-1 & 2023/7 & 2023/9 & 1.8e-8 & 4.3e37 & \citet{Alabarta23a,Alabarta23b,Yan25},MAXI \\
        %Aql X-1 & 2023/12 & 2024/1 & 1.4e-9 & 3.3e36 & MAXI \\-Sun
        Aql X-1 & 2024/9 & 2024/10 & 1.1e-8 & 2.7e37 & \citet{Liu24_Aql,Mandal25}, MAXI \\
        %Aql X-1 & 2024/12 & 2025/1 & 2.0e-9 & 4.9e36 & MAXI \\-Sun
	\enddata
	\tablecomments{Outbursts of Aquila X-1. $F_X$ in 2-10 keV ergs \revise{cm$^{-2}$ s$^{-1}$}, $L_X$ in 2-10 keV ergs s$^{-1}$. $N_H$ is about $4\times10^{21}$ cm$^{-2}$.  D=4.5 kpc \citep{Galloway08}. Reference to an instrument (e.g. MAXI) indicates we have made use of that instrument's archive to generate some information for that line.}
    \label{tab:aquila}
\end{deluxetable}
\end{longrotatetable}

\begin{figure}
   % \centering
    \includegraphics[width=3.5in]{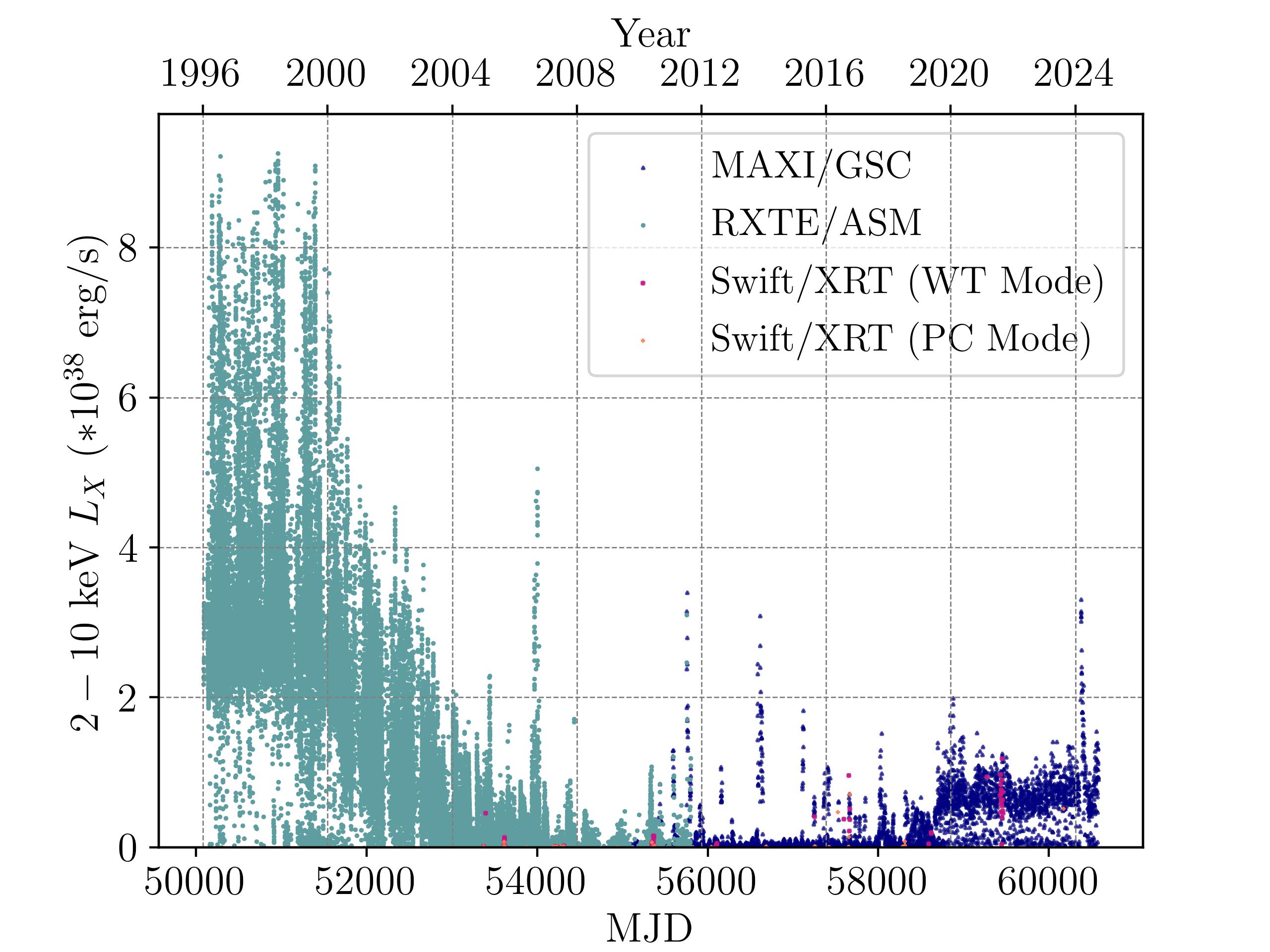}
 \caption{The RXTE/ASM, MAXI/GSC, and Swift/XRT lightcurve of Cir X-1 since 1996. 
 }
    \label{fig:CirX-1}
\end{figure}

\onecolumngrid
{\bf Cir X-1} has a distance estimate of $9.4^{+0.8}_{-1.0}$ kpc from the geometry of light echoes \citep{Heinz15}. Its eccentric orbit makes its behavior different from other systems here, and it is arguable whether it is a low-mass or high-mass X-ray binary. Fig.~\ref{fig:CirX-1}  illustrates its overall behavior since 1996, using RXTE/ASM, MAXI/GSC and Swift/XRT data. (See \citealt{SazParkinson03} for an excellent summary of earlier data.). 
In Table~\ref{tab:cir} we give details on its short outbursts.

\startlongtable
\centerwidetable
%\begin{longrotatetable}
\begin{deluxetable}{lccccl}
\tablehead{		\colhead{NS XRB}  & \colhead{Date}& \colhead{MJD}  &  \colhead{Peak $F_X$}  & \colhead{$L_X$} &   \colhead{References} \\
       \colhead{} & \colhead{} & \colhead{} &  \colhead{2-10 keV} &  \colhead{2-10 keV} & }
    \startdata		
        Cir X-1 & 1965/4 & 38875 &  5.8e-9 & 6.1e37 & \citet{Amnuel82} (rocket) \\
        Cir X-1 & 1969/6 & 40386 & 1.2e-8   & 1.3e38 & \citet{Amnuel82} (rocket) \\
        Cir X-1 & 1970/5 & 40719 & 1.2e-8   & 1.3e38 & \citet{Amnuel82} (rocket) \\
        Cir X-1 & 1971/2 & 40996 & 2.8e-8   & 3.0e38 & \citet{Forman76_var} \\
        Cir X-1 & 1972/5 & 41459 & 1.5e-8 & 1.6e38 &  \citet{Canizares74} \\
        Cir X-1 & 1973/8  & - & 6.3e-10 & 6.7e36 & \citet{Amnuel82} \\
        Cir X-1 & 1974/4  & - & 5.8e-9 & 6.1e37 & \citet{Amnuel82,DavisonTuohy75} \\
        Cir X-1 & 1976/1  & 42793 & 2.0e-8 & 2.1e38 & \citet{Buff77} \\
        Cir X-1 & 1976/2  & 42810 & 2.6e-8 & 2.7e38 & \citet{Kaluzienski76} \\
        Cir X-1 & 1976/2  & 42825 & 1.4e-8 & 1.5e38 & \citet{WilsonCarpenter76} \\
        Cir X-1 & 1976/10  & 43075 & 1.6e-8 & 1.7e38 & \citet{Dower82} \\
        Cir X-1 & 1976/12  & 43124 & 6.7e-9 & 7.1e37 & \citet{Dower82} \\
        Cir X-1 & 1977/1  & 43158 & 1.4e-8 & 1.5e38 & \citet{Dower82} \\
        Cir X-1 & 1977/5  & 43273 & 4.5e-9 & 4.8e37 & \citet{Dower82} \\
        Cir X-1 & 1977/8  & 43383 & 3.1e-8 & 3.3e38 & \citet{Kaluzienski77_cir} \\
        Cir X-1 & 1977/9  & 43397 & 2.3e-8 & 2.4e38 & \citet{KaluzienskiHolt77} \\
        Cir X-1 & 1977/9  & 43407 & 3.2e-8 & 3.4e38 & \citet{Dower82} \\
        Cir X-1 & 1977/10  & 43423 & 1.2e-8 & 1.3e38 & \citet{Dower82} \\
        Cir X-1 & 1978/1  & 43524 & 6.8e-8 & 7.2e38 & \citet{KaluzienskiHolt78} \\
        Cir X-1 & 1978/2  & 43541 & 9.1e-9 & 9.6e37 & \citet{KaluzienskiHolt78,Murdin80} \\
        Cir X-1 & 1978/8  & 43737 & 9.2e-9 & 9.7e37 & \citet{Dower82} \\
        Cir X-1 & 1979/1 &  43888 & 4.6e-8 & 4.9e38 & \citet{KaluzienskiHolt79_cir} \\
        Cir X-1 & 1984/5 &  45843 & 2.3e-8 & 2.4e38 & \citet{Inoue89,Miyamoto85} \\
        Cir X-1 & 1987/2  &  - &    3.0e-8 & 3.2e38 & \citet{Tsunemi89} \\
        Cir X-1 & 1989/4  & 47617 & 8.6e-8 & 9.1e38 & \citet{Stewart91} \\
        Cir X-1 & 2006/1  & 53761 & 3.4e-9 & 3.6e37 & RXTE/ASM \\
        Cir X-1 & 2006/3  & 53805 & 8.3e-9 & 8.8e37 & RXTE/ASM \\
        Cir X-1 & 2006/6  & 53890 & 4.2e-9 & 4.4e37 & \citet{Jonker06_CirX1},RXTE/ASM \\
        Cir X-1 & 2006/7  & 53938 & 1.5e-8 & 1.6e38 & RXTE/ASM \\
        Cir X-1 & 2006/8  & 53968 & 3.4e-8 & 3.6e38 & RXTE/ASM \\
        Cir X-1 & 2006/9  & 54001 & 4.8e-8 & 5.0e38 & RXTE/ASM \\
        Cir X-1 & 2006/12  & 54074 & 5.0e-9 & 5.2e37 & RXTE/ASM \\
        Cir X-1 & 2006/12  & 54085 & 8.1e-9 & 8.6e37 & RXTE/ASM \\
        Cir X-1 & 2007/1  & 54106 & 4.9e-9 & 5.2e37 & \citet{Nicolson07},RXTE/ASM \\
        Cir X-1 & 2007/11  & 54430 & 1.6e-8 & 1.7e38 & RXTE/ASM \\
        Cir X-1 & 2008/4  & 54558 & 4.6e-9 & 4.9e37 & RXTE/ASM \\
        Cir X-1 & 2009/9  & 55100 & 4.4e-10 & 4.7e36 & MAXI \\
        Cir X-1 & 2009/11  & 55156 & 8.0e-10 & 8.5e36 & \citet{Asai14}, MAXI \\
        Cir X-1 & 2010/5  & 55324 & 6.6e-9 & 7.0e37 & \citet{Asai14,D'Ai12,Nakajima10} \\
        Cir X-1 & 2010/6  & 55360 & 1.6e-9 & 1.7e37 & \citet{D'Ai12,Asai14,Isobe10} \\
        Cir X-1 & 2010/7  & 55382 & 5.8e-10 & 6.1e36 & \citet{Krimm11}, MAXI \\
        Cir X-1 & 2010/8  & 55438 & 8.0e-9 & 8.4e37 & MAXI, RXTE/ASM \\
        Cir X-1 & 2010/11  & 55520 & 6.7e-10 & 7.1e36 & MAXI \\
        Cir X-1 & 2011/2  & 55593 & 1.4e-8 & 1.5e38 & \citet{Asai14},RXTE/ASM,\citet{Krimm11} \\
        Cir X-1 & 2011/3  & 55642 & 1.2e-9 & 1.3e37 &  \citet{Asai14},MAXI \\
        Cir X-1 & 2011/5  & 55679 & 2.9e-10 & 3.1e36 & MAXI,\citet{Krimm11} \\
        Cir X-1 & 2011/6  & 55732 & 9.5e-10 & 1.0e37 & MAXI \\
        Cir X-1 & 2011/7  & 55753 & 3.2e-8 & 3.4e38 &  \citet{Asai14,Kimura12},RXTE/ASM,MAXI \\
        Cir X-1 & 2011/8  & 55800 & 9.7e-9 & 1.0e38 &  \citet{Asai14},RXTE/ASM \\
        Cir X-1 & 2011/10  & 55840 & 5.8e-10 & 6.1e36 & \citet{Asai14},MAXI \\
        Cir X-1 & 2011/10  & 55862 & 1.7e-9 & 1.8e37 & \citet{Asai14},MAXI \\
        Cir X-1 & 2011/12  & 55908 & 4.7e-9 & 5.0e37 & \citet{Asai14} \\
        Cir X-1 & 2012/1  & 55936 & 3.4e-9 & 3.6e37 & \citet{Asai14} \\
        Cir X-1 & 2012/4  & 56021 & 9.9e-10 & 1.0e37 & MAXI \\
        Cir X-1 & 2012/6  & 56090 & 1.1e-9 & 1.2e37 & MAXI \\
        Cir X-1 & 2012/6  & 56105 & 5.1e-10 & 5.4e36 & Swift/XRT \\
        Cir X-1 & 2012/7  & 56112 & 8.6e-10 & 9.1e36 & \citet{Asai14} \\
        Cir X-1 & 2012/8  & 56154 & 1.0e-8 & 1.1e38 & \citet{Asai14} \\
        Cir X-1 & 2012/12  & 56280 & 3.5e-9 & 3.7e37 & MAXI,\citet{Kimura12} \\
        Cir X-1 & 2012/12  & 56287 & 1.6e-9 & 1.7e37 & \citet{Asai14} \\
        Cir X-1 & 2013/3  & 56379 & 1.3e-9 & 1.4e37 & MAXI \\
        Cir X-1 & 2013/8  & 56528 & 1.2e-9 & 1.3e37 & MAXI \\
        Cir X-1 & 2013/10  & 56576 & 2.0e-9 & 2.1e37 & \citet{Asai14} \\
        Cir X-1 & 2013/10  & 56585 & 2.4e-8 & 2.5e38 & \citet{Asai14} \\
        Cir X-1 & 2013/11  & 56612 & 3.2e-8 & 3.4e38 & \citet{Asai14} \\
        Cir X-1 & 2014/8  & 56891 & 1.1e-9 & 1.1e37 & MAXI \\
        Cir X-1 & 2014/11  & 56964 & 1.0e-9 & 1.1e37 & MAXI \\
        Cir X-1 & 2015/1  & 57033 & 1.1e-9 & 1.2e37 & MAXI \\
        Cir X-1 & 2015/3  & 57111 & 9.5e-9 & 1.0e38 & \citet{Zhang15},MAXI \\
        Cir X-1 & 2015/4  & 57119 & 1.7e-8 & 1.8e38 & MAXI \\
        Cir X-1 & 2015/8  & 57248 & 6.4e-9 & 6.8e37 & MAXI \\
        Cir X-1 & 2015/12  & 57367 & 9.4e-9 & 9.9e37 & MAXI \\
        Cir X-1 & 2016/1  & 57404 & 1.0e-8 & 1.1e38 & MAXI \\
        Cir X-1 & 2016/2  & 57436 & 8.8e-9 & 9.3e37 & MAXI \\
        Cir X-1 & 2016/5  & 57533 & 5.8e-9 & 6.1e37 & MAXI \\
        Cir X-1 & 2016/7  & 57598 & 3.4e-9 & 3.5e37 & Swift/XRT \\
        Cir X-1 & 2016/9  & 57655 & 9.0e-9 & 9.5e37 & \citet{SannaMotta16},Swift/XRT \\
        Cir X-1 & 2016/12  & 57731 & 3.8e-9 & 4.0e37 & MAXI \\
        Cir X-1 & 2017/1  & 57777 & 3.9e-9 & 4.1e37 & MAXI \\
        Cir X-1 & 2017/4  & 57846 & 6.3e-9 & 6.6e37 & MAXI \\
        Cir X-1 & 2017/8  & 57976 & 1.4e-9 & 1.4e37 & MAXI \\
        Cir X-1 & 2017/9  & 58024 & 1.2e-8 & 1.3e38 & MAXI \\
        Cir X-1 & 2017/10  & 58037 & 1.4e-8 & 1.5e38 & MAXI \\
        Cir X-1 & 2017/12  & 58112 & 2.1e-9 & 2.2e37 & MAXI \\
        Cir X-1 & 2018/1  & 58135 & 1.4e-9 & 1.5e37 & MAXI \\
        Cir X-1 & 2018/2  & 58172 & 4.0e-9 & 4.2e37 & MAXI \\
        Cir X-1 & 2018/3  & 58193 & 1.3e-9 & 1.4e37 & MAXI \\
        Cir X-1 & 2018/7  & 58317 & 7.0e-9 & 7.4e37 & MAXI \\
        Cir X-1 & 2018/8  & 58358 & 1.6e-9 & 1.6e37 & MAXI \\
	\enddata
	\tablecomments{Recorded outbursts of Cir X-1. $F_X$ in 2-10 keV ergs \revise{cm$^{-2}$ s$^{-1}$}, $L_X$ in 2-10 keV ergs s$^{-1}$. $N_H$ is about $2\times10^{22}$ cm$^{-2}$.  D=9.4 kpc \citep{Heinz15}. Reference to an instrument (e.g. MAXI) indicates we have made use of that instrument's archive to generate some information for that line.
    }
    \label{tab:cir}
\end{deluxetable}

%\twocolumngrid
\section{Bursters with no outbursts}

In Table~\ref{tab:no_outbursts} we give details on the four known X-ray bursters which have no recorded outbursts.

{\bf SAX J1752.3-3138}; the distance is estimated at 9.2$\pm0.4$ kpc by \citet{Cocchi01}.

{\bf SAX J1818.7+1424}; \citet{Cornelisse02} place a limit of $<$9.4 kpc from an X-ray burst.

{\bf AX J1824.5-2451}; \citet{Gotthelf97} identified an X-ray burst from M28 using ASCA in 1997 (identified as AX J1824.5-2451), at a time when the total X-ray luminosity of the cluster was only $\sim10^{33}$ ergs s$^{-1}$.

{\bf SAX J2224.9+5421}; the distance upper limit of 7.1 kpc comes from the discovery burst \citep{Cornelisse02}.

\begin{deluxetable*}{lcccccl}
  	\label{tab:no_outbursts}
	%\begin{deluxetable*}{lcccccccr} %
\tablehead{		\colhead{NS XRB} & \colhead{Year/Month} &  \colhead{Peak $F_X$}  & \colhead{$N_H$} &  \colhead{$L_X$} & \colhead{d}  &  \colhead{References}\\
% \tablehead{ 
\colhead{}   & \colhead{} &  \colhead{2-10 keV} & \colhead{cm$^{-2}$} & \colhead{2-10 keV} & \colhead{kpc}  & \colhead{} } 
	%\centering
	%\begin{tabular}{lcccccr} %
	%	NS XRB & Year/Month & Peak $F_X$  & $N_H$ & $L_X$ &  Distance & References \\
         %& & 2-10 keV & cm$^{-2}$ & 2-10 keV &   \\
    \startdata
     %   SAX J1324.5-6313 & 1997 &  $<$3.4e-11 &    & $<$1.6e35 & $<$6.2 &   \citet{Cornelisse02} \\
     %   SAX J1324.5-6313 & 2015 & $<$1.6e-10 & & $<$7.3e35 & $<$6.2 & \citet{Negoro15,Bahramian15atel} \\
     %   \hline 
        SAX J1752.3-3138 & 1999/9 & $<$1.3e-10 & 6e21  & $<$1.3e36 & 9.2 &  \citet{Cocchi01} \\
        \hline 
     %   SAX J1753.5-2349 & 1996/8  & $<$1e-10 &  & $<$7.6e35 & 8 &  \citet{intZand99_bursters} \\
    %\hline
    SAX J1818.7+1424 & 1997/8  & $<$1.1e-10 & 1e21  & $<$1.2e36 & $<$9.4 &  \citet{Cornelisse02}  \\
        \hline
        AX J1824.5-2451 & 1995/3 & $<$3e-13 & 6e21 &  $<$1e33 & 5.5 &  \citet{Gotthelf97} \\
        \hline
    SAX J2224.9+5421 & 1999/11 & $<$1.3e-13 & $\sim$5e21 &  $<$8e32 & 7.1 & \citet{Cornelisse02,Degenaar14_SAX_J2224} \\
        \hline
	%\end{tabular}
    \enddata
	\caption{Objects that have shown X-ray bursts, but have no recorded clear outbursts. Note that AX J1824.5-2451, in M28, may be the burster and transitional MSP IGR J18245-2452, or may not.
    }
 \end{deluxetable*}

\section{Known persistent neutron star low-mass X-ray binaries}

In %Table~\ref{tab:persistent} 
Table 10 
we list the neutron star X-ray binaries we take to be persistent systems.

\begin{deluxetable*}{lcr}
\tablehead{ 
\colhead{Persistent LMXB} & \colhead{Alt. name} & \colhead{Nature} 
}
\startdata
4U 0513-40 & & burster \\
4U 0614+09 & & burster \\ 
2S 0918-549 & 4U 0919-54 & burster \\ 
4U 1246-588 & 1A 1246-588 & burster \\
4U 1254-69 & & burster \\
4U 1323-62 & & burster \\
4U 1543-624 & & burster \\
UW CrB & 1E 1603.6+2600 & burster \\
4U 1626-67 & &  pulsar \\
4U 1636-536 & & burster \\
4U 1702-429 & & burster \\
4U 1708-23 & & burster \\
4U 1705-32 & \revise{1RXS J170854.4-321857} & burster \\
4U 1705-44 & & burster \\
4U 1708-40 & & burster \\
XTE J1710-281 & & burster \\
SAX J1712.6-3739 & & burster \\
RX J1718.4-4029 & & burster \\
% very faint
IGR J17254-3257 & & burster \\
4U 1722-30 & 4U 1724-307 & burster \\ 
4U 1728-34 & & burster \\
1RXH J173523.7-354013 & & burster  \\
% very faint
SLX 1735-269 & & burster \\
4U 1735-444 & & burster \\
SLX 1737-282 & & burster \\
1A 1742-294 & & burster \\
GX 3+1 & 4U 1744-26 &  burster \\ 
SLX 1744-299 & & burster \\
SLX 1744-300 & & burster \\
4U 1746-37 & & burster \\
%IGR J17597-2201 & XTE J1759-220 &  burster \\
GX 13+1 & V5512 Sgr, 4U 1811-17 &  burster \\ 
4U 1812-12 & & burster \\
GX 17+2 & NP Ser, 4U 1813-14 &  burster \\
4U 1820-303 & 4U 1820-30 &  burster \\
4U 1822-000 & 4U 1823-00 &  burster \\
2A 1822-371 & &  pulsar \\
GS 1826-24  &  4U 1826-24, GS 1826-238 &  burster \\ 
% V4634 Sgr,
XB 1832-330 & NGC 6652 A &  burster \\
Ser X-1 & 4U 1837+04, 3A 1837+049 &  burster \\
4U 1850-086 & 4U 1850-087 &  burster \\
XB 1916-053 & 4U 1916-05 & burster \\
M15 X-2  & 4U 2129+12\tablenotemark{a} &  burster \\
Cyg X-2 & 4U 2142+38 &  burster \\
\enddata
\caption{Known persistent low-mass X-ray binaries. These are the objects that appear to be persistent sources, from the MINBAR catalog of X-ray bursters \citep{Galloway2020}, along with  persistent X-ray pulsar LMXBs from \citet{PatrunoWatts}.}
\tablenotetext{a}{4U 2129+12 refers to both bright X-ray sources in M15. AC 211 in M15 is generally believed to contain a NS as well \citep{vanZyl04}, but we do not include it.}
\label{tab:persistent}
\end{deluxetable*}

\twocolumngrid 

\bibliography{outbursts}{}
\bibliographystyle{aasjournal}


\end{document}